\documentclass[twocolumn,floatfix,amsmath,amssymb,aps,prc,superscriptaddress,showpacs,preprintnumbers,nofootinbib]{revtex4-2}
\usepackage{graphicx,color}

\usepackage{bm,bbm}
\usepackage{comment}
\usepackage[colorlinks = true, linkcolor = blue]{hyperref}
\usepackage{color}
\usepackage{ulem}

\newcommand{\boat}[1]{{\color[rgb]{0.8,0.3,0.3}\ifmmode\text{\uwave{$#1$}}\else\uwave{#1}\fi}}

\newcommand{\Remove}[1]{{\color[rgb]{0.8,0.3,0.3}\ifmmode\text{\sout{$#1$}}\else\sout{#1}\fi}}

\DeclareMathOperator{\tr}{tr}

\newcommand{\phis}{\phi}
\newcommand{\phiv}{\phi}

\begin{document}
\title{
Covariant formulation of relativistic quantum molecular dynamics
for a system of interacting wave packets
}
\author{Yasushi Nara}
\affiliation{
Akita International University, Yuwa, Akita-city 010-1292, Japan}
\author{Asanosuke Jinno}
\affiliation{
Department of Physics, Faculty of Science,
Kyoto University, Kyoto 606-8502, Japan}

\author{Koichi Murase}
\affiliation{Department of Physics, Tokyo Metropolitan University,
Hachioji 192-0397, Japan}

\date{\today}
\preprint{KUNS-3064}
\begin{abstract}
We present a new formulation for the mean-field propagation part of
relativistic quantum molecular dynamics, simulating an $N$-body system of
interacting Gaussian wave packets via Lorentz scalar and vector potentials.
Covariant equations of motion are derived based on the principle of least
action with a weak form of mass-shell conditions and time-fixation constraints
defined with respect to a chosen foliation.
However, as is common with traditional relativistic quantum molecular dynamics,
the dynamics exhibits a residual dependence on the chosen foliation,
which is unavoidable because a finite number of interacting degrees of freedom
is not strictly compatible with relativity.
Nevertheless, we show that this dependence remains small for physically
reasonable choices of the foliation in practical applications.
By introducing a new approximation method to the spatial
integral in the equations of motion,
these covariant equations of motion can be solved with a computational
cost comparable to that of conventional noncovariant quantum molecular
dynamics.  Furthermore, the new equations of motion accurately estimate the
density-dependent potential, as demonstrated through comparison of the forces
with the numerical integration.  We apply them to $N$-body systems interacting
via the Skyrme-type potentials or the relativistic mean field to simulate
heavy-ion collisions.  Our results show that the derived equations of motion
provide a robust approximation to the dynamics of the full numerical
integrations.
\end{abstract}

\maketitle

\section{Introduction}

Microscopic transport models based on kinetic theories have been successful
tools for investigating heavy-ion collisions across a wide range of collision
energies.  The evolution of the one-body phase-space distribution function is
described by transport equations such as the Boltzmann--Uehling--Uhlenbeck
(BUU)~\cite{Bertsch:1984gb,Aichelin:1985zz,Kruse:1985hy,Kruse:1985pg,Bertsch:1988ik,Cassing:1990dr,Danielewicz:1991dh,SMASH:2016zqf}
and its relativistic version (RBUU), which is formulated within the
relativistic mean-field
theory~\cite{Ko:1987gp,Blattel:1988zz,Weber:1992qc,Blaettel:1993uz,Fuchs:1995fa,Cassing:2009vt,Buss:2011mx}.
Quantum molecular dynamics
(QMD)~\cite{Aichelin:1986wa,Aichelin:1991xy,Aichelin:2019tnk,Hartnack:1997ez,Bass:1998ca,Aichelin:2019tnk,Maruyama:1990zz,Maruyama:1997rp}
and its relativistic version
(RQMD)~\cite{Sorge:1989dy,Maruyama:1991bp,Maruyama:1996rn,Mancusi:2009zz,Marty:2012vs,Isse:2005nk}
follow the spacetime evolution of the $N$-body phase-space distribution,
enabling event-by-event simulations of heavy-ion collisions.  RQMD models
incorporating relativistic mean-field theory have also been
developed~\cite{Fuchs:1996uv,Nara:2019qfd,Nara:2020ztb}.  Furthermore, antisymmetrized
molecular dynamics (AMD)~\cite{Ono:1991uz,Ono:1992uy} takes into account the
fermionic nature of the nucleons.

These microscopic transport approaches include a Boltzmann-type collision term
and the mean-field propagation to describe the nonequilibrium dynamics of
nuclear reactions in a semiclassical way.  For detailed comparisons of these
transport models, we refer the reader to the series of TMEP collaboration
papers~\cite{TMEP:2016tup,TMEP:2017mex,TMEP:2019yci,TMEP:2021ljz,TMEP:2022xjg,TMEP:2023ifw}.

In this work, we focus on the development of the mean-field propagation in
transport models.  The BUU-type approach solves the time evolution of the
one-body phase-space distribution function by using the test particle method,
in which the distribution function is represented by an ensemble of
oversampled particles~\cite{Bertsch:1988ik}.  These test particles follow
Hamilton's equations of motion (EoMs).  In the limit of an infinite number of test
particles, the numerical solution converges to the exact solution.  In
practical numerical implementations of the BUU approach, a spatial grid is
often introduced to evaluate the local density and its derivatives.  However,
due to the limited number of test particles, unphysical numerical fluctuations
can arise, necessitating additional smearing of
the particle distribution~\cite{Cassing:1990dr,Sorensen:2020ygf,Oliinychenko:2022uvy}.
In the QMD approach, particles are represented by Gaussian wave packets, and their
centroid coordinates and corresponding momenta evolve in time following the
canonical EoMs.  In the QMD approach, the total momentum is exactly conserved
by construction in each event.  However, in the traditional QMD calculations,
an approximation is introduced in the evaluation of the density-dependent
potential for nonlinear terms.  This approximation results in a weaker gradient
of the repulsive mean-field potential~\cite{TMEP:2021ljz,TMEP:2023ifw}.
Consequently, the equation of state (EoS) of nuclear matter may not be
accurately reflected during the dynamical evolution of the system in QMD\@.
Several methods are proposed to improve the treatment of the density dependence
of the potential in
QMD~\cite{Maruyama:1990zz,Maruyama:1997rp,Aichelin:2019tnk}.

In both the BUU and QMD approaches, more accurate calculation methods have been
developed.  In the BUU framework, the lattice Hamiltonian
method~\cite{Lenk:1989zz,Persram:1999ec,Persram:2001dg,Wang:2019ghr}
was introduced to solve the EoMs for the test particles, in which excellent
energy and momentum conservation are achieved.  For QMD models, the
density-dependent potential is calculated exactly by numerical integration as
implemented in ImQMD-L~\cite{Yang:2021gwa} and AMD~\cite{Ono:1993ac}.  However,
no such calculations have been performed within relativistic models.

In this work, we propose a new consistent formulation of Lorentz-covariant EoMs
with the finite-size Gaussian wave packets interacting via the scalar and
vector potentials based on the manifestly covariant variational principle.  
Instead of imposing the mass-shell constraints on the centroids of the wave packets,
we impose the mass-shell constraints in a weak form where each mass-shell constraint is
satisfied after the phase-space integration over the Gaussian profile of the
corresponding wave packet.  This choice of the mass-shell constraints enables us
to derive the consistent form of EoMs throughout different levels of treatment
of the mean fields: We first derive the EoMs in the case of the mean fields
being externally specified as external fields in
Sec.~\ref{sec:covariant.external} and show that the dynamical-field case
produces consistent EoMs in Sec.~\ref{sec:covariant.dynamical}. This produces
the ``BUU-like'' covariant EoMs.  In Sec.~\ref{sec:covariant.static}, by
explicitly considering the field part of the action including the effective
self-interaction of the fields, we obtain the ``BUU-like'' EoMs also in the
case of the mean fields determined by local densities through the
self-consistent equations.

We also show that a ``BUU-like'' EoM, which is written with weighted spatial
integration of a potential (typically a single-particle potential), can be
transformed into ``QMD-like'' covariant EoMs, which is written with spatial
integration of density derivatives of an integrated potential per particle
(typically a one-particle potential energy~\cite{Nara:2021fuu}), in the case
that the mean fields are evaluated from local densities.  Since this is
equivalent transformation, the ``QMD-like'' EoMs maintain energy--momentum
conservation of the total system explicitly.  The density-dependent potential
written by spatial integration in the ``QMD-like'' EoMs will accurately be
calculated by applying the Monte Carlo integration with the relativistic
Gaussian weight.
This establishes a basis for a better estimation of the EoS than previous RQMD
approaches.

Based on this, to reduce the computational cost of the Monte Carlo integration
in evaluating the forces for each particle at every time step, we propose an
approximation to the spatial integration in the EoMs while controlling the
approximation error.
We demonstrate that the approximation provides a good agreement with the exact
solution obtained through numerical integration.  We note that our EoMs can
also be used to propagate test particles in the BUU and RBUU models.


Although it is not the primary purpose of this work, another important aspect
in telling the history of RQMD is the choice of the time fixations of temporal
coordinates appearing in the formulation.  In the RQMD-type approach, quantum
dynamics is approximated using a finite number of degrees of freedom, while an
infinite number of local degrees such as dynamical fields are integrated out.
To formulate relativistic $N$-particle system in a covariant manner, an
$8N$-dimensional phase space is introduced. In order to extract physical
trajectories, $2N$ constraints are imposed, consisting of $N$-mass shell
constraints and $N$-time fixation conditions.  The time fixations fix the
temporal coordinates of particles, but their choice is not unique.  Ideally,
since the choice of the time fixations is gauge degrees of freedom of
specifying temporal coordinates of particles, it should not affect the physics,
which is the time-reparametrization invariance.  However, the idea of RQMD
describing a system with a finite number of interacting degrees of freedom is
explicitly incompatible with relativity and prohibits the exact
time-reparametrization invariance because the interaction cannot be local with
a finite number of degrees of freedom located on different spatial points.

In this perspective, one of the interests in formulating RQMD has been the
design of reasonable time fixations that effectively capture the dynamics of
real systems, while keeping violation of time-reparametrization invariance
numerically small.  Different prescriptions for determining the time fixations
correspond to different models of RQMD.
In the original RQMD approach~\cite{Sorge:1989dy,Maruyama:1991bp}, the time
fixations are defined so that when two particles are close in space, their
times coincide in their center-of-mass frame.  Although the EoMs are formally
given under this condition, issues with this formulation become clear in 
numerical implementations~\cite{Marty:2012vs, Nara:2023vrq}.  For example,
at each two-body collision or decay, time-fixation conditions are violated,
requiring the solution of a set of $N$-dimensional non-linear equations.  Such
equations do not always admit solutions, or even when solutions exist,
unphysical backward evolution may occur.
Consequently, actual implementations are forced to introduce ad hoc
modifications~\cite{Maruyama:1991bp, Nara:2023vrq} at the EoM level.
Later, simpler time-fixation schemes were proposed
in Ref.~\cite{Maruyama:1996rn}, which
yields results similar to those obtained with the original constraints~\cite{Sorge:1989dy},
and have been widely adopted in subsequent
studies~\cite{Mancusi:2009zz, Marty:2012vs,Isse:2005nk,Nara:2019qfd,Nara:2020ztb,Nara:2021fuu}.
In these schemes, the times of all particles are set equal
in the global center-of-mass frame.

In the present paper, we introduce the spacetime foliation.
A foliation is the family of spacelike hypersurfaces that defines simultaneity.
The time-fixation condition is implemented as the constraint that assigns the particle coordinates
to one hypersurface in the chosen foliation.
Different choices of foliation may lead to residual foliation dependence.

It is useful to distinguish three related but different concepts.
First, Lorentz covariance means that the EoMs retain the same
form under Lorentz transformations when all four-vector quantities,
including the particle coordinates, momenta, and the foliation vector, are transformed consistently.
Second, Lorentz symmetry refers to the invariance of the physical dynamics
under Lorentz transformations without changing the prescription that defines the theory.
Third, the dependence on the choice of spacetime foliation is conceptually distinct
from both Lorentz covariance and Lorentz symmetry.
A formulation may, in general, exhibit a residual dependence on
the chosen foliation while remaining Lorentz covariant.

The prescriptions of Sorge~\cite{Sorge:1989dy,Maruyama:1991bp}
 and others~\cite{Sudarshan:1981pp,Samuel:1982jk,Samuel:1982jn,Marty:2012vs}
 define the time-fixation condition
through dynamical variables,
although they have other well-known drawbacks such as difficulties with cluster separability
or practical implementation.
Maruyama's prescription~\cite{Maruyama:1996rn} provides a simpler alternative by employing
an externally specified foliation vector.
Consequently, the resulting dynamics depends on the chosen foliation.

In the present work, we adopt the time-fixation prescription of Maruyama.
The EoMs are Lorentz covariant
in the sense that all four-vector quantities,
including the foliation vector, are transformed consistently under Lorentz transformations
and the equations retain the same form in any inertial frame.
In interpretations where the foliation vector is regarded as
an externally specified background structure,
the residual foliation dependence corresponds to a violation of Lorentz symmetry.
The present work does not attempt to construct a completely
foliation-independent formulation.
Instead, our objective is to establish a Lorentz-covariant formulation
for a specified foliation and to quantify the residual foliation dependence.
To this end, we perform a quantitative study of the foliation dependence
and show that it remains weak for physically reasonable choices of the foliation.
This residual foliation dependence should therefore be regarded as a controlled model dependence.

A covariant cascade method has been developed to simulate the Boltzmann-type
collision term based on the constrained Hamiltonian formulation in
Ref.~\cite{Nara:2023vrq}.  Combining this covariant cascade method with the new
mean-field EoMs enables us to perform event-by-event simulations of the
spacetime evolution of heavy-ion collisions in a fully covariant way within
the RQMD approach.  We have implemented our new EoMs into the event generator
\texttt{JAM2}~\cite{JAM2}, and performed simulations of heavy-ion collisions,
which are compared with the results from the previous RQMD\@.

This paper is organized as follows.  In Sec.~\ref{sec:qmd}, we introduce the
new QMD EoMs, which provide a good approximation to full integration and
exactly conserve total momentum.  In Sec.~\ref{sec:covariant}, we derive the
covariant EoMs based on the variational principle.  After briefly
summarizing the EoMs in Sec.~\ref{sec:numerics}, the covariant EoMs are applied
to the $\sigma$-$\omega$ type relativistic mean-field theory in
Sec.~\ref{sec:rmf} and Lorentz-vector Skyrme-type potentials in
Sec.~\ref{sec:skyrme}\@.  We present the results of our RQMD simulations for
Au+Au collisions in Secs.~\ref{sec:numericaltest.rmf}
and~\ref{sec:numericaltest}\@.  Finally, a summary is provided in
Sec.~\ref{sec:summary}.

\section{Mean-field part of quantum molecular dynamics}
\label{sec:qmd}

We first discuss the nonrelativistic case.
We consider the system of $N$ nucleon system interacting with the density $n$ dependent
single-particle potential $U(n)$ by the following Hamiltonian,
\begin{equation}
 H = \frac{g_N}{(2\pi)^3}\int d^3x d^3p \frac{\bm{p}^2}{2m}f(\bm{x},\bm{p}) + \int d^3x \int_0^{n(\bm{x})} U(n)dn,
 \label{eq:Hnonrel}
\end{equation}
where $g_N$ is the degeneracy factor for spin
and isospin, and we assume the Skyrme-type single-particle potential,
\begin{equation}
 U(n)= \alpha\frac{n}{n_0}
  + \beta\left(\frac{n}{n_0}\right)^\gamma.
  \label{eq:single-particle-potential}
\end{equation}

In the quantum molecular dynamics (QMD)
approach~\cite{Aichelin:1991xy,Aichelin:2019tnk,Hartnack:1997ez,Bass:1998ca,Aichelin:2019tnk},
the total $N$-body wave function is assumed to be a direct product of Gaussian
wave packets.  Hamilton's EoMs of the $N$-body system are obtained from
the variational principle.  The zero-point kinetic energy appears due to
Gaussian wave packets satisfying the uncertainty principle.  In QMD, this
zero-point kinetic energy is often dropped, which means that the Gaussian in
the momentum space is replaced by the $\delta$-function~\cite{Ono:2019jxm}.  We
take the distribution function to be the sum of a Gaussian wave packet,
\begin{equation}
 f(\bm{x},\bm{p}) = \frac{(2\pi)^3}{g_N}\sum_{i=1}^N
 g(\bm{x}-\bm{x}_i) \delta(\bm{p}-\bm{p}_i).
\label{eq:QMDGauss}
\end{equation}
The Gaussian for a particle is defined as
\begin{equation}
  g(\bm{x}-\bm{x}_i)=\left(\frac{1}{2\pi L}\right)^{3/2} \exp\biggl(-\frac{[\bm{x}-\bm{x}_i(t)]^2}{2L}\biggr).
\end{equation}
The local density $n(\bm{x})$ in Eq.~\eqref{eq:Hnonrel} is evaluated by the sum of Gaussian profiles,
\begin{equation}
  n(\bm{x})=\frac{g_N}{(2\pi)^3} \int d^3p f(\bm{x},\bm{p})
 = \sum_{i=1}^N g(\bm{x}-\bm{x}_i).
\label{eq:particledensity}
\end{equation}
Hamilton's equation for the $i$th particle yields
\begin{equation}
 \frac{d\bm{x}_i}{dt} =  \frac{\bm{p}_i}{m_i}, \qquad
 \frac{d\bm{p}_i}{dt}
   = - \int d^3x\, U\big(n(\bm{x})\big) \frac{\partial g(\bm{x}-\bm{x}_i)}{\partial\bm{x}_i}.
\label{eq:ImQMDL}
\end{equation}
After integrating Eq.~\eqref{eq:ImQMDL} by parts or taking the differentiation of the
Gaussian, the equation can be viewed as an integral with the Gaussian weight.
In the harmonic approximation~\cite{Fuchs:1995fa}, the integration of a
function $F(\bm{x})$ with a Gaussian weight is estimated by the value of the
function at the center of the Gaussian:
\begin{align}
 \label{eq:HarmonicApprox}
 \int d^3 x F(\bm{x}) g(\bm{x}-\bm{x}_i) \approx F(\bm{x}_i).
\end{align}
This approximation yields the \textit{BUU-like} equation
within the QMD framework
\begin{equation}
 \frac{d\bm{p}_i}{dt}
  =  -\sum_{j=1}^N \frac{dU(n(\bm{x}_i))}{dn} \frac{\partial
g(\bm{x}_i-\bm{x}_j)}{\partial\bm{x}_i}.
  \label{eq:BUU-like}
\end{equation}
This EoM is the same as the one in the relativistic
Landau-Vlasov method~\cite{Fuchs:1995fa} in the nonrelativistic limit.
In this approach,  a particle's momentum changes by the mean-field potential $U(n)$ at its
position.

We now consider the QMD framework.
The key part of this study includes a new method to evaluate the potential
integral in Eq.~\eqref{eq:ImQMDL} numerically.  Different existing models of
QMD and BUU can be concisely summarized as the respective methods to evaluate
the potential integral.  After reviewing the existing methods, we propose an
efficient and accurate way to evaluate the potential integral.

We first note that Eq.~\eqref{eq:ImQMDL} includes the self-interaction of the
$i$th particle through $n(\bm{x})$.  In the Hartree-Fock method, where the
wave function for fermions is antisymmetrized, the self-interaction vanishes
due to the cancellation of the Hartree and Fock terms.  This is not enforced in the
QMD framework, where the wave function is given by a direct product of Gaussian
wave packets.  To partially take account of the antisymmetrization of the
fermion wave function, the self-interaction is typically excluded by hand%
~\footnote{The actual EoMs are unaffected by this replacement only in the case
  of linear $U(n)$ [i.e., $\beta=0$ in
  Eq.~\eqref{eq:single-particle-potential}].  In general, the exclusion of the
  self-interaction affects the EoMs in the presence of the nonlinear potential
  term, which originates in microscopic three-body (or many-body) forces.},
replacing $n(\bm{x})$ with $n(\bm{x}) - g(\bm{x}-\bm{x}_i) = \sum_{j(\ne i)}
g(\bm{x} - \bm{x}_j)$ in Eq.~\eqref{eq:ImQMDL}.
However, this manual tweak of the EoMs introduces a subtlety in extension to
the relativistic case.  The fact that each particle feels a distinct density
$n(\bm{x}) - g(\bm{x}-\bm{x}_i) = \sum_{j(\ne i)} g(\bm{x} - \bm{x}_j)$
($i=1,\dots,N$) implies that there are effectively $N$-independent local
densities, which is
incompatible with the relativistic picture of the interaction mediated by
common dynamical fields as discussed in Sec.~\ref{sec:rmf}\@.
To avoid subtleties arising from the manual tweak and to be more consistent
with the dynamical-field case, we fully consider the self-interaction in the
main part of this study following the approach of the ImQMD
model~\cite{Zhang:2014sva,Zhang:2020dvn}.

The QMD model uses the one-particle potential $V(n)$~\cite{Nara:2021fuu}, which
is related with the single-particle potential
by~\cite{OmanaKuttan:2022the,Steinheimer:2022gqb}
\begin{equation}
  V(n)  = \frac{1}{n}\int U(n)dn.
 \label{eq:one-particle-potential}
\end{equation}
Using this one-particle potential, the Hamiltonian~\eqref{eq:Hnonrel} can be rewritten as
\begin{align}
 H = \frac{g_N}{(2\pi)^3}\int d^3x d^3p\, \frac{\bm{p}^2}{2m}f(\bm{x},\bm{p}) + \int d^3x\, n V(n).
\label{eq:Hqmd}
\end{align}
In the traditional QMD,
the potential term in the Hamiltonian~\eqref{eq:Hqmd} is evaluated by
what we refer to as the \textit{QMD approximation}.
In this approximation, the spatial integral of the density-dependent potential becomes
\begin{equation}
 \int d^3x\, n  V(n)
   = \sum_{i=1}^N \int d^3x\, g(\bm{x}-\bm{x}_i) V(n)
  \approx \sum_{i=1}^N V(\langle n\rangle_i ),
  \label{eq:qmdapp}
\end{equation}
where the \textit{interaction density} $\langle n\rangle_i$ is defined as the sum of Gaussian overlap,
\begin{equation}
  \langle n\rangle_i = \sum_{j\ne i}^N \int d^3x\, g(\bm{x}-\bm{x}_i)
  g(\bm{x}-\bm{x}_j)
 = \sum_{j\ne i}^N g_{ij},
\label{eq:interaction_density}
\end{equation}
where
\begin{equation}
  g_{ij} = \frac{1}{(4\pi L)^{3/2}}\exp\Bigl[-\frac{(\bm{x}_i-\bm{x}_j)^2}{4L}\Bigr].
\end{equation}
The approximation~\eqref{eq:qmdapp} is symbolically stated as $\langle
n^\gamma\rangle\approx\langle n\rangle^\gamma$, with $\langle A\rangle := \int
d^3x'\,g(\bm{x}'-\bm{x})A(\bm{x}')$, for a term in $V(n)$ proportional to
$n^\gamma$ (see also Eq.~(44) of Ref.~\cite{Nara:2021fuu} and Eq.~(11) of
Ref.~\cite{Maruyama:1997rp}).
Within this QMD approximation, the QMD EoM
for the $i$th particle leads to
\begin{equation}
 \frac{d\bm{p}_i}{dt}
= -\sum_{j\ne i}^N
  \biggl[\frac{dV(n)}{dn}\biggr|_{n = \langle n\rangle_i}
 + \frac{dV(n)}{dn}\biggr|_{n = \langle n\rangle_j}
  \biggr]
   \frac{\partial g_{ij}}{\partial\bm{x}_i}.
\label{eq:QMDEoM}
\end{equation}
The QMD EoMs take the form of the symmetric interaction between two particles
so that the total momentum is exactly conserved.  However, the
density-dependent potential is not correctly estimated due to the approximation
$\langle n^\gamma\rangle\approx\langle n\rangle^\gamma$ in the nonlinear case
$\gamma \ne 1$.

In this paper, ``BUU-like'' EoMs are defined as the ones using the
single-particle potential $U(n)$,
such as Eqs.~\eqref{eq:ImQMDL} and~\eqref{eq:BUU-like}, while
``QMD-like'' EoMs are defined as the ones using the one-particle potential
$V(n)$, such as Eqs.~\eqref{eq:QMDEoM}, \eqref{eq:qmdlike},
and~\eqref{eq:ImQMDLmod}.

It is practically possible to integrate the EoM~\eqref{eq:ImQMDL} numerically.
The lattice-Hamiltonian method in the BUU approach~\cite{Lenk:1989zz} divides
the coordinate space into cells $\Delta x$ and evaluates the derivatives as
\begin{equation}
 \int d^3x\, U(n)\frac{\partial g(\bm{x}-\bm{x}_i)}{\partial\bm{x}_i} \approx
\sum_\alpha (\Delta x)^3\, U(n_\alpha)
 \frac{\partial g(\bm{x}_\alpha - \bm{x}_i)}{\partial\bm{x}_i},
 \end{equation}
where $\alpha$ is a grid label, and $n_\alpha$ is the density in the grid cell $\alpha$. The lattice-Hamiltonian method typically uses a
triangle profile for $g(\bm{x}-\bm{x}_i)$.  The energy-momentum has been confirmed to be
conserved well in this method~\cite{Lenk:1989zz,Persram:1999ec}.  Within the
QMD approaches, the numerical integration with the Gauss-Legendre quadrature
was performed in ImQMD-L~\cite{Yang:2021gwa}.  The Monte Carlo integration is
used in the AMD model~\cite{Ono:1993ac}.  The Monte Carlo integration of
Eq.~\eqref{eq:ImQMDL} can be performed as
\begin{align}
\frac{d\bm{p}_i}{dt} &= -\int d^3x\,
U(n(\bm{x}))\frac{\bm{r}-\bm{r}_i}{L}g(\bm{x}-\bm{x}_i) \nonumber\\
&\approx -\frac{1}{N_\mathrm{MC}}\sum_{k=1}^{N_\mathrm{MC}}
U(n(\bm{x}_k))\frac{\bm{r}_k-\bm{r}_i}{L},
\label{eq:MCQMD}
\end{align}
where we sample the coordinates $\bm{x}_k$ according to the Gaussian
distribution $g(\bm{x}-\bm{x}_i)$, and $N_\mathrm{MC}$ is the number of Monte Carlo
sampling points.

Since the integration is numerically expensive, we consider another approximation
for the potential integration.  Instead of using the single-particle potential
in Eq.~\eqref{eq:ImQMDL}, we use one-particle potential: from the Hamiltonian
\eqref{eq:Hqmd}, we obtain
\begin{align}
  \frac{d\bm{p}_i}{dt}
  &=-\int d^3x \biggl[V(n) + n\frac{dV(n)}{dn}\biggr]
  \frac{\partial g(\bm{x}-\bm{x}_i)}{\partial\bm{x}_i}.
 \label{eq:qmdlike}
\end{align}
After integrating the first term by parts and replacing the total density $n(x)$
by the sum of the Gaussian~\eqref{eq:particledensity}, we obtain
\begin{align}
\frac{d\bm{p}_i}{dt}&=
-\sum_{j=1}^N\int d^3x
  \biggl[\frac{dV(n)}{dn}\frac{\partial
g(\bm{x}-\bm{x}_j)}{\partial\bm{x}} g(\bm{x}-\bm{x}_i) \nonumber\\
 &+ \frac{dV(n)}{dn} \frac{\partial
g(\bm{x}-\bm{x}_i)}{\partial\bm{x}_i} g(\bm{x}-\bm{x}_j)
  \biggr].
\label{eq:ImQMDLmod}
\end{align}
We now approximate the derivatives of the potential term at the center of the
Gaussian, and then take the Gaussian overlap integration, which yields the
EoMs,
\begin{equation}
 \frac{d\bm{p}_i}{dt}
\approx -\sum_{j=1}^N
  \left[\frac{dV(n(\bm{x}_i))}{dn}
 + \frac{dV(n(\bm{x}_j))}{dn}
  \right]
   \frac{\partial g_{ij}}{\partial\bm{x}_i}.
   \label{eq:ourQMDEoM}
\end{equation}
We refer to this procedure as the \textit{QMD2 approximation}.
These equations may be regarded as a variant of the Monte Carlo sampling with
two sample points, where the points are fixed to be the representative points
in Eq.~\eqref{eq:ImQMDLmod}, $\bm{x}_i$ and $\bm{x}_j$.  These equations have the same
structure as the EoMs for the traditional QMD approximation~\eqref{eq:QMDEoM}.
They have the interaction term from the position of the other particles, and
the total momentum is exactly conserved.  These equations are exact for
$\gamma=1$, and as shown below, a good approximation for the density-dependent
potential term is achieved for $\gamma \ne 1$.  We note that the derivative of
the one-particle potential is obtained by
\begin{equation}
  \frac{dV(n)}{dn}= \frac{1}{n}[U(n) - V(n)].
\end{equation}

\begin{figure}[tbhp]
\includegraphics[width=8.5cm]{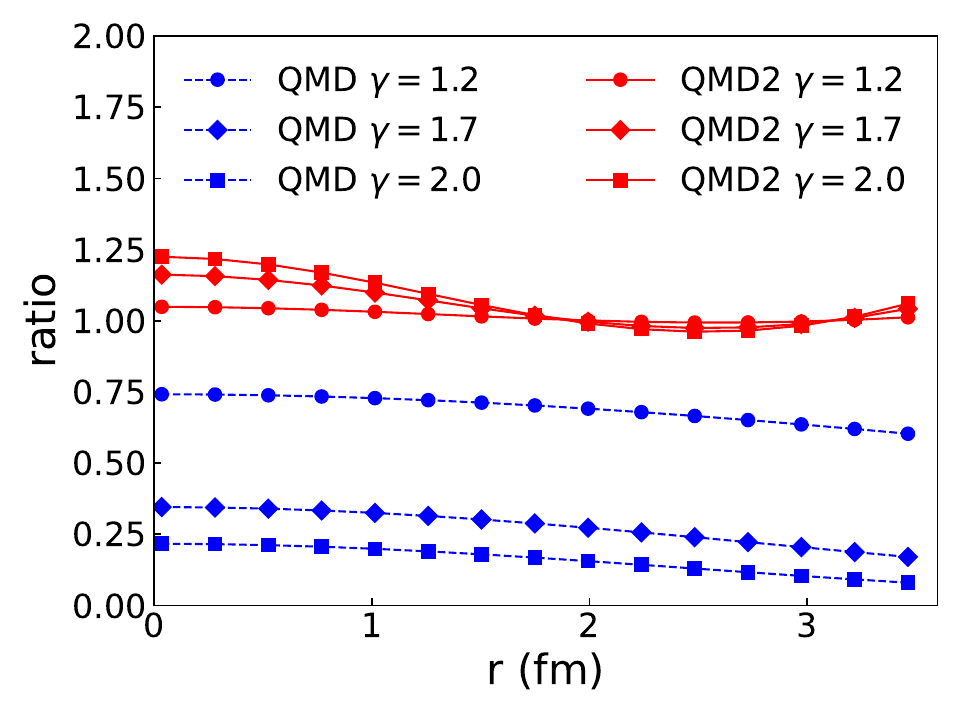}
\caption{Ratios of the absolute value of the force for the potential
$U(n)=n^\gamma$ to the numerical integration in the two-particle case are
plotted as a function of the separation between two Gaussians.  The results
from the traditional QMD are shown by the dashed lines, while those from the
new method (denoted as QMD2) are shown by the solid lines.  A Gaussian width
$L=2.0~\text{fm}^2$ is used.  The circles, diamonds, and squares correspond to
results for $\gamma=1.2$, $1.7$, and $2.0$, respectively.  }
\label{fig:gauss2}
\end{figure}

For the potential $U=n^\gamma$, the traditional QMD EoM becomes
\begin{equation}
 \frac{d\bm{p}_i}{dt}
 = -\sum_{j\neq i}^N \frac{\gamma}{\gamma+1}\left(\langle n\rangle_i^{\gamma-1}
  + \langle n\rangle_j^{\gamma-1}\right)\frac{\partial g_{ij}}{\partial\bm{x}_i}.
\label{eq:QMDg}
\end{equation}
The difference between the traditional QMD EoM and our EoM is the following.  The
traditional QMD approach uses the interaction
density~\eqref{eq:interaction_density}, while we use the particle
density~\eqref{eq:particledensity}.  Thus, for the potential $U=n^\gamma$, the
explicit expression of our EoM~\eqref{eq:ourQMDEoM} is obtained by replacing
the interaction density $\langle n\rangle_i$ in Eq.~\eqref{eq:QMDg} with the
particle density $n(\bm{x}_i)$:
\begin{equation}
 \frac{d\bm{p}_i}{dt}
 = -\sum_{j\neq i}^N \frac{\gamma}{\gamma+1}
  \left[n(\bm{x}_i)^{\gamma-1} + n(\bm{x}_j)^{\gamma-1}\right]
  \frac{\partial g_{ij}}{\partial\bm{x}_i}.
\label{eq:QMD2g}
\end{equation}

In Fig.~\ref{fig:gauss2}, we compare the ratio of the absolute value of the
force to the results from the numerical integration~\eqref{eq:MCQMD} for the
potential $U(n)=n^\gamma$ in a two-particle system for QMD~\eqref{eq:QMDg} and
a new QMD (QMD2)~\eqref{eq:QMD2g}.  The traditional QMD significantly
underestimates the absolute value of force for large $\gamma$,
mainly because of the absence of self-interaction (i.e.,
the contribution from the wave packet of itself in the density evaluation), which is relevant in the two-particle
system.  Indeed, when self-interaction is included in the evaluation of the
density in QMD, the ratio improves significantly from about 0.2 to 0.8 in the $\gamma=2$ case.
In contrast, our method, denoted by ``QMD2'', provides a good
approximation, as confirmed by comparison with numerical integration using the
Monte Carlo integration with the Gaussian weight~\eqref{eq:MCQMD}.  We have
checked that the Monte Carlo integration yields results practically identical
to those obtained with the Gauss-Legendre quadrature method, while the
Monte Carlo integration is much faster than the Gauss-Legendre quadrature
method.

\begin{figure}[tbhp]
\includegraphics[width=8.5cm]{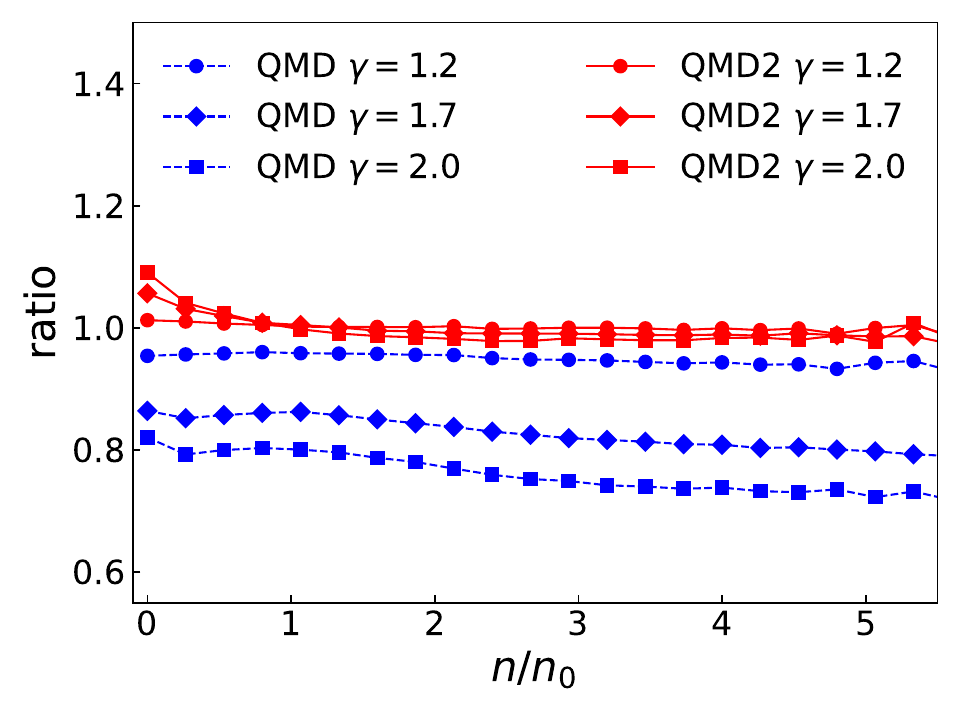}
\caption{ Same as Fig.~\ref{fig:gauss2} but as a function of normalized baryon
density for a system of 400 particles sampled randomly inside a sphere.
}
\label{fig:gauss3}
\end{figure}
In the case of heavy-ion collisions such as Au + Au collisions, the total
number of nucleons is typically 400.  To mimic such situations, we sample 400
nucleons inside a sphere, where the density of the system is controlled by the
radius of the sphere.  Figure~\ref{fig:gauss3} represents the baryon density
dependence of the ratio of the absolute value of the force to the Monte Carlo
integration result for the 400-nucleon system.
We randomly chose one particle and computed its force and average over
events.
We compare the traditional QMD
results with our results.  Our formula is found to provide a good approximation
over a wide range of densities.
We have checked that when particles  are chosen from the center or surface,
our approximation method still performs well.
The traditional QMD results agree well with
the exact solution for $\gamma=1.2$, which corresponds to the value typically
used for a soft EoS\@.  However, for larger values of $\gamma$, the traditional QMD
underestimates by about 20\%, in contrast to the two-particle system.  This is
because the self-interaction is less relevant in systems with a large
number of particles.
When self-interaction is included in QMD, the results does not change
much for the 400 particle case.
The main reason for the underestimation in the QMD model
is the use of the interaction density when evaluating the local density.  Note
that the Gaussian width of the interaction
density~\eqref{eq:interaction_density} is $\sqrt{2}$ times larger than that
of the local density~\eqref{eq:particledensity}.

\begin{figure}[tbhp]
\includegraphics[width=8.5cm]{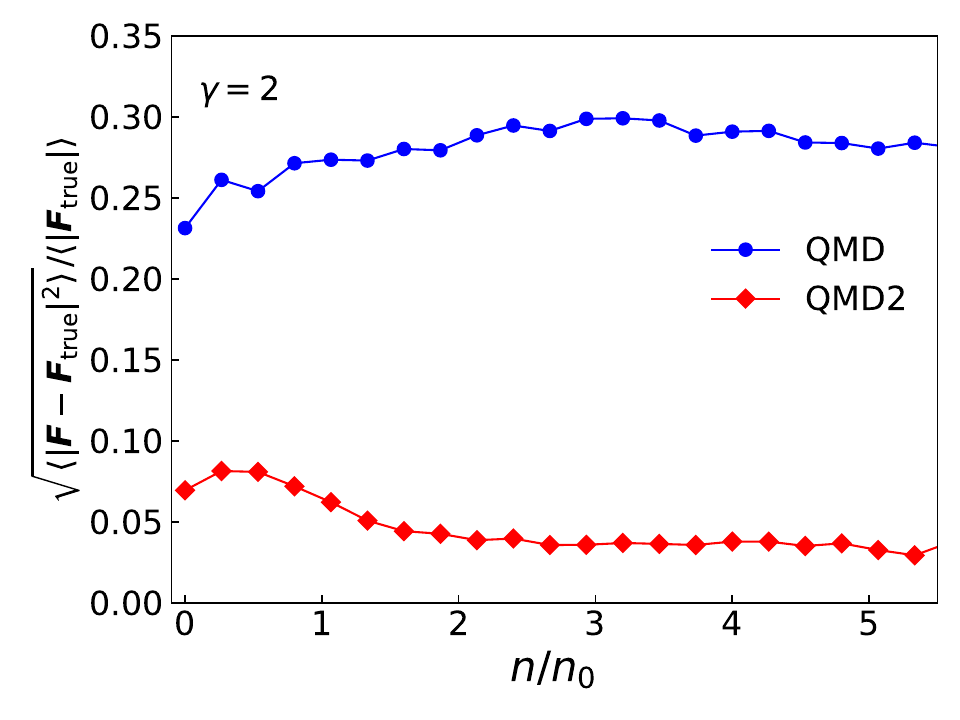}
\caption{ Same as Fig.~\ref{fig:gauss3} but
the root-mean-square deviation of the force is plotted.
}
\label{fig:rmsd}
\end{figure}
To further see the accuracy of the approximation,
we plot in Fig.~\ref{fig:rmsd}, the relative root-mean-square deviation (RMSD),
\begin{equation}
  \frac{\sqrt{\langle |\bm{F}-\bm{F}_\mathrm{true}|^2 \rangle}}{\langle |\bm{F}_\mathrm{true}|\rangle},
\end{equation}
as a function of the baryon density for the potential parameter $\gamma=2$.
One can see that the event-by-event deviation observed with the QMD approximation of
the spatial integral is significantly reduced by the new QMD2 approximation.


\section{Variational formulation of RQMD}
\label{sec:covariant}

We present a new covariant formulation of the mean-field part of the EoMs
in RQMD, which extends QMD to systems of relativistically
interacting wave packets via the scalar and vector potentials.

In this work, we provide a manifestly covariant formulation of RQMD based on
the variational principle.  Specifically, we adopt the modified Hamilton's
principle (also known as the Weiss action principle), which is the variational
principle within the Hamiltonian formalism to ensure Lorentz covariance.  A
more detailed discussion of this formulation, including its relation to the
standard Hamiltonian formulation and Dirac's constrained Hamiltonian method,
will be presented in a separate paper.

For simplicity, we consider a single species of particles; however, the
extension to multispecies is straightforward.  We study an $N$-particle system
interacting through the single-particle Lorentz scalar and vector potentials,
$U_s(x,p)$ and $U^\mu(x,p)$, which generally depend on the eight-dimensional
phase space $(x,p)$, where $x^\mu$ and $p^\mu$ are the spacetime coordinates
and the corresponding canonical momentum. The effective mass $m^*(x, p)$ and
kinetic momentum $p^*(x, p)$ are defined as
\begin{align}
  m^{*}(x, p) &= m + U_s(x, p),
  \label{def:effective-mass} \\
  p^{*\mu}(x, p) &= p^\mu - U^\mu(x, p),
  \label{def:kinetic-momentum}
\end{align}
where $m$ is the bare mass. We discuss three different treatments of the
single-particle potentials: the cases with ``external fields,''
``dynamical fields,'' and ``local density approximation'' in
Secs.~\ref{sec:covariant.external}--\ref{sec:covariant.static}.  The
actual meanings of the three terms will be explained in the respective
sections.

\subsection{External fields}
\label{sec:covariant.external}

For the external-field case, we assume the following form of the Weiss action
for an interacting $N$-particle system:
\begin{multline}
  S_\mathrm{part}[\{x_i, p_i\}_i] =
  \sum_{i=1}^N \int p_i(s) \cdot \frac{dx_i(s)}{ds} ds \\
    - \int d^4xd^4p\, W(x,p)f(x,p),
  \label{eq:covariant.action.part}
\end{multline}
where $W(x, p)$ is the \textit{generalized potential}~\cite{Weber:1992qc}:
\begin{equation}
  W(x,p) = \frac{p^{*2}(x,p) - m^{*2}(x,p)}{2}.
  \label{def:W}
\end{equation}
The physical state of $f(x,p)$ is constrained by the mass-shell
condition~\cite{Weber:1992qc,Blaettel:1993uz} as
\begin{equation}
  W(x,p) f(x, p) = 0, \qquad\text{for any $x, p \in \mathbb{R}^{3,1}$},
  \label{eq:mass-shell-constraintW}
\end{equation}
which requires $f(x,p)$ to vanish outside the physical hypersurface specified
by $W(x,p)=0$.  However, Eq.~\eqref{eq:mass-shell-constraintW} is too strong
when we approximate the distribution function $f(x, p)$ with a parametrized
form such as the Gaussian form.  Equation~\eqref{eq:mass-shell-constraintW} imposes an
infinite number of constraints, which cannot be satisfied in general by
adjusting only a finite number of parameters $\{\bm{x}_i,p_i\}_i$, unless the
parametrized form is intentionally designed to satisfy
Eq.~\eqref{eq:mass-shell-constraintW} in advance.  Later in this section, we
will find a weaker form of the mass-shell
constraints~\eqref{eq:mass-shell-constraint.weak}.

The action~\eqref{eq:covariant.action.part} is a functional of the phase-space
particle trajectories $\{x_i(s), p_i(s)\}_{i=1}^N$, where the particle
trajectories are the functions of a Lorentz-invariant evolution parameter $s$.
In the Weiss action principle, independent variations of four-coordinates
$x_i(s)$ and four-momenta $p_i(s)$ are considered.  It is straightforward to
extend the action for the case of multiple particle species by giving
$S_\mathrm{part}$ as the sum over the different species: $\sum_B
S^B_\mathrm{part}$.

To formulate the dynamics based on the variational principle for the
action~\eqref{eq:covariant.action.part}, we first specify the phase-space
density $f(x, p)$ as a functional of the phase-space trajectories $\{x_i(s),
p_i(s)\}_i$.  We assume that the distribution function $f(x ,p)$ is expressed
by a sum of single-particle distributions:
\begin{align}
  f(x,p) &= \sum_{i=1}^N f_i(x,p).
  \label{def:f-as-fi-sum}
\end{align}
Here, we define the single-particle phase-space density $f_i(x,p)$ for the
$i$th particle as a functional of its phase-space trajectory $(x_i(s),
p_i(s))$:
\begin{align}
  f_i(x, p) =
  \int ds\, \lambda_i(s) G(x, p; x_i(s), p_i(s)),
  \label{def:f.with-ds-v2}
\end{align}
with
\begin{align}
  G(x, p; x_i, p_i)
  := \delta\bigl(\hat{s}(x) - \hat{s}(x_i)\bigr)
  g(x, p; x_i, p_i),
  \label{def:gi}
\end{align}
where $\hat{s}(x)$ is a Lorentz-scalar field~\footnote{%
  The statement that $\hat{s}(x)$ is a Lorentz-scalar field means that
  the scalar field $\hat{s}(x)$ experiences the proper transformation
  $\hat{s}(x)\mapsto \hat{s}'(x') = \hat{s}(\Lambda^{-1} x')$ on the coordinate
  transformation for the change of the reference frame $x\mapsto x' = \Lambda
  x$, where $\Lambda$ is a Lorentz transformation.  It also means that the
  field $\hat{s}(x)$ undergoes the proper transformation $\hat{s}(x)\mapsto
  \hat{s}'(x') = \hat{s}(x - \epsilon)$ for the translation transformation
  $x^\mu \mapsto x'^\mu = x^\mu + \epsilon^\mu$, with $\epsilon^\mu$ being a
  constant displacement.  In this sense, strictly speaking, it may be more
  appropriate to call it a Poincar\'e-scalar field instead of a Lorentz-scalar
  field, but we here simply refer to it as a Lorentz-scalar field by
  convention.}
externally specified to foliate
spacetime into a series of spacelike equal-time hypersurfaces: $\Sigma_s =\{x
\in \mathbb{R}^{3,1} \mid \hat{s}(x) = s\}$.  The foliation dependence of the
resulting dynamics will be discussed later in
Sec.~\ref{sec:covariant.foliation}.  The proper choice of the functional
form of the particle profile $g(x, p; x_i, p_i)$ depends on the choice of
$\hat{s}(x)$, as the normalization must hold on each hypersurface $\Sigma_s$:
\begin{align}
  \int d^4x d^4p\,
  \delta(\hat{s}(x) - s)
  g(x, p; x', p') = 1,
\end{align}
for any $s \in \mathbb{R}$, $p' \in \mathbb{R}^{3,1}$, and $x' \in \Sigma_s$.
The undetermined coefficient $\lambda_i(s)$ will be specified later to ensure
that $x_i^\mu(s)$ lies on the hypersurface $\Sigma_s$ of the same time $s$,
i.e., $\hat{s}(x_i(s)) = s$~\footnote{
  Using this dynamical constraint $\hat{s}(x_i(s)) = s$ retroactively, one
  might think it would be useful to simplify the $\delta$ function in
  Eq.~\eqref{def:f.with-ds-v2} as $\delta(\hat{s}(x) - s)$, and thus $f_i(x, p)
  = \lambda(\hat s(x))g(x,p;x_i(\hat s(x)), p_i(\hat s(x)))$.  However, if such
  rewriting is performed on $f_i(x,p)$ in the action, then it breaks the dynamics
  because it effectively introduces an explicit dependence on the time
  parameter $s$ into the Hamiltonian, i.e., temporal translation symmetry is
  broken.  For example, the weak mass-shell
  constraint~\eqref{eq:mass-shell-constraint.weak} is broken by an extra term
  $\partial \bar W_i/\partial s$ in Eq.~\eqref{eq:mass-shell-constraint.proof}.
}.
The form~\eqref{def:f.with-ds-v2} of the distribution function can be regarded
as an extension of the one in Ref.~\cite{Fuchs:1995fa}, with the identification
$\hat s(x) = x \cdot u_i$, but an important difference is that we use a single
common time function $\hat s(x)$ for all the particles to foliate the spacetime
consistently in the presence of interaction between particles.

As in Sec.~\ref{sec:application} for the application, the particle profile
is often taken to be the Lorentz-contracted Gaussian
form~\eqref{eq:gx}--\eqref{eq:vardyn.contractedGauss}, which depends on the
velocity of the particle $u_i$.  In the noninteracting case, the velocity can
be written as $u_i^\mu = p_i^\mu/m_i$, where $m_i = \sqrt{p_i^\mu p_{i\mu}}$.
However, in the interacting case, the expression of the velocity $u_i$ as a
function of $(x_i,p_i)$ is unknown until we obtain EoMs, which implies the
necessity of the special treatment of the $u_i$ dependence of the profile
$g(x-x_i,p-p_i;u_i)$ in the derivation step of the EoMs.  A full treatment of
the $u_i$ dependence would require the modification in the variational
principle because the ``Hamiltonian'' becomes a function of the velocity $\dot
x_i$ in addition to $x_i$ and $p_i$.  In the present study, we ignore the $u_i$
dependence of the profile in the derivation step, as if we already know the
expression of the velocity in terms of $(x_i, p_i)$.  As will be discussed
in~Sec.~\ref{sec:application}, this approximate treatment leaves a small problem in
the derived EoMs, but it is expected to be negligible in the slow-acceleration
regime.  A full treatment of the variation with explicit velocity dependence is
left for future studies.

The principle of least action yields the following Hamilton's EoMs for the
$i$th particle:
\begin{align}
  \frac{dx^\mu_i}{ds} = \frac{\partial H}{\partial p_{i\mu}}, \qquad
  \frac{dp^\mu_i}{ds} = -\frac{\partial H}{\partial x_{i\mu}},
  \label{eq:WeissEoM}
\end{align}
where the Hamiltonian is identified as
\begin{align}
  H(\{x_i, p_i\}_i) &= \sum_{i=1}^N \lambda_i \bar W(x_i, p_i),
  \label{eq:constraintH} \\
  \bar W(x_i, p_i) &= \int d^4x d^4p\, W(x, p)
    G(x, p; x_i, p_i).
  \label{eq:Weiss.weak-mass-shell}
\end{align}
Thus, the EoMs can be transformed into
\begin{align}
  \frac{dx^\mu_i}{ds} = \lambda_i\bar\Pi_i^\mu, \qquad
  \frac{dp^\mu_i}{ds} = -\lambda_i\bar Q_i^\mu,
  \label{eq:covariant.eom}
\end{align}
where
\begin{align}
  \bar\Pi_i^\mu :=
    \int d^4x d^4p\, W(x, p) \frac{\partial G(x, p; x_i, p_i)}{\partial p_{i\mu}},
  \label{def:barPi} \\
  \bar Q_i^\mu  :=
    \int d^4x d^4p\, W(x, p) \frac{\partial G(x, p; x_i, p_i)}{\partial x_{i\mu}}.
  \label{def:barQ}
\end{align}
The coefficients $\lambda_i$ are determined to satisfy
\begin{align}
  \hat{s}(x_i(s)) = s,
  \label{eq:covariant.time-fixation}
\end{align}
i.e., $d\hat{s}(x_i(s))/ds = 1$:
\begin{align}
  \lambda_i = \frac1{\bar\Pi_i \cdot a_i}, \qquad
  a_i^\mu := \frac{\partial\hat{s}(x_i)}{\partial x_{i\mu}},
  \label{eq:Lagrange-multiplier-determined}
\end{align}
where we note that $a_i^\mu$ forms a contravariant Lorentz
vector.~\footnote{The statement that $a_i^\mu$ is a contravariant Lorentz vector
specifically means that $a_i^\mu$ transforms as a contravariant Lorentz vector
under changes of the reference frame.  This is by construction because the
field $\hat s(x)$ is defined to be a Lorentz-scalar field and the derivative
with respect to the coordinates $x_{i\mu}$ with the lower Lorentz index,
$\partial/\partial x_{i\mu}$, forms a contravariant vector.}

To solve the dynamics as an initial-value problem consistently, it is
practically necessary to use the above form of EoMs~\eqref{eq:covariant.eom},
which are expressed with the common evolution parameter $s$.  Nevertheless, it
is useful to examine the form of the EoMs that uses the proper time of
each particle, $\tau_i = \int \sqrt{dx_i^\mu dx_{i\mu}}$:
\begin{align}
  \frac{dx_i^\mu}{d\tau_i} = \frac{\bar\Pi_i^\mu}{\bar m_i} \equiv u_i^\mu, \qquad
  \frac{dp_i^\mu}{d\tau_i} = -\frac{\bar Q_i^\mu}{\bar m_i},
 \label{eq:propertime}
\end{align}
where $\bar m_i = \sqrt{\bar\Pi_i^\mu\bar\Pi_{i\mu}}$, and $u_i^\mu =
dx_i^\mu/d\tau_i$ is the four-velocity of the $i$th particle.
Here, $\bar{\Pi}_i^\mu$ and $\bar{m}_i$ are the \textit{generalized momentum
and mass}, respectively. When the potentials do not depend on the momentum,
they are identical to the kinetic ones, $\bar{\Pi}_i=p_i^*$ and
$\bar{m}_i=m_i^*$ within the harmonic approximation~\eqref{eq:HarmonicApprox}
of Eq.~\eqref{eq:Weiss.weak-mass-shell}.
Using
\begin{align}
  \frac{d\tau_i}{ds}
  = \sqrt{\frac{dx_i^\mu}{ds}\frac{dx_{i\mu}}{ds}}
  = \lambda_i\bar m_i = \frac1{u_i\cdot a_i},
\end{align}
we may obtain another representation of the EoMs:
\begin{equation}
  \frac{dx^\mu_i}{d\tau_i} = u_i \cdot a_i \frac{\partial H}{\partial p_{i\mu}}, \qquad
  \frac{dp^\mu_i}{d\tau_i} = -u_i \cdot a_i \frac{\partial H}{\partial x_{i\mu}}.
\end{equation}
If we choose the time function to be the time in the inertial frame of the
$i$th particle (i.e., $\hat s(x) = u_i \cdot x$), the factor $u_i \cdot a_i$
reduces to unity.  If we choose the time function to be the time in the
reference frame (i.e., $\hat s(x) = x_\mathrm{ref}^0$), the factor $u_i \cdot
a_i$ reduces to the Lorentz factor $u_i^0$.

A weak form of the mass-shell constraints~\eqref{eq:mass-shell-constraintW},
\begin{align}
  \bar W(x_i, p_i) = 0,\qquad (i = 1, \ldots, N),
  \label{eq:mass-shell-constraint.weak}
\end{align}
is satisfied for any time $s$ for the solution of EoMs if we assume it holds at
the initial time $s_0$ because
\begin{align}
  \frac{d\bar W_i}{ds}
  &= [H, \bar W_i] \notag \\
  &= \sum^N_{j=1} ([\lambda_j, \bar W_i]\bar W_j + \lambda_j [\bar W_j, \bar W_i]) = 0,
  \label{eq:mass-shell-constraint.proof}
\end{align}
where the Poisson bracket is defined as
\begin{align}
  [A,B] = \sum_{i=1}^N \left(
  \frac{\partial A}{\partial p_i}
  \cdot\frac{\partial B}{\partial x_i}
 -\frac{\partial A}{\partial x_i}
  \cdot\frac{\partial B}{\partial p_i}
  \right).
  \label{def:Poisson-bracket}
\end{align}

Similar to the constrained Hamiltonian
formulation~\cite{Sudarshan:1981pp,Samuel:1982jk,Samuel:1982jn,Sorge:1989dy},
the $2N$-constraints~\eqref{eq:covariant.time-fixation}
and~\eqref{eq:mass-shell-constraint.weak} reduce the phase space from $8N$
dimensions to the physical $6N$ dimensions.

\subsection{Dynamical fields}
\label{sec:covariant.dynamical}

In the dynamical-field case, the action is written as the sum of the particle
part~\eqref{eq:covariant.action.part} and the field part $S_\mathrm{field}$:
\begin{equation}
  S[\{x_i, p_i\}_i, \Sigma] = S_\mathrm{part}[\{x_i, p_i\}_i, \Sigma] + S_\mathrm{field}[\Sigma].
\end{equation}
This is a functional of the phase-space particle trajectories $\{x_i(s),
p_i(s)\}_i$ and the field degrees of freedom collectively denoted by
$\Sigma(x)$.  The actual set of fields in $\Sigma$ depends on the system.
For example, in the case of the $\sigma$-$\omega$ model (which we will discuss
in Sec.~\ref{sec:rmf}), the dynamical fields are identified to be the scalar
and vector fields: $\Sigma(x) = \bigl(\sigma(x), \omega_\mu(x)\bigr)$.
Consequently, $p^*[\bm{x}, p, \Sigma]$, $m^*[\bm{x}, p, \Sigma]$, $W[\bm{x}, p,
\Sigma]$, and $\bar W[\bm{x}, p, \Sigma]$ become functionals of $\Sigma$.
The particle part $S_\mathrm{part}$ is given by
Eq.~\eqref{eq:covariant.action.part} and includes their interaction with the
fields, where $U_s(x,p)$ and $U_\mu(x,p)$ appearing in $m^*(x,p)$ and
$p^*(x,p)$ of $W(x,p)$ are given by the field degrees $\Sigma(x)$.
The field part $S_\mathrm{field}[\Sigma]$ is a normal action for the fields
times an extra negative sign\footnote{The extra negative sign is present to be
consistent with the sign in $S_\mathrm{part}$ written by $\bar W_i$; the
mass-shell constraint $W_i$ for the $i$th particle in the current action
effectively has a negative sign relative to the single-particle energy $E_i$ in
the normal Hamiltonian, as observed in $(1/E_i^*)\partial W_i/\partial \bm{p}_i
\sim -\bm{p}_i^*/E_i^*$ and $\partial E_i/\partial \bm{p}_i \sim
\bm{p}_i^*/E_i^*$. \label{fn:extra-negative-sign}};
we consider the Weiss action principle for the particle part and the normal
variation for the dynamical fields $\Sigma(x)$.  The field part contains only
the self-interactions of the fields.  In this section, we omit the explicit
form of the field part of the action because it is irrelevant in deriving the
EoMs for particles.

The EoMs for the particles turn out to be identical to the
EoMs~\eqref{eq:covariant.eom}--\eqref{eq:Lagrange-multiplier-determined}
derived for the external-field case.  This is because the new term
$S_\mathrm{field}[\Sigma]$ in the action does not depend on the particle
degrees of freedom, and thus it does not contribute to the variation with
respect to $\delta x_i(s)$ and $\delta p_i(s)$.

Consequently, the discussions regarding the particle EoMs for the
external-field case in Sec.~\ref{sec:covariant.external} remain valid for the
dynamical-field case, except for the weak mass-shell
constraints~\eqref{eq:mass-shell-constraint.weak}.  The original
proof~\eqref{eq:mass-shell-constraint.proof} of the weak mass-shell constraints
in the external-field case is no longer applicable because the field degrees of
freedom are involved.  The proof for the dynamical-field case is the following:
First, we note that
\begin{align}
  \frac{d\bar W_i}{ds}
  &= \int d^4xd^4p W(x,p) \frac{dG_i}{ds},
  \label{eq:covariant.dynamical.mass-shell.proof1}
\end{align}
with $G_i := G(x,p;x_i,p_i)$ being defined in Eq.~\eqref{def:gi}.  It should be
noted that $dW(x,p)/ds = 0$ because the evolution parameter in the generalized
potential $W(x,p)$~\footnote{$W(x,p)$ is written in terms of $p^*(x,p)$ and
$m^*(x,p)$.  Then, $p^*(x,p)$ and $m^*(x,p)$ are determined by the field degrees of
freedom $\Sigma$, and $\Sigma$ is determined by field EoMs as a function of the evolution
parameter $s$ in the dynamical-field case.}
is specified by the integral variable $x$ through $\hat
s(x)$ and is independent of the evolution parameter $s$ specified externally.
Since $G_i$ is a function of $x_i$ and $p_i$ and independent of the field
degrees of freedom $\Sigma$, its time derivative can be expanded as
\begin{align}
  \frac{dG_i}{ds} &=
    \frac{dx_i^\mu}{ds} \frac{\partial G_i}{\partial x_i^\mu} +
    \frac{dp_i^\mu}{ds} \frac{\partial G_i}{\partial p_i^\mu}.
  \label{eq:mass-shell-constraint.mass-shell.proof2}
\end{align}
Plugging this into Eq.~\eqref{eq:covariant.dynamical.mass-shell.proof1} and
using EoMs~\eqref{eq:covariant.eom}, we obtain
\begin{align}
  \frac{d\bar W_i}{ds}
  &=
    \frac{dx_i^\mu}{ds} \int d^4xd^4p W(x,p) \frac{\partial G_i}{\partial x_i^\mu} \notag \\ &\quad +
    \frac{dp_i^\mu}{ds} \int d^4xd^4p W(x,p) \frac{\partial G_i}{\partial p_i^\mu} \notag \\
  &=
    (\lambda_i \bar\Pi_i^\mu) \bar Q_{i\mu} +
    (-\lambda_i \bar Q_i^\mu) \bar \Pi_{i\mu} = 0,
  \label{eq:covariant.dynamical.mass-shell.proof3}
\end{align}
and hence $\bar W_i \equiv 0$ if it initially vanishes.

Finally, we note that one can recover the same EoMs in the RBUU approach when
the harmonic approximation is applied to both the coordinate and momentum parts
of the particle profile.  In this case, the weak mass-shell
constraints~\eqref{eq:mass-shell-constraint.weak} become exact at the center of
the profile:
\begin{align}
  W(x_i, p_i) = 0,
  \label{eq:mass-shell-constraint}
\end{align}
which leads to the EoMs,
\begin{align}
  \frac{dx_i^\mu}{d\tau_i} = \frac{\Pi^\mu(x_i,p_i)}{\tilde m_i}, \qquad
  \frac{dp_i^\mu}{d\tau_i} = -\frac{Q^\mu(x_i,p_i)}{\tilde m_i},
\end{align}
where $\tilde m_i = \sqrt{\Pi^\mu(x_i,p_i)\Pi_\mu(x_i,p_i)}$.  These equations
are identical to those derived in the RBUU approach in
Ref.~\cite{Weber:1992qc}. Furthermore, they extend the equations presented in
Ref.~\cite{Fuchs:1995fa} by including the momentum-dependent potential.

\subsection{Local density approximation}
\label{sec:covariant.static}

In the previous section, Sec.~\ref{sec:covariant.dynamical}, we have considered
the dynamical-field case, where the particle and field degrees of freedom,
$\{x_i,p_i\}_i$ and $\Sigma(x)$, are independent variables.  The
single-particle potentials, $U_s(x, p)$ and $U_\mu(x, p)$, appearing in
$m^*(x,p)$~\eqref{def:effective-mass} and
$p^*_\mu(x,p)$~\eqref{def:kinetic-momentum}, are determined by $\Sigma(x)$ but
do not directly depend on $\{x_i, p_i\}_i$.

However, in the numerical implementation, the derivatives of the meson fields,
namely, the kinetic terms of the fields, are often neglected.  As a result, the
meson fields are determined from the present configuration of particles,
$\{x_i, p_i\}_i$. This approximation is called the local density
approximation~\cite{Weber:1992qc}.  In other words, to formulate the dynamics
purely in terms of the particle degrees of freedom, we may specify $U_s(x, p)$
and $U_\mu(x, p)$ using the local quantity determined by the particle degrees of freedom
$\{x_i, p_i\}_i$.  In this work, we consider the general case of the local
density approximation, where the effective mass and the kinetic momentum
depend not only on the local scalar and vector densities but also on the local
momentum distribution $f(x,p)$ because of the momentum-dependent
potential~\cite{Weber:1992qc}.

Under the local density approximation, $U_s(x, p)$ and $U_\mu(x, p)$ are
functions of phase-space variables of particles, $\{x_i, p_i\}$, and subject to
variation by the particle trajectories.
We first introduce the free-energy functional $F[\phi, \phiv_\mu]$ to give
$U_s(x, p)$ and $U_\mu(x, p)$ consistently.  We then show that the form of
the particles' EoMs matches the dynamical-field case,
Eqs.~\eqref{eq:WeissEoM}--\eqref{eq:Weiss.weak-mass-shell}. This owes to the
exact cancellation of the contributions from the variation of $U_s(x, p)$ and
$U_\mu(x, p)$ and that of the field energy.

To define the free-energy functional, we first introduce new fields, the scalar
and vector distributions, $\phis(x,p)$ and $\phiv_\mu(x,p)$, with $x, p \in
\mathbb{R}^{3,1}$, and $\mu = 0$, $1$, $2$, $3$.  We consider the free energy
functional $F[\phis, \phiv_\mu]$ as an arbitrary functional of $\phis(x,p)$ and
$\phiv_\mu(x,p)$.  This generalizes the total potential energy in the
nonrelativistic case in Refs.~\cite{Bertsch:1988ik,Nara:2021fuu} (see Eq.~(E.2)
of Ref.~\cite{Bertsch:1988ik} and Eq.~(9) in Ref.~\cite{Nara:2021fuu})
to the relativistic case.

Using the free-energy functional $F[\phis, \phiv_\mu]$, the single-particle
potentials are written as
\begin{align}
  U_s(x, p) &= \frac{\delta F[\phis, \phiv_\mu]}{\delta \phis(x,p)}, \label{eq:sec3.spot} \\
  U_\mu(x, p) &=  g_{\mu\nu} \frac{\delta F[\phis, \phiv_\mu]}{\delta \phiv_\nu(x,p)}, \label{eq:sec3.vpot}
\end{align}
where $\phis(x, p)$ and $\phiv_\mu(x, p)$ are determined by solving
the self-consistent equations,
\begin{align}
  \phis(x, p) &= m^*(x, p) f(x, p),
  \label{eq:self-consistent.s} \\
  \phiv_\mu(x, p) &= p^*_\mu(x, p) f(x, p),
  \label{eq:self-consistent.v}
\end{align}
combined with Eqs.~\eqref{def:effective-mass}, \eqref{def:kinetic-momentum},
\eqref{eq:sec3.spot}, and~\eqref{eq:sec3.vpot}.  The distribution function
$f(x, p)$ is specified externally but can optionally depend on $m^*(x, p)$ and
$p^*_\mu(x, p)$ in solving the self-consistent equations.

To consider the $N$-particle dynamics, we give the distribution function
$f(x,p)$ in terms of the phase-space variables $\{x_i, p_i\}_{i=1}^N$.  In this
case, $m^*(x,p)$, $p^*(x,p)$, $\phis(x,p)$, and $\phiv_\mu(x,p)$ become
functions of $\{x_i, p_i\}_i$.  We give the action by
\begin{equation}
  S[\{x_i, p_i\}_i] = S_\mathrm{part}[\{x_i, p_i\}_i] + S_\mathrm{field}[\{x_i, p_i\}_i],
\end{equation}
with the particle part $S_\mathrm{part}$ being
Eq.~\eqref{eq:covariant.action.part}.  The field part of the action is
identified~\hyperref[footnotelabel]{\textsuperscript{\ref{fn:extra-negative-sign}}} as
\begin{align}
  S_\mathrm{field} &= V_\mathrm{field},
\end{align}
with the field potential functional $V_\mathrm{field}$ being the Legendre
transform of the free-energy functional:
\begin{multline}
  V_\mathrm{field}[\phis, \phiv_\mu]
  = F[\phis, \phiv_\mu] - \int d^4x d^4p \\
  \times \biggl\{\phis(x,p) \frac{\delta F[\phis, \phiv_\mu]}{\delta \phis(x,p)}
  + \phiv_\mu(x,p) \frac{\delta F[\phis, \phiv_\mu]}{\delta \phiv_\mu(x,p)} \biggr\}.
  \label{def:covariant.static.V}
\end{multline}
It should be noted that the fields $\Sigma(x)$ in
Eq.~\eqref{eq:covariant.action.part} are no longer dynamical degrees in this
setup and determined by the particle degrees of freedom, $\{x_i, p_i\}_i$.

The variation of the action is expressed as
\begin{align}
  \delta S
  &= \delta \int ds \sum_i p_i \cdot \frac{dx_i}{ds}
  - \int d^4x d^4p\, W(x,p) \delta f(x,p) \notag \\
  &\quad - \int d^4x d^4p\, \delta W(x,p) f(x,p)
  + \delta V_\mathrm{field}.
  \label{eq:covariant.static.action-variation}
\end{align}
The first line matches the variation of the action with respect to the particle
degrees discussed in Sec.~\ref{sec:covariant.dynamical}\@.  The second line
corresponds to the variation through the fields specified by the particle
degrees.  We can show that two terms on the second line exactly cancel with
each other (see Appendix~\ref{app:totpot.cancel}):
\begin{align}
  -\int d^4x d^4p\, \delta W(x,p) f(x,p) + \delta V_\mathrm{field} = 0.
  \label{eq:totpot.cancel}
\end{align}
As a result, the substantial part of the variation of the action is only the
first line of Eq.~\eqref{eq:covariant.static.action-variation}, which matches
the variation of the particle part of the action in the external-field case.

For this reason, with the local density approximation, the least action
principle yields exactly the same form of EoMs as
Eqs.~\eqref{eq:covariant.eom}--\eqref{eq:Lagrange-multiplier-determined} for
the external- and dynamical-field cases in Secs.~\ref{sec:covariant.external}
and~\ref{sec:covariant.dynamical}\@.
The only difference is that the potentials $U_s(x,p)$ and
$U_\mu(x,p)$ are given by the particle degrees, while they are given
independently of the particle degrees in Sec.~\ref{sec:covariant.dynamical}\@.

This result is reasonable: If we first derive the EoMs for the dynamical-field
case and then take the limit where the kinetic terms of the dynamical fields
vanish, the form of the particle EoMs remains unchanged during taking the
limit.  Therefore, it is not surprising that we obtain the same form of the
EoMs even when we instead first apply the local density approximation to the
action and then derive the EoMs for the particles.

Finally, we shall check the weak mass-shell
constraints~\eqref{eq:mass-shell-constraint.weak} in the case of the local
density approximation.  Since the EoMs have the same structure as the
dynamical-field case,
Eqs.~\eqref{eq:covariant.dynamical.mass-shell.proof1}--\eqref{eq:covariant.dynamical.mass-shell.proof3}
also apply to the present case with the local density approximation.  Thus, the
weak form of the mass-shell constraint~\eqref{eq:mass-shell-constraint.weak}
is preserved by the local density approximation.

\subsection{Foliation dependence and uniform foliation}
\label{sec:covariant.foliation}

We finally discuss the foliation dependence of the obtained dynamics.  So far,
we discussed the formalism with the general foliation $\hat s(x)$.  However, we
shall note that the dynamics depends on the choice of the foliation
$\hat{s}(x)$ in this formulation in general, which means that different physics
results may be obtained for different choices of the foliation.  We have
discussed the time-fixation dependence of the RQMD frameworks in the
introduction.  The foliation dependence is the manifestation of the
time-fixation dependence within the present formulation, and this explicitly
breaks the time-reparametrization invariance of the dynamics despite the
covariant forms of the EoMs.  This comes from effective nonlocal interactions
in the RQMD framework, where interaction happens between two spatially distant
coordinates on a time hypersurface in the current choice of the foliation.  As
discussed in the introduction, this is an unavoidable consequence of in
relativistic $N$-body particle dynamics of a finite number of degrees of
freedom interacting with each other~\cite{Sudarshan:1981pp}.  In practical
applications, it needs to be checked that the results do not largely depend on
different choices of reasonable foliations, as will be done in
Sec.~\ref{sec:validation} in the present study.  It is noted that the foliation
dependence is eliminated in the limit of the delta particle profile with an
infinite number of test particles as in the RBUU approach or in the case of the
point particles interacting through dynamical fields.

To perform actual calculations, we need to specify an appropriate foliation
that roughly traces the proper time of the matter.  In the subsequent
discussions, motivated by the time-fixation conditions used in
Refs.~\cite{Maruyama:1996rn, Mancusi:2009zz, Marty:2012vs, Isse:2005nk,
Nara:2019qfd, Nara:2020ztb, Nara:2021fuu}, we suppose the uniform foliation
$\hat s(x) = \hat a \cdot x$ with $\hat a^\mu$ normalized as $\hat a^\mu\hat
a_\mu = 1$.  However, it should be emphasized here that we define $\hat{a}^\mu$
as a Lorentz vector\footnote{The statement that $\hat{a}^\mu$ is defined to be
a Lorentz vector means that $\hat{a}^\mu$ transforms as a Lorentz vector
properly under changes of the reference frame.  This definition of
$\hat{a}^\mu$ is chosen to be consistent with the Lorentz vector $a_i^\mu$ in
Eq.~\eqref{eq:Lagrange-multiplier-determined} so that $a_i^\mu$ becomes uniform
in the spacetime: $a_i^\mu = \hat a^\mu$.} so that $\hat{s}(x)$ is a
Lorentz-scalar field.  This choice implies that the times $x_i^0$ of all
particles are the same in the specific inertial frame in which $\hat{s}(x) =
\hat{a}\cdot x = x^0$ is satisfied.~\footnote{The apparently noncovariant
expression $\hat{s}(x) = \hat{a}\cdot x = x^0$ is only valid in a specific
frame, and this does not imply $\hat{a}$ is fixed to be $(1, 0, 0, 0)$ in an
arbitrary reference frame. The function $\hat{s}(x)$ and the vector $\hat a$
are still considered to transform appropriately as a Lorentz-scalar field
and a Lorentz-vector, respectively, under changes of the reference frame.}
In other words, we impose the $N$ time-fixation conditions:
\begin{equation}
 \hat{a}\cdot x_i - s = 0, \qquad (i=1,\ldots,N),
\end{equation}
where $s$ is the Lorentz invariant evolution parameter.  The normal vector
$\hat a^\mu$ can be chosen arbitrarily in principle, depending on the concrete
setup.
With this choice of time function, $a_i^\mu$ in
Eq.~\eqref{eq:Lagrange-multiplier-determined} becomes constant $a_i^\mu \equiv
\hat a^\mu$, and thus the EoMs read
\begin{equation}
  \frac{dx^\mu_i}{ds} = \frac{\bar{\Pi}^\mu(x_i,p_i)}{\bar{\Pi}_i\cdot\hat{a}}, \qquad
  \frac{dp^\mu_i}{ds} = -\frac{\bar{Q}^\mu(x_i,p_i)}{\bar{\Pi}_i\cdot\hat{a}},
  \label{eq:covariant.eom.ahat}
\end{equation}
which has the same structure as obtained in the constrained Hamiltonian
formulation~\cite{Nara:2023vrq}.  We note that the free particle version of the
EoMs~\eqref{eq:covariant.eom.ahat} is used for the Lorentz covariant cascade
method proposed in Ref.~\cite{Nara:2023vrq}.

Under this spatially uniform foliation, we may consider the translationally
invariant particle profile:
\begin{align}
 G(x,p;x_i, p_i)
 = G(x-x_i,p-p_i),
 \label{def:particle-profile-d}
\end{align}
where the dependence on the positions and momenta appears in the combinations
$x-x_i$ and $p-p_i$.  With this particle profile, Eqs.~\eqref{def:barPi}
and~\eqref{def:barQ} can be explicitly written down as
\begin{align}
 &\bar\Pi_i^\mu
   = \int d^4x d^4p\, \Pi^\mu(x,p) G(x-x_i, p-p_i),
   \label{eq:ceomx} \\
 & \bar Q_i^\mu
   = \int d^4x d^4p\, Q^\mu(x,p) G(x-x_i, p-p_i),
   \label{eq:ceom}
\end{align}
where the derivatives of the generalized potential are defined as
\begin{align}
 \Pi^\mu(x,p) &:= \frac{\partial W(x,p)}{\partial p_\mu}, \\
 Q^\mu(x,p) &:= \frac{\partial W(x,p)}{\partial x_\mu}.
\end{align}

We also comment on energy--momentum conservation in the present framework.
Because of the presence of the externally specified foliation $\hat s(x)$, if
the foliation is not considered subject to the translation, the translational
symmetry of the system is apparently broken in general.  However, the
translational symmetry of the system is preserved at least when the uniform
foliation in the form $\hat s(x) =\hat a \cdot x$ is considered, in which case
the energy--momentum conservation holds.  In other cases, the energy--momentum
conservation is not necessarily satisfied by simply considering the momenta of
the particles and mean fields.  Detailed discussions for the implication will
be discussed in future studies.

\section{Applications}
\label{sec:application}

In this section, for the numerical demonstration with physical setups, we shall
apply our relativistic quantum molecular dynamics (RQMD) formulation to the
interaction by the relativistic mean field and the Skyrme-type potentials
implemented as a Lorentz vector.  We assume the uniform foliation $\hat s(x) =
\hat a\cdot x$ with $\hat a_\mu \hat a^\mu = 1$.

Let us first summarize the covariant equations~\eqref{eq:covariant.eom.ahat}
formulated in Secs.~\ref{sec:covariant.dynamical}
and~\ref{sec:covariant.static}:
\begin{equation}
  \frac{dx_i^\mu}{ds} = \frac{\bar{\Pi}_i^\mu}{\bar{\Pi}_i\cdot\hat{a}}, \qquad
  \frac{dp_i^\mu}{ds} = -\frac{\bar{Q}_i^\mu}{\bar{\Pi}_i\cdot\hat{a}},
  \label{eq:covariant.eom.ahat2}
\end{equation}
where
\begin{align}
  &\bar\Pi_i^\mu
    = \int d^4x d^4p\, \Pi^\mu(x,p) G(x-x_i, p-p_i),
    \label{eq:ceomx2} \\
  & \bar Q_i^\mu
    = \int d^4x d^4p\, Q^\mu(x,p) G(x-x_i, p-p_j),
    \label{eq:ceom2}
\end{align}
and the derivatives of the generalized potential are defined as
\begin{align}
  & \Pi^\mu(x,p) = \frac{\partial W(x,p)}{\partial p_\mu} \notag \\ & \qquad
    = p^*(x,p) \cdot \frac{\partial p^*(x,p)}{\partial p_\mu}
    - m^*(x,p) \frac{\partial m^*(x,p)}{\partial p_\mu}, \\
  & Q^\mu(x,p) = \frac{\partial W(x,p)}{\partial x_\mu}\notag \\ & \qquad
    = p^*(x,p) \cdot \frac{\partial p^*(x,p)}{\partial x_\mu}
    - m^*(x,p) \frac{\partial m^*(x,p)}{\partial x_\mu}.
\end{align}

In the numerical implementation, we assume the factorization of the particle
profile into the spatial and momentum parts, and the momentum-space profile is
taken to be a $\delta$ function:
\begin{align}
  G(x,p;x_i,p_i) = G(x-x_i) \delta(p-p_i),
\end{align}
as in the QMD case in Sec.~\ref{sec:qmd}\@.  We also assume the following
Lorentz-contracted Gaussian for the particle profile:
\begin{align}
  G(x-x_i;u_i)
    &= (u_i\cdot\hat a) \tilde{G}(x-x_i;u_i),
    \label{eq:gx} \\
  \tilde{G}(x-x_i;u_i)
    &= \delta((x - x_i)\cdot\hat a) \tilde{g}(x-x_i;u_i),
    \label{eq:Gtilde} \\
  \tilde{g}(x-x_i;u_i)
    &= \frac{1}{(2\pi L)^{3/2}} \exp\frac{R(x-x_i;u_i)^2}{2L},
    \label{eq:gtilde} \\
  R(x-x_i;u_i)^\mu
    &= \Delta_i^{\mu\nu} (x_\nu - x_{i\nu}),
    \label{eq:vardyn.contractedGauss}
\end{align}
where $L$ is the Gaussian width parameter, and $\Delta_i^{\mu\nu} =
g^{\mu\nu} - u_i^\mu u_i^\nu$ is the spatial projector orthogonal to the
velocity $u_i^\mu$ of the $i$th particle.
As already discussed in Sec.~\ref{sec:covariant.external}, this form of the
profile $G(x-x_i;u_i)$ has an additional dependence on $u_i = dx_i/d\tau_i$,
which was not taken into account when deriving the
EoMs~\eqref{eq:covariant.eom.ahat2} and obtaining the
expressions~\eqref{eq:ceomx2} and~\eqref{eq:ceom2}.  Although the Lorentz-contracted
Gaussian~\eqref{eq:gx}--\eqref{eq:vardyn.contractedGauss} violates the weak
mass-shell constraints~\eqref{eq:mass-shell-constraint.weak} due to the $u_i$
dependence, this violation is expected to be small in the slow-acceleration
regime ($du_i/d\tau_i \ll 1/L$).  In the subsequent discussions, unless we
intend to emphasize, we omit the explicit dependence on $u_i$ in the argument
of $G(x-x_i)$~\eqref{eq:gx}, $\tilde{G}(x-x_i)$~\eqref{eq:Gtilde},
$\tilde{g}(x-x_i)$~\eqref{eq:gtilde}, $g(x,p;x_i,p_i)=(u_i\cdot\hat
a)g(x-x_i)\delta(p-p_i)$, but they implicitly depend on $u_i$, where $i$ shall
be identified by the symbol $x_i$ appearing in an explicit argument.

Under this approximation, Eqs.~\eqref{eq:ceomx2} and~\eqref{eq:ceom2} are
evaluated by
\begin{align}
  \bar{\Pi}^\mu_i &= \int d^4x\, \Pi^\mu(x,p_i) G(x-x_i),
  \label{eq:ceom_s1} \\
  \bar{Q}^\mu_i &= \int d^4x\, Q^\mu(x,p_i) G(x-x_i).
  \label{eq:ceom_s2}
\end{align}
Since $G(x-x_i;u_i)$ is a normalized Gaussian profile, Eqs~\eqref{eq:ceom_s1}
and~\eqref{eq:ceom_s2} can be evaluated by the Monte Carlo integration as
\begin{align}
 \bar{\Pi}_i^\mu &= \frac{1}{N_\mathrm{MC}}\sum_{k=1}^{N_\mathrm{MC}} \Pi_i^\mu(x_k,p_i),\\
 \bar{Q}_i^\mu  &= \frac{1}{N_\mathrm{MC}}\sum_{k=1}^{N_\mathrm{MC}}  Q_i^\mu(x_k,p_i),
\end{align}
with $N_\mathrm{MC}$ being the number of Monte Carlo points for the coordinates
$x_k$, which are sampled according to the relativistic Gaussian distribution
$G(x-x_i;u_i)$.  The detailed procedure can be found in
Appendix~\ref{sec:MCrelgauss}.

The EoMs are explicitly written as
\begin{align}
\frac{dx_i^\mu}{ds}
  &=\int d^4x\, (u_i\cdot\hat a) \frac{p^{*\mu}(x, p_i)}{\bar \Pi_i\cdot\hat a} \tilde{G}(x-x_i)\nonumber\\
   &-\int d^4x
    \frac{m^*(x,p_i)}{\bar{m}_i}\frac{\partial m^*(x,p_i)}{\partial p_{i\mu}}
    \tilde{G}(x-x_i) \notag \\
   &-\int d^4x
    \frac{p^{*\nu}(x,p_i)}{\bar{m}_i}\frac{\partial U_\nu(x,p_i)}{\partial p_{i\mu}}
   \tilde{G}(x-x_i),\label{eq:eoms1}\\
\frac{dp_i^\mu}{ds}
    &= \int d^4x
      \frac{m^*(x,p_i)}{\bar{m}_i}\frac{\partial m^*(x,p_i)}{\partial x_{\mu}}
      \tilde{G}(x-x_i) \notag \\
    &+ \int d^4x
      \frac{p^{*\nu}(x,p_i)}{\bar{m}_i}\frac{\partial U_\nu(x,p_i)}{\partial x_{\mu}}
      \tilde{G}(x-x_i).\label{eq:eoms2}
\end{align}

The scalar and vector potentials are determined by the scalar density and the
vector current,
\footnote{
We note that the conserved current may be given by
$
 J^\mu(x)= \int d^4p\, \Pi^\mu(x,p) f(x,p),
$
when momentum-dependent potentials are included.
}
\begin{align}
  \rho_s(x) &= \int d^4p\, m^*(x,p) f(x,p),
  \label{def:scalar-density} \\
  J^\mu(x) &= \int d^4p\, p^{*\mu}(x,p) f(x,p).
  \label{def:vector-current}
\end{align}
To solve the EoMs, we need to evaluate the distribution function $f(x,p)$
appearing in the scalar density~\eqref{def:scalar-density} and baryon
current~\eqref{def:vector-current}.  The distribution
function~\eqref{def:f.with-ds-v2} involves the time integration from the past to
the future, which would be inconvenient in solving the EoMs.  However, within the
current setup, using the dynamical constraint $\hat s(x_i(s)) = s$, the time
integration in Eq.~\eqref{def:f.with-ds-v2} can be eliminated as
\begin{align}
  f_i(x, p) = \lambda_i(\hat{s}(x))
  g\bigl(x, p; x_i(\hat{s}(x)), p_i(\hat{s}(x))\bigr).
\end{align}
In the case of the Gaussian~\eqref{eq:gx} for the coordinate space and the
$\delta$ function for the momentum space, the distribution function is given by
\begin{align}
  f_i(x, p)
    &= \frac{u_i\cdot\hat{a}}{\bar{\Pi}_i\cdot\hat{a}}\tilde{g}(x-x_i(\hat{s}(x)))\delta(p-p_i(\hat{s}(x)))\nonumber\\
    &= \frac{1}{\bar{m}_i}\tilde{g}(x-x_i(\hat{s}(x)))\delta(p-p_i(\hat{s}(x))),
\end{align}
where $u_i = \bar{\Pi}_i/\bar{m}_i$ was used.
Then, the scalar density and the vector current are evaluated by the following
expressions:
\begin{align}
  \rho_s(x) &= \sum_{i=1}^N  \frac{m^*(x,p_i)}{\bar{m}_i} \tilde{g}(x-x_i(\hat{s}(x))),\\
  J(x) &= \sum_{i=1}^N \frac{p^*(x,p_i)}{\bar{m}_i} \tilde{g}(x-x_i(\hat{s}(x))).
\label{eq:current}
\end{align}

\subsection{Numerical implementation}
\label{sec:numerics}

In the first numerical implementation, we make simplifications,
\begin{align}
  \frac{m^*(x,p_i)}{\bar m_i} & \approx 1, \\
  \frac{p^*(x,p_i)}{\bar m_i} & \approx \frac{p^*(x,p_i)}{\sqrt{p^{*2}(x,p_i)}} =: u(x,p_i),
\end{align}
to avoid numerical complications in solving the RQMD
EoM\@.  Then, the EoMs become
\begin{align}
\frac{dx_i^\mu}{ds} &=\frac{p_i^{*\mu}}{\Pi_i\cdot\hat{a}}
   -\int d^4x \biggl[
    \frac{\partial m^*(x,p_i)}{\partial p_{i\mu}} \notag \\&\qquad
    +u^{\nu}(x,p_i)\frac{\partial U_\nu(x,p_i)}{\partial p_{i\mu}}
   \biggr] \tilde{G}(x-x_i),\label{eq:eom_rqmdsv1} \\
\frac{dp_i^\mu}{ds} &= \int d^4x \biggl[
    \frac{\partial m^*(x,p_i)}{\partial x_{\mu}} \notag \\&\qquad
    +u^{\nu}(x,p_i)\frac{\partial U_\nu(x,p_i)}{\partial x_{\mu}}
   \biggr]\tilde{G}(x-x_i),\label{eq:eom_rqmdsv2}
\end{align}
where the first term of Eq.~\eqref{eq:eom_rqmdsv1} is obtained by using the
harmonic approximation that the argument $x$ of $p^*(x,p_i)$ and $m^*(x,p_i)$
is replaced by the coordinates of the center of the Gaussian, i.e., $x_i$.
Also, $u_i$ in the Gaussian profile is approximated by $u_i \approx u(x_i,p_i)$.
Accordingly, the scalar density and vector current become
\begin{equation}
 \rho_s(x) =\sum_{i=1}^N \tilde{g}(x-x_i), \qquad
  J^\mu(x) =\sum_{i=1}^N
    u_i^\mu \tilde{g}(x-x_i).
  \label{eq:current2}
\end{equation}
In the next sections, we apply these equations to the relativistic mean-field
theory and the Skyrme-type potentials.

The relativistic molecular dynamics (RQMD) model combines the Boltzmann-type
collision term for hadrons and the propagation of hadrons according to the
mean-field potential.  We implemented the covariant EoMs derived in the
following sections in the Monte Carlo event generator
\texttt{JAM2}~\cite{JAM2}\@. \texttt{JAM2} is a C++ version of
\texttt{JAM1}~\cite{Nara:1999dz}.  The collision term includes the hadronic
resonance and string excitation and their decay to simulate particle production
processes.  \texttt{JAM2} uses
\texttt{Pythia8}~\cite{Sjostrand:2014zea,PYTHIA8} for the string fragmentation
with the formation time.  During the formation time, hadrons from strings do
not scatter, but the leading hadrons (which carry original constituent quarks)
can scatter within the formation time with reduced cross sections.


\subsection{RQMD for the relativistic mean-field theory (RQMD.RMF2)}
\label{sec:rmf}

We now apply the formulation for the system of interacting wave packets via the
relativistic mean fields.  Specifically, we consider the $\sigma$ and $\omega$
meson exchange.  The Lagrangian density is given by
\begin{align}
\mathcal{L}&= \sum_{B}\bar{\psi}_B\{\gamma^\mu[i\partial_\mu - U_{B,\mu}(\omega)]
  - m^*_B(\sigma)\}\psi_B \nonumber\\
 &+ \frac{1}{2}\partial_\mu\sigma\partial^\mu
\sigma-V_\sigma(\sigma)
 -\frac{1}{4}\omega_{\mu\nu}\omega^{\mu\nu} + V_\omega(\omega_\mu),
\end{align}
where $B$ runs over different baryon species, $U_{B,\mu} :=
g_{\omega,B}\omega_\mu$, $m^*_B(\sigma) := m_B + U_{\sigma,B}(\sigma)$, and
$\omega_{\mu\nu} := \partial_\mu\omega_\nu - \partial_\nu\omega_\mu$.  In the
case of the Walecka-type model, the single-particle scalar potential is given
by $U_{\sigma,B} = -g_{\sigma,B} \sigma$.

In the $\sigma$-$\omega$ model, the following potentials are often used:
\begin{align}
 V_\sigma(\sigma) &=  \frac{m_\sigma^2}{2}\sigma^2 + \frac{g_2}{3}\sigma^3 + \frac{g_3}{4}\sigma^4,
\label{eq:sigma4} \\
 V_\omega(\omega_\mu) &=  \frac{m_\omega^2}{2}\omega_\mu\omega^\mu + \frac{c_4}{4}(\omega_\mu\omega^\mu)^2.
\label{eq:omega4}
\end{align}
We assume the mean-field approximation, where the meson-field operators are
replaced by the expectation value of the field.  Furthermore, we take the
positive energy solutions of the Dirac equations (the no-sea approximation).  A
variation of the Lagrangian density with respect to the fields $\sigma$ and
$\omega$ yields the following EoMs,
\begin{align}
\partial_\mu\partial^\mu\sigma   + \frac{\partial V_\sigma}{\partial\sigma}
 &= \sum_{B}y_B(x) \rho_{s,B}, \label{eq:gap1}\\
\partial_\nu \omega^{\nu\mu}+\frac{\partial V_\omega}{\partial\omega_\mu}
  &= \sum_{B} g_{\omega,B} J_B^\mu, \label{eq:gap2}
\end{align}
where $y_B(x)=-\partial m^*_B/\partial\sigma =-\partial
U_{\sigma,B}/\partial\sigma$.  The scalar density and the baryon current are
given by the expectation values of the field operators: $\rho_{s,B}=\langle
\bar\psi_B\psi_B\rangle$ and
$J_B^\mu=\langle\bar\psi_B\gamma^\mu\psi_B\rangle$.

In the transport description of a system, one switches to the phase space
picture~\cite{Weber:1992qc,Blaettel:1993uz}, and within the RQMD approach, the
scalar density and the vector current are given by the sum of the Gaussian,
\begin{align}
  \sum_{B}y_B(x)\rho_{s,B}(x)
  &= \sum_{j=1}^N
    y_j \tilde{g}(x-x_j)
  \equiv
  n_s(x),
  \label{def:application.ns} \\
  \sum_{B} g_{\omega,B} J_B^\mu(x)
  &= \sum_{j=1}^N g_{\omega,j}u_j^\mu \tilde{g}(x-x_j)
  \equiv j^\mu(x),
  \label{def:application.j}
\end{align}
where the summation index $j$ runs over $N$ particles of all species, and
$g_{\omega,j} = g_{\omega,B_j}$~\footnote{Here, we do not consider the
contributions from antibaryons.} with $B_j$ being the baryon species of the
$i$th particle.  Here, we made the same approximations as in
Eq.~\eqref{eq:current2}: $y_{B_j}(x) \approx y_{B_j}(x_j) =: y_j$ and $u(x,p_j)
\approx u_j$.  The spatial part of the BUU-like EoM~\eqref{eq:eom_rqmdsv2}
becomes
\begin{equation}
  \frac{d\bm{p}_i}{ds} =
  \int d^4x \biggl[y_i\frac{\partial \sigma(x)}{\partial\bm{x}}
    - g_{\omega,i}u_i^\mu \frac{\partial \omega_\mu(x)}{\partial\bm{x}} \biggr]
    \tilde{G}(x-x_i),
  \label{eq:application.mf.eom1}
\end{equation}
where we replaced $y_{B_i}(x)$ and $u(x,p_i)$ with $y_i$ and $u_i$.

Below, we assume the local density approximation that neglects the kinetic terms
of the meson fields in the field EoMs~\eqref{eq:gap1} and~\eqref{eq:gap2}.  The
field potential~\eqref{def:covariant.static.V} is given by
\begin{equation}
  V_\mathrm{field}[\sigma, \omega]
  = \int d^4x [V_\sigma(\sigma(x)) - V_\omega(\omega_\mu(x))].
\end{equation}
The corresponding free-energy functional (in the formulation of
Sec.~\ref{sec:covariant.static}) is found to be
\begin{align}
  F[\rho_{s,B}, J_B^\mu]
  &= V_\mathrm{field} + \sum_B \int d^4x (\rho_{s,B} U_{\sigma,B} + J_B^\mu\cdot U_{B\mu})
    \notag \\
  &= V_\mathrm{field} +
    \int d^4x \biggl(\sum_B \rho_{s,B} U_{\sigma,B} + \omega\cdot j\biggr).
\end{align}

We follow the same approach as done in the previous sections to obtain the
QMD-like EoMs from the BUU-like ones.  In a similar way to Sec.~\ref{sec:qmd},
we consider rewriting Eq.~\eqref{eq:application.mf.eom1} by defining
``one-particle'' scalar and vector fields:
\begin{equation}
  \tilde{\sigma} = \frac{1}{n_s}\int \sigma dn_s,
  \qquad \tilde{\omega}_\mu = \frac{j_\mu}{n^2}\int \omega \cdot dj,
\end{equation}
where $n=\sqrt{j_\mu j^\mu}$.  One can show that the integral $\int \omega
\cdot dj$ is well-defined using the fact that the vector current can be
expressed by the $\omega^\mu$ field as
\begin{align}
  j^\mu = \frac{\partial V_\omega}{\partial \omega_\mu} = 2 \frac{\partial V_\omega}{\partial(\omega_\mu\omega^\mu)} \omega^\mu \equiv f(\omega_\mu\omega^\mu)\omega^\mu.
\end{align}
Since $\omega^2$ can be solved as a function of $n$ using $n^2 = j^2 =
f(\omega^2)\omega^2$, $\omega\cdot dj = [n/f(\omega^2)] dn$ is integrable.  The
one-particle scalar field is evaluated as
\begin{equation}
  \tilde{\sigma}
 = \frac{1}{n_s}\left(\sigma n_s - \int n_s d\sigma \right)
 = \sigma - \frac{V_\sigma}{n_s},
\end{equation}
and the derivative of the $\tilde{\sigma}$ may be evaluated as
\begin{equation}
  \frac{\partial\tilde{\sigma}}{\partial n_s}
  = \frac{V_\sigma}{n^2_s}.
\end{equation}
In the same way, $\tilde{\omega}$ is expressed by the function of $\omega$
field
\begin{align}
  \tilde{\omega}_\mu
  &= \frac{j_\mu}{n^2}\left(\omega \cdot j - \int j\cdot
  d\omega\right)
   = \omega_\mu -  \frac{j_\mu}{n^2} V_\omega.
\end{align}
The derivative of $\tilde{\omega}$ with respect to the vector current reads
\begin{equation}
\frac{\partial \tilde{\omega}_\mu}{\partial j^\nu}
 = \frac{\partial \omega_\mu}{\partial j^\nu}
 - \frac{j_\mu}{n^2}\left(
     j^\lambda
    \frac{\partial \omega_\lambda}{\partial j^\nu}
    - \frac{2V_\omega}{n^2}j_\nu
     \right)
 -  \frac{V_\omega}{n^2}g_{\mu\nu}.
\end{equation}
As worked out in Appendix~\ref{sec:nlvector}, the derivative of the vector
field with respect to the vector current for the quartic
interaction~\eqref{eq:omega4} is evaluated as
\begin{equation}
\frac{\partial \omega_\mu}{\partial j^\nu}
= \frac{1}{m^2+c\omega^2} \biggl(
  g_{\mu\nu} - \frac{2c\omega_\mu \omega_\nu}{m^2+3c\omega^2}\biggr).
\end{equation}

We are now in a position to rewrite the BUU-like
EoMs~\eqref{eq:application.mf.eom1} to derive the QMD-like EoMs for the scalar
and vector fields.
The EoM for the momentum is written as the sum of two forces from the
scalar and vector potentials:
\begin{equation}
\frac{d\bm{p}_i}{ds}
  = \bm{F}_{\sigma,i} + \bm{F}_{\omega,i}.
\end{equation}
After replacing the $\sigma$ and $\omega$ fields with the
$\tilde{\sigma}$ and $\tilde{\omega}$ fields by using the relations,
\begin{align}
  \sigma &= \frac{\partial (n_s\tilde{\sigma})}{\partial n_s}
  = \tilde{\sigma} + n_s\frac{\partial \tilde{\sigma}}{\partial n_s}, \\
  \omega_\mu &= \frac{\partial (\tilde{\omega}\cdot j)}{\partial j^\mu}
  = \tilde{\omega}_\mu + j^\nu\frac{\partial \tilde{\omega}_\nu}{\partial j^\mu},
\end{align}
and replacing the scalar density and vector current by the sum of the
Gaussian using Eqs.~\eqref{def:application.ns} and~\eqref{def:application.j},
we obtain
\begin{align}
\bm{F}_{\sigma,i} &=
\sum_{j=1}^N \int d^4x\, y_{ij}\biggl[
    \frac{\partial\tilde{\sigma}(x)}{\partial n_s}
    \frac{\partial\tilde{g}(x-x_j)}{\partial\bm{x}} \tilde{G}(x-x_i)\nonumber \\
  &+
   \frac{\partial\tilde{\sigma}(x)}{\partial n_s}
   \frac{\partial \tilde{G}(x-x_i)}{\partial\bm{x}_i}
   \tilde{g}(x-x_j)
  \biggr],
\end{align}
and
\begin{align}
\bm{F}_{\omega,i} &=
-\sum_{j=1}^N \int d^4x\, u_{ij}^{\mu\nu}
    \biggl[
     \frac{\partial\tilde{\omega}_\mu(x)}{\partial j^\nu}
  \frac{\partial\tilde{g}(x-x_j)}{\partial\bm{x}} \tilde{G}(x-x_i)\nonumber \\
  &+ \frac{\partial\tilde{\omega}_\nu(x)}{\partial j^\mu}
   \frac{\partial \tilde{G}(x-x_i)}{\partial\bm{x}_i}
   \tilde{g}(x-x_j)
  \biggr],
\end{align}
where $y_{ij}=y_iy_j$ and $u^{\mu\nu}_{ij}= g_{\omega,i}g_{\omega,j} u_{i}^\mu u_{j}^\nu$.
We approximate the integral by the QMD2 approximation~\eqref{eq:ourQMDEoM},
which replaces the argument $x$ of the potential in the integral with the
center of a Gaussian wave packet and performs the integral for the overlap of
two Gaussians.
Then, we obtain EoMs that are symmetric (in the sense of the law of action and
reaction for each particle pair) and exactly conserve the total momentum,
with the scalar part of the force being
\begin{equation}
  \bm{F}_\sigma
  = \sum_{j=1}^N  y_{ij} \left[
      \frac{V_\sigma(x_i)}{n_s^2(x_i)}
      +\frac{V_\sigma(x_j)}{n_s^2(x_j)}
            \right]
     \frac{\partial \tilde{g}_{ij}}{\partial\bm{x}_i},
\end{equation}
and the vector part being
\begin{equation}
  \bm{F}_{\omega,i}
  = -\sum_{j=1}^N
     u_{ij}^{\mu\nu}
   \left[
      \frac{\partial\tilde{\omega}_\mu(x_i)}{\partial j^\nu}
      + \frac{\partial\tilde{\omega}_\nu(x_j)}{\partial j^\mu}
            \right]
     \frac{\partial \tilde{g}_{ij}}{\partial\bm{x}_i}.
\end{equation}
The symbol $\tilde{g}_{ij}$ denotes the relativistic Gaussian overlap integral:
\begin{align}
  \tilde{g}_{ij} &:= \int d^4x \delta((x-x_i)\cdot \hat{a})\tilde{g}(x-x_i)\tilde{g}(x-x_j) \notag \\
  &=\frac{1}{(4\pi L)^{3/2}\sqrt{d}}
    \exp\frac{(x_i-x_j)^\mathrm{T} A(x_i-x_j)}{4L},
\end{align}
where
\begin{align}
  A^{\mu\nu} &=g^{\mu\nu}
    - \frac{u_{ij}^{\mu\nu}}{2d},
\end{align}
with
\begin{align}
u_{ij}^{\mu\nu} &=   b u_i^\mu u_i^{\nu} + a u_j^{\mu} u_j^{\nu}
  + c(u_{i}^{\mu}u_j^\nu +u_j^{\mu}u_i^\nu),\\
 a &= 1+\frac{1}{2}[(u_i\cdot \hat{a})^2 - u_i^2],\\
 b &= 1+\frac{1}{2}[(u_j\cdot \hat{a})^2 - u_j^2],\\
 c& = \frac{1}{2}[(u_i\cdot\hat{a})(u_j\cdot\hat{a}) - u_i\cdot u_j],\\
 d& = ab - c^2.
\end{align}
A derivation of the above expressions is given in
Appendix~\ref{sec:GaussianOverLap}\@.  We finally note that the exact energy
conservation is not maintained by the QMD2 approximation in general unlike the
momentum conservation, and thus the violation of the energy conservation based
on these EoMs should be checked carefully as we do in
Sec.~\ref{sec:validation}.

\begin{figure}[tbhp]
\includegraphics[width=8.5cm]{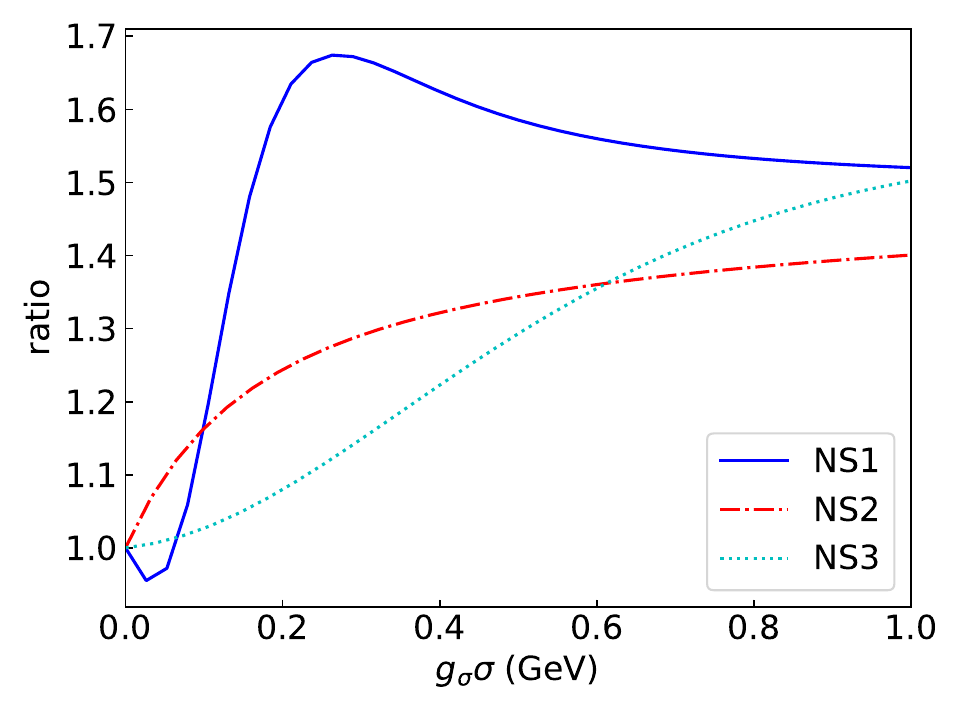}
\caption{ Ratio of Eq.~\eqref{eq:mbarg} to Eq.~\eqref{eq:mstar} as a
function of $\sigma$ for the parameter sets of NS1, NS2, and NS3 taken from
Ref.~\cite{Nara:2020ztb}.  }
\label{fig:mbarg}
\end{figure}

In the previous RQMD.RMF model~\cite{Nara:2019qfd,Nara:2020ztb},
the mass-shell condition is taken to be
\begin{equation}
 (m_i + V_s)^2 - (p_i -  V_i)^2=0,
\end{equation}
and we simply assumed $V_s=-g_\sigma \sigma/2$
and $V_i = g_\omega \omega/2$
to symmetrize the EoMs to conserve total
momentum.  When fields do not contain the nonlinear terms, these choices
are identical to the one-particle fields $\tilde{\sigma}$ and
$\tilde{\omega}$.
In the previous RQMD.RMF model,
the relevant term for the scalar part of the EoMs is
$\partial\sigma/\partial n_s = (\partial^2 V_\sigma/\partial \sigma^2)^{-1}$.
Thus, it is interesting to compare the differences of the present and
previous models.  We compare them for the scalar potential~\eqref{eq:sigma4}
\begin{equation}
 \frac{V_\sigma}{n_s^2}=
  V_\sigma
    \left(
   \frac{\partial V_\sigma}{\partial\sigma}
   \right)^{-2}
 = \frac{\frac{m_\sigma^2}{2} + \frac{g_2}{3}\sigma + \frac{g_3}{4} \sigma^2}
 {(m_\sigma^2 + g_2\sigma + g_3 \sigma^2)^2},
\label{eq:mbarg}
\end{equation}
and
\begin{equation}
 \frac{1}{2}\left(\frac{\partial^2
V_\sigma}{\partial \sigma^2}\right)^{-1}
 = \frac{1}{2}(m_\sigma^2 + 2g_2\sigma + 3g_3 \sigma^2)^{-1}.
\label{eq:mstar}
\end{equation}

In Fig.~\ref{fig:mbarg}, we compare the ratio of Eq.~\eqref{eq:mbarg} to
Eq.~\eqref{eq:mstar} as a function of $\sigma$ field for the EoS parameter sets
of NS1, NS2, and NS3 taken from Ref.~\cite{Nara:2020ztb}.  It is seen that the
new approach predicts larger values than the previous model for all parameter
sets at the large $\sigma$ field.  We may make an order estimate for the
typical values of $\sigma$ field by neglecting the nonlinear term in the
$\sigma$ field, which yields $g_\sigma\sigma = (g_\sigma/m_\sigma)^2n_s\approx
0.18$, $0.27$, and $0.3$ GeV for NS1, NS2, and NS3 at the normal nuclear
density, respectively.  In this range of the $\sigma$ field, the difference is
within 30\%.

\begin{figure}[tbhp]
\includegraphics[width=8.0cm]{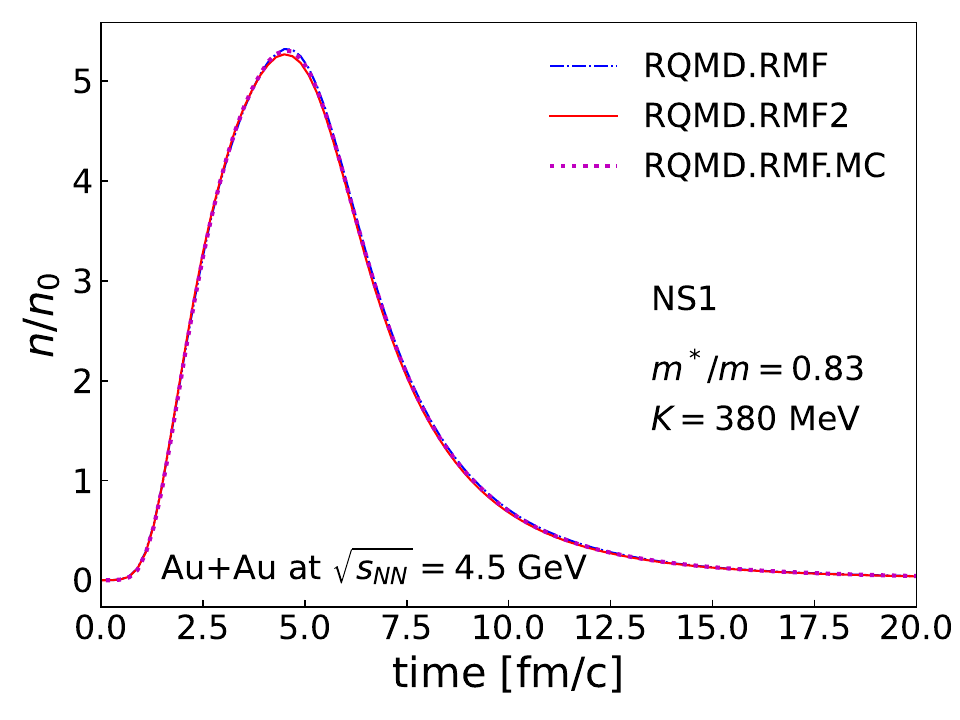}
\caption{Time evolution of the baryon density at the center of mid-central
Au + Au collisions at $\sqrt{s_{NN}}=4.5$ GeV from the RQMD with the
$\sigma$-$\omega$ model.  Impact parameter range of $4.6 < b < 9.4$ fm is
selected.  The RQMD.RMF model results are compared with the new model, RQMD.RMF2,
and the RQMD using the Monte Carlo integration (RQMD.RMF.MC).  }
\label{fig:dens45rmf}
\end{figure}

\subsection{Application to heavy-ion collisions with RQMD.RMF2}
\label{sec:numericaltest.rmf}


To investigate the effects of different treatments in the nonlinear term for
the scalar interaction, the time evolution of the central density in
mid-central Au + Au collisions at $\sqrt{s_{NN}}= 4.5\, \text{GeV}$ is shown in
Fig.~\ref{fig:dens45rmf} for the NS1 EoS\@.  For the specific calculations in
this subsection, we have chosen the uniform foliation with $\hat a = (1,0,0,0)$
in the initial global CM frame defined by the initial total momentum.  The
impact parameter range is
selected as $4.6 < b < 9.4$ fm, which approximately corresponds to the 10--40\%
centrality class in STAR experiments.  The results are compared for three
models: the previous model (dubbed RQMD.RMF, shown by the dash-dotted line),
the new model (RQMD.RMF2, the solid line), and the Monte Carlo integration
(RQMD.RMF.MC, the dotted line),
where we solve Eqs.~\eqref{eq:eom_rqmdsv1} and \eqref{eq:eom_rqmdsv2}.

The comparison reveals no significant
differences in the density evolution among the three models.  We also note that no
significant differences are observed in the 0--5\% central Au + Au collisions.

\begin{figure}[tbhp]
  \includegraphics[width=8.0cm]{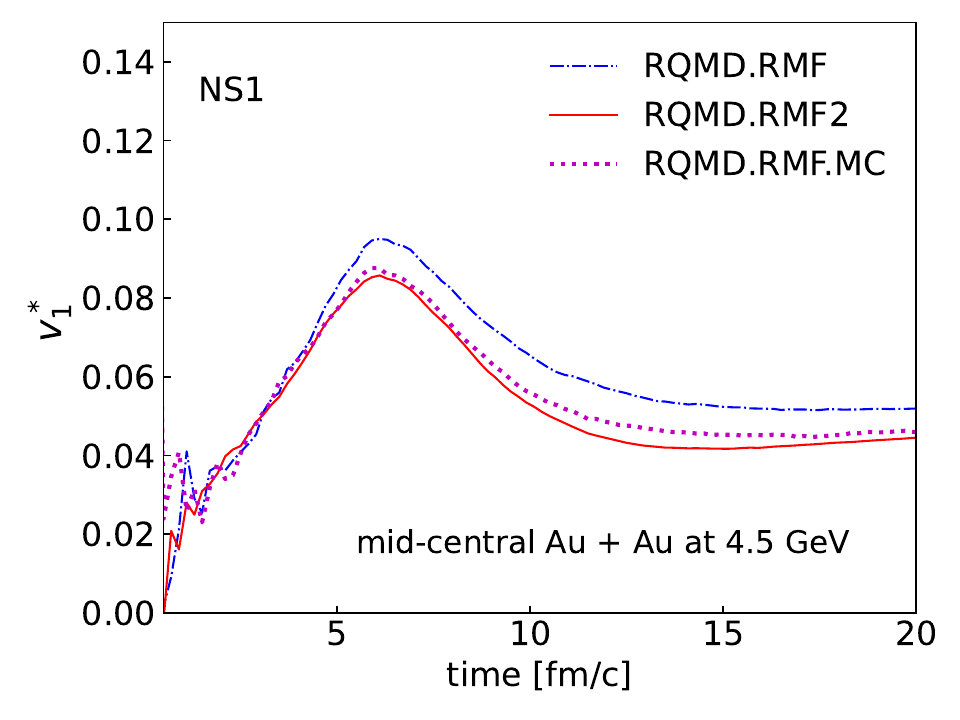}
  \caption{Time evolution of the sign-weighted directed flow $v_1^*$ in
  mid-central Au + Au collisions at $\sqrt{s_{NN}}=4.5$ GeV are shown from the
  RQMD with the $\sigma$-$\omega$ model.  The RQMD.RMF model results are
  compared with the new model, RQMD.RMF2, and the RQMD using the Monte Carlo
  integration (RQMD.RMF.MC).  }
\label{fig:timeevolv1NS}
\end{figure}

The directed flow $v_1 = \langle \cos\phi \rangle$ is considered to be a
sensitive probe of the EoS, where $\phi$ is the azimuthal angle with respect to
the reaction plane.  Figure~\ref{fig:timeevolv1NS} shows the time evolutions of
the sign-weighted directed flow of baryons defined as
\begin{align}
  v^*_1
  &= \frac{\int_{-1}^1 dy \int d\phi \frac{dN}{dyd\phi} \cos\phi \operatorname{\mathrm{sgn}} y}{\int_{-1}^1 dy \int d\phi \frac{dN}{dyd\phi}},
 \label{eq:sign-weightedv1}
\end{align}
where the range of the integration is taken to be $|y|<1$.  As pointed out in
Ref.~\cite{Nara:2021fuu}, the directed flow at mid-rapidity increases during
the compression stage of the collisions and then decreases in the expansion
stages due to a tilted shape of the matter, which is also observed with the
interaction by the relativistic mean-field.  We observe that RQMD.RMF2 provides
a good approximation to RQMD.RMF.MC\@.  RQMD.RMF predicts a larger directed
flow than the new models due to a less attractive scalar potential.

\begin{figure}[tbhp]
  \includegraphics[width=8.0cm]{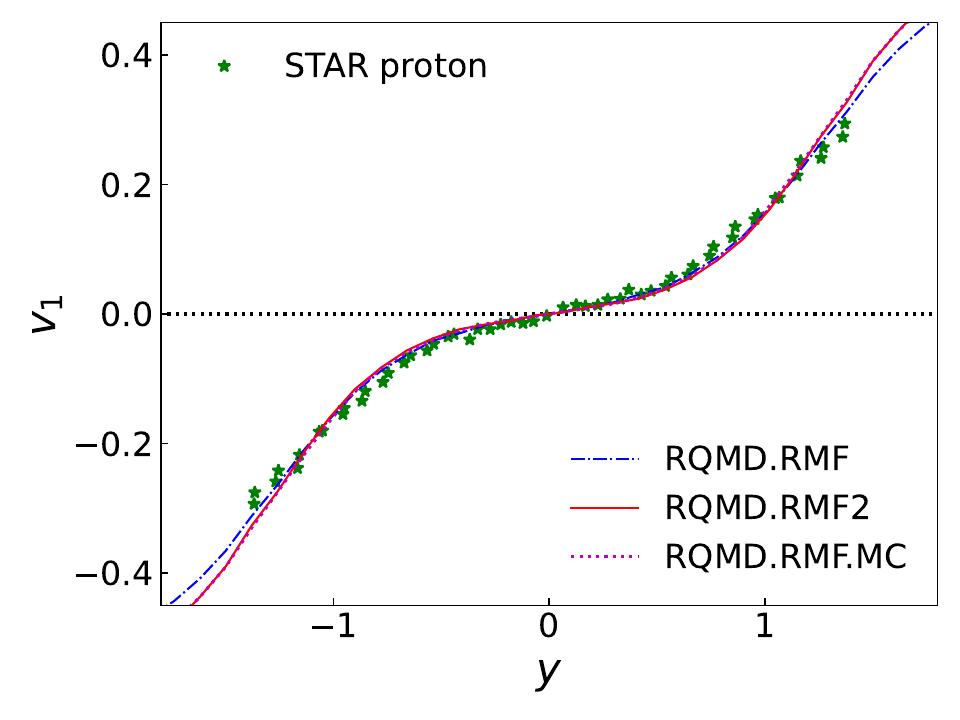}
  \caption{Rapidity dependence of directed flow for protons with transverse
  momentum $0.4 < p_T < 2.0$ GeV in mid-central Au + Au collisions at
  $\sqrt{s_{NN}}=4.5$ GeV is compared with three models within RQMD with the
  $\sigma$-$\omega$ potential.  The result from the RQMD.RMF model is shown as
  a dash-dotted line, the RQMD.RMF2 result is plotted by a solid line, and
  the RQMD.RMF.MC result is plotted by a dotted line.  STAR data were taken
  from~\cite{STAR:2020dav}.  }
\label{fig:v1_NS145}
\end{figure}

In Fig.~\ref{fig:v1_NS145}, we compare the rapidity dependence of the proton
$v_1$ among the three models in mid-central Au + Au collisions at
$\sqrt{s_{NN}}=4.5$ GeV\@.  We select ``free'' protons using nuclear coalescence
with the phase space cutoff parameters $r_0 < 3$ fm and $p_0 < 0.3$ GeV/$c$
at the simulation time of 40~fm/$c$.
We do not observe any significant differences among the three models.
It is seen at high rapidity region that the RQMD.RMF2 results are in good agreement with the
RQMD.RMF.MC results.
We note that this agreement is also seen in the rapidity distribution.
In these models, the treatment of the $\omega$ field is the same, because only the
linear dependence on the baryon density is included.  The effects of the
nonlinear density dependence in the vector potential will be examined in the
next section.

We have checked that the main differences between the old model and the new
models for density evolution and directed flow from other parametrization of
NS2 and NS3 predictions are the same as those of NS1.

\subsection{RQMD for Skyrme potentials (RQMDs2 and RQMDv2)}
\label{sec:skyrme}

In this section, we consider the Skyrme-type density-dependent
potential together with the Lorentzian-type momentum-dependent potential.
In the following discussion, we consider the specific structure of the
free-energy functional:
\begin{align}
  F[\phis, \phiv_\mu] = \int d^4x \mathcal{F}(x),
\end{align}
with
\begin{multline}
  \mathcal{F}(x)
  = \int_0^{\rho_s[\phi(x,p)]} U_s^d(z) dz + \int_0^{\rho[\phiv_\mu(x,p)]} U_v^d(z) dz \\
  + \frac{C}{2\rho_0}\int d^4pd^4p'\, D(p-p') \phis(x,p) \phis(x,p') \\
  + \frac{C}{2\rho_0}\int d^4pd^4p'\, D(p-p') \phiv_\mu(x,p) \phi^\mu(x,p'),
  \label{eq:skyrme.free}
\end{multline}
where $C$ is fixed by fitting the optical potential and $\rho_0$ is the normal
nuclear density.  The functions $U_s^d(z)$ and $U_v^d(z)$ specify the
density-dependent parts of the potential, and $D(p-p')$ specifies the
momentum-dependent part.
The functional form of the momentum-dependence is taken to be
\begin{equation}
  D(p-p') = \frac1{1-(p-p')^2/\mu^2},
  \label{def:D}
\end{equation}
where $\mu$ is a parameter for the momentum width.
The scalar density $\rho_s[\phi]$ and the current density $\rho[\phiv_\mu]$ are
given by
\begin{align}
  \rho_s(x) &= \int d^4p\, \phis(x, p), \\
  J_\mu(x) &= \int d^4p\, \phiv_\mu(x, p), \\
  \rho(x) &= \sqrt{J^\mu(x) J_\mu(x)}.
\end{align}
The single-particle potentials are obtained as
\begin{align}
  U_s(x, p)
    &= U_s^d(\rho_s(x)) + U_s^\mathrm{MD}(x,p),
    \label{def:skyrme.Us} \\
  U_\mu(x, p)
    &= U^d_\mu(J^\mu(x)) + U_\mu^\mathrm{MD}(x, p),
    \label{def:skyrme.Uv}
\end{align}
with Skyrme-type density-dependent single-particle potential as
\begin{equation}
U_{s,v}^d(n(x)) = \alpha \frac{n(x)}{\rho_0}
   + \beta \biggl[\frac{n(x)}{\rho_0}\biggr]^\gamma,
\label{eq:SkyrmePotential}
\end{equation}
where $n(x)$ can be either the scalar density or the baryon density
corresponding to the notations $U_s(\rho_s)$ and $U_v(\rho)$, respectively.
The density-dependent part of the vector potential is
defined as $U^d_\mu := (J_\mu/\rho) U_v^d(\rho)$.
According to Eqs.~\eqref{eq:sec3.spot} and \eqref{eq:sec3.vpot},
the single-particle momentum-dependent potentials become
\begin{align}
  U_s^\mathrm{MD}(x,p) &= \frac{C}{\rho_0}\int d^4p'\, D(p-p')\phis(p')\nonumber \\
 &=\frac{C}{\rho_0}\sum_{i=1}^N
 D(p-p_i)\tilde{g}(x-x_i), \label{eq:Usm} \\
  U_\mu^\mathrm{MD}(x, p) &= \frac{C}{\rho_0} \int d^4p'\, D(p-p')\phiv_\mu(p')\nonumber\\
 &=\frac{C}{\rho_0}\sum_{i=1}^N
  u_{i,\mu} D(p-p_i)\tilde{g}(x-x_i),\label{eq:Uvm}
\end{align}
where we make simplifications as stated in Sec.~\ref{sec:numerics}:
$\bar{m}_i=m^*(x,p_i)$ and $u_i=p^*(x,p_i)/m^*(x,p_i)$.
We here note that in general the kernel $D(p-p')$ appearing in the single-particle
momentum-dependent potentials~\eqref{eq:Usm} and~\eqref{eq:Uvm} need to be
symmetric with respect to its arguments, $D(p-p') = D(p'-p)$, which is required
to ensure conservation laws of current and energy-momentum~\cite{Weber:1992qc}.
In the present formulation starting with the free energy
functional~\eqref{eq:skyrme.free}, the kernel $D(p-p')$ in the free energy
functional does not have this restriction for the symmetry because the kernel
in Eqs.~\eqref{eq:Usm} and~\eqref{eq:Uvm} becomes $[D(p-p') + D(p'-p)]/2$,
which automatically satisfies the symmetry.
For simplicity, we replaced $[D(p-p') +
D(p'-p)]/2$ with $D(p-p')$ using the symmetry of Eq.~\eqref{def:D}, $D(p-p') =
D(p'-p)$.
The field potential is given by
\begin{align}
  V[\phis, \phiv_\mu] &= \int d^4x \mathcal{V}_\mathrm{field}(x),
\end{align}
with
\begin{align}
  \mathcal{V}_\mathrm{field}(x) &= \rho_s U_s^d(\rho_s) - \int_0^{\rho_s} U_s^d(z) dz \notag \\
    & \quad + \rho U_v^d(\rho) - \int_0^{\rho} U_v^d(z) dz \notag \\
    & \quad + \frac12 \int d^4p\, \phis(x, p) U_s^\mathrm{MD}(x,p) \notag \\
    & \quad + \frac12 \int d^4p\, \phi^\mu(x, p) U_\mu^\mathrm{MD}(x, p).
\end{align}

We now consider the QMD2 approximation~\eqref{eq:ourQMDEoM} of the spatial integral.  We apply the same
procedure as discussed in the previous section for the integration of
Eqs.~\eqref{eq:eom_rqmdsv1} and~\eqref{eq:eom_rqmdsv2} to obtain the covariant
QMD-like EoMs.  We eliminate the $p^0$ dependence of the momentum-dependent
potentials $U_s^\mathrm{MD}(x,p)$ and $U_\mu^\mathrm{MD}(x,p)$ by substituting
the mass-shell constraint for $p^0$ so that $\Pi_i^0=p_i^{*0}$.  In the
center-of-mass frame where $\hat{a}=(1,0,0,0)$, $\Pi_i\cdot\hat{a}= p^{*0}_i$.
Let us consider below in the Lorentz frame specified by $\hat{a}=(1,0,0,0)$,
where the times of all particles are the same $t=x_i^0$.  The time integration
can be trivially done, and the Gaussian becomes the following form:
\begin{align}
  \tilde{g}(\bm{x};u_i)
  &= \int dx^0\, \tilde{G}(x) \notag \\
  &= \frac{1}{(2\pi L)^{3/2}}
     \exp\left[-\frac{\bm{x}^2 + (\bm{x}\cdot\bm{u}_i)^2}{2L}
        \right],
\end{align}
It is sufficient to propagate the spatial components of momentum $\bm{p}_i$,
since the time component $p^0_i$ can be determined by the mass-shell
constraint. In this work, we use a mass-shell condition within the harmonic
approximation for simplicity, i.e., $p_i^0$ is determined by solving
$W(x_i,p_i) = 0$ instead of $\bar W(x_i,p_i) = 0$.

The EoMs~\eqref{eq:eom_rqmdsv1} and~\eqref{eq:eom_rqmdsv2} receive
contributions from the four terms in the scalar and vector
potentials~\eqref{def:skyrme.Us} and~\eqref{def:skyrme.Uv}:
\begin{align}
  \frac{d\bm{x}_i}{dt}
  &= \frac{\bm{p}_i^*}{p_i^{*0}}
  + \bm{E}_{s,i}^\mathrm{MD}
  + \bm{E}_{v,i}^\mathrm{MD}, \\
  \frac{d\bm{p}_i}{dt}
  &= \bm{F}_{s,i}^d
  + \bm{F}_{v,i}^d
  + \bm{F}_{s,i}^\mathrm{MD}
  + \bm{F}_{v,i}^\mathrm{MD},
\end{align}
where $\bm{F}_{s,i}^d$ and $\bm{F}_{v,i}^d$ denote the forces from the
density-dependent parts of the scalar and vector potentials, respectively. The
symbols $\bm{F}_{s,i}^\mathrm{MD}$ and $\bm{F}_{v,i}^\mathrm{MD}$ denote the
forces from the momentum-dependent parts, while $\bm{E}_{s,i}^\mathrm{MD}$
and $\bm{E}_{v,i}^\mathrm{MD}$ denote similar contributions to the spatial part
of the EoM\@.

To obtain the density-dependent parts of the EoM, we introduce one-particle
potentials for the density-dependent potential defined as
\begin{align}
  V_s(x) &= \frac{1}{\rho_s}\int U_s^d(\rho_s)d\rho_s,
    \label{eq:QMDscalar}\\
  V^\mu(x)
    &= \frac{J^\mu}{\rho^2}\int U_\alpha^d(\rho) dJ^\alpha
    = \frac{J^\mu}{\rho^2}\int U_v^d(\rho) d\rho.
    \label{eq:QMDvector}
\end{align}
Then, the single-particle potentials can be replaced with the one-particle
potentials by using the relations
\begin{equation}
  U_s^d(x)=  V_s + \rho_s \frac{\partial V_s}{\partial \rho_s}, \qquad
  U_\mu^d(x)= V_\mu + J^\alpha\frac{\partial V_\alpha}{\partial J^\mu}.
\end{equation}
The density-dependent scalar potential part of the EoMs is given by
\begin{align}
  \bm{F}_{s,i}^d
    &= - \int d^3x \frac{\partial U_s^d(x)}{\partial \bm{x}} \tilde{g}(x-x_i) .
\end{align}
We replace the single-particle potential $U^d_s(x)$ with the one-particle
potential $V_s(x)$, and replace the total scalar density $\rho_s(x)$ with the
sum of the Gaussian profiles.  We obtain
\begin{align}
  \bm{F}_{s,i}^d
  &= -\sum_{j=1}^N  \biggl[ \int d^3x\,
       \frac{dV_s(x)}{d\rho_s}\frac{\partial \tilde{g}(x-x_j)}{\partial\bm{x}}
          \tilde{g}(x-x_i)\nonumber\\
     &+ \int d^3x \frac{dV_s(x)}{d\rho_s}
     \frac{\partial \tilde{g}(x-x_i)}{\partial\bm{x}_i} \tilde{g}(x-x_j)
            \biggr].
\end{align}
We approximate this integral by replacing the argument of the potential by the
center of the Gaussian $x_i$ and $x_j$, and perform integration, which yields
\begin{equation}
  \bm{F}_{s,i}^d \approx
  -\sum_{j=1}^N  \left[
       \frac{dV_s(x_i)}{d\rho_s}
     +  \frac{dV_s(x_j)}{d\rho_s}
            \right]
     \frac{\partial \tilde{g}_{ij}}{\partial\bm{x}_i}\,,
\end{equation}
where
\begin{equation}
  \tilde{g}_{ij} = \int d^3x\,\tilde{g}(x-x_i)\tilde{g}(x-x_j).
\end{equation}

Similarly, the vector potential part is obtained as
\begin{equation}
  \bm{F}_{v,i}^d \approx
  - \sum_{j=1}^N
  u_{i\mu} u_{j\nu}
  \biggl[
    \frac{\partial V^\mu(x_i)}{\partial J_\nu}
    + \frac{\partial V^\nu(x_j)}{\partial J_\mu}
    \biggr]
  \frac{\partial \tilde{g}_{ij}}{\partial\bm{x}_i}.
\end{equation}
The derivative of the vector potential can be evaluated by using $V^\mu=
V_v(\rho)J^\mu/\rho$, with $V_v(\rho)=(1/\rho)\int U_v^d(\rho) d\rho$,
\begin{align}
\frac{\partial V^\mu}{\partial J_\nu}u_\nu
= \biggl[\frac{\partial V_v(\rho)}{\partial \rho}
  - \frac{V_v(\rho)}{\rho} \biggr]
     \frac{J^\nu u_\nu}{\rho}\frac{J^\mu}{\rho}
    + \frac{V_v(\rho)}{\rho} u^\mu.
\end{align}
The contributions from the momentum-dependent scalar potentials are
expressed as
\begin{align}
  \bm{E}_{s,i}^\mathrm{MD}
  &= \frac{C}{\rho_0}\sum_{j=1}^N
      \frac{\partial D(p_i-p_j)}{\partial\bm{p}_i}
         \tilde{g}_{ij}, \\
  \bm{F}_{s,i}^\mathrm{MD}
  &=-\frac{C}{\rho_0}\sum_{j=1}^N
    D(p_i-p_j)
      \frac{\partial \tilde{g}_{ij}}{\partial \bm{x}_i}.
\end{align}
For the momentum-dependent vector potential, we have the expression
\begin{align}
  \bm{E}_{v,i}^\mathrm{MD}
  &= \frac{C}{\rho_0}\sum_{j=1}^N
    (u_i\cdot u_j)
      \frac{\partial D(p_i-p_j)}{\partial\bm{p}_i}
         \tilde{g}_{ij}, \\
  \bm{F}_{v,i}^\mathrm{MD}
  &=-\frac{C}{\rho_0}\sum_{j=1}^N (u_i\cdot u_j)
     D(p_i-p_j) \frac{\partial \tilde{g}_{ij}}{\partial \bm{x}_i}.
\end{align}
In the numerical simulation, to evaluate the momentum-dependent function
$D(p_i-p_j)$, we use the relative momentum $p_i-p_j$ seen in the two-particle
center-of-mass system:
\begin{equation}
 (\bm{p}_i - \bm{p}_j)^2_\mathrm{c.m.}=
  -(p_i-p_j)^2 + \frac{[(p_i-p_j)\cdot(p_i + p_j)]^2}{(p_i+p_j)^2}.
\end{equation}

To summarize, RQMD2 EoMs are obtained as
\begin{align}
\frac{dx^\mu_i}{ds} &= \frac{p^*_i}{\Pi_i\cdot \hat{a}}\nonumber\\
    &- \frac{C}{\rho_0}
     \sum_{j=1}^N
    \left(
    u_i\cdot u_j
  + 1
   \right)
   \frac{\partial D(p_i-p_j)}{\partial p_{i\mu}}
     \tilde{g}_{ij},
  \label{eq:RMQD2.x} \\
\frac{dp_i^\mu}{ds} &=  \sum_{j=1}^N
 \left(  u_{i\mu}u_{j\nu}\, V^{\mu\nu}_{ij}
  + V^s_{ij} \right)
   \frac{\partial \tilde{g}_{ij}}{\partial x_{i\mu}},
  \label{eq:RMQD2.p}
\end{align}
where
\begin{align}
 V^s_{ij} &=
   \frac{\partial V_s(x_i)}{\partial n_{s}}
   +\frac{\partial V_s(x_j)}{\partial n_{s}}
   +\frac{C}{\rho_0}D(p_i-p_j), \\
 V^{\mu\nu}_{ij} &=
   \frac{\partial V^\mu(x_i)}{\partial J_\nu}
   +\frac{\partial V^\nu(x_j)}{\partial J_\mu}
   +\frac{C}{\rho_0}D(p_i-p_j)g^{\mu\nu}.
\end{align}
We define RQMDs2 to be the version only with the scalar part of the Skyrme
potential in Eqs.~\eqref{eq:RMQD2.x} and~\eqref{eq:RMQD2.p}, and RQMDv2 to be
the version only with the vector part.

\subsection{Application to heavy-ion collisions with RQMDv2}
\label{sec:numericaltest}

In this section, we focus on the vector potentials
as they have a stronger influence on the dynamics than the scalar
potentials~\cite{Nara:2021fuu}.
We compare the results of the previous RQMDv approach \cite{Nara:2021fuu} with those
obtained using the newly proposed equations of motion (RQMDv2),
as well as with the exact solutions calculated via numerical integration
using the Monte Carlo method with a Gaussian weight (RQMDv.MC,
which solves Eqs.~\eqref{eq:eom_rqmdsv1} and
\eqref{eq:eom_rqmdsv2}
by simulating Au+Au collisions with the \texttt{JAM2} event generator.

\begin{figure*}[tbhp]
  \includegraphics[width=8.0cm]{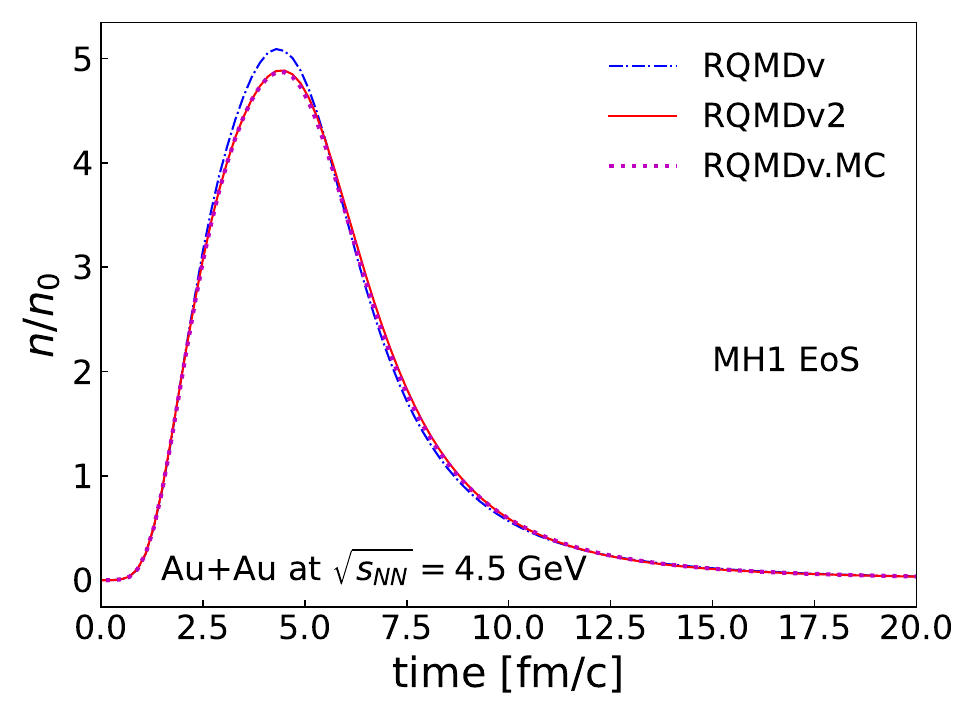}
  \includegraphics[width=8.0cm]{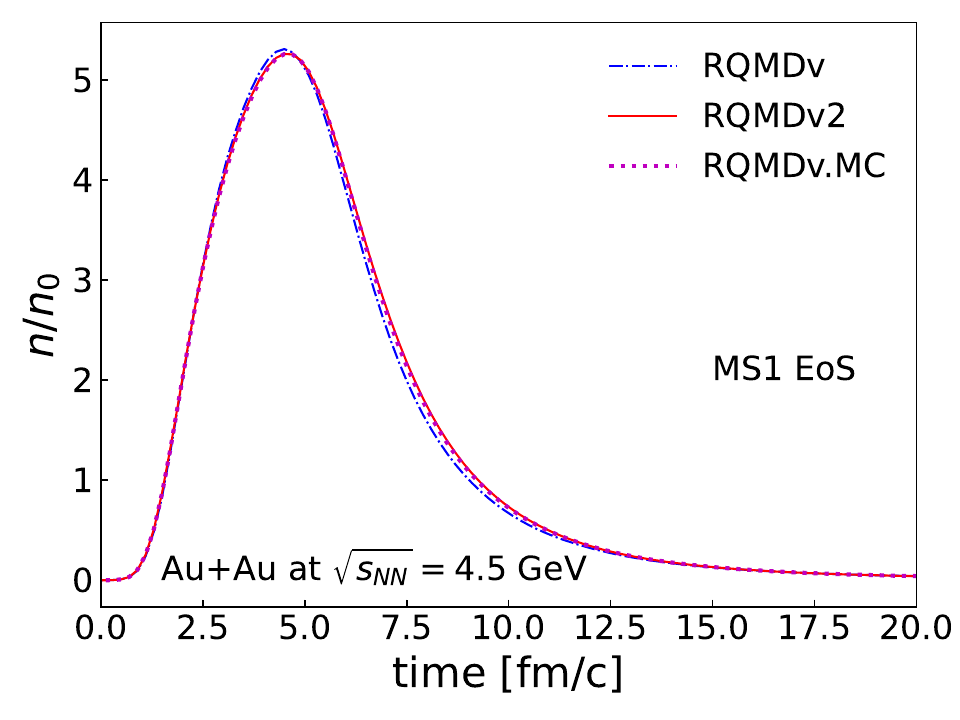}
  \caption{Time evolution of the density at the center of mid-central Au +
  Au collisions at $\sqrt{s_{NN}}=4.5\,\text{GeV}$ from RQMD with the Skyrme
  vector potential.  The left panel shows the results from hard
  momentum-dependent EoS (MH1) and the right panel shows the results from the soft momentum-dependent
  EoS (MS1).  The dashed-dotted line corresponds to the result from the previous
  RQMDv model. The RQMDv2 model prediction is shown by the solid line, and
  the RQMDv.MC result is shown by the dashed line.  }
  \label{fig:dens45hard}
\end{figure*}

In Fig.~\ref{fig:dens45hard}, we plot the time evolution of baryon density at
the center of the matter in mid-central Au + Au collisions at
$\sqrt{s_{NN}}=4.5$ GeV\@.  For the calculations in this subsection, we have
chosen the uniform foliation with $\hat a = (1,0,0,0)$ in the initial global CM
frame the same as in Sec.~\ref{sec:numericaltest.rmf}.  We compare two
different EoS taken from
Ref.~\cite{Nara:2021fuu}.  The hard momentum-dependent Skyrme EoS (MH1) uses a
repulsive density-dependent potential~\eqref{eq:SkyrmePotential} with the
exponent $\gamma=2.273$, while the soft Skyrme EoS (MS1) uses $\gamma=1.109$.
The left and right panels show the results from the MH1 and MS1, respectively.
For the MH1 EoS, the numerical estimate by the Monte Carlo integration
(RQMDv.MC) suppresses the density compared with the traditional RQMD results,
indicating that the traditional QMD approximation underestimates the repulsive
density-dependent potential. This is consistent with the findings in
Ref.~\cite{TMEP:2023ifw}.  We observe in Fig.~\ref{fig:dens45hard} that our new
approximation method for computing the density-dependent potential (RQMDv2)
reproduces the RQMDv.MC results, confirming that RQMDv2 can be reliably used
for the simulations of heavy-ion collisions, which is numerically much more
efficient than the Monte Carlo integration.  On the other hand, in the case of
the MS1 EoS, there is no significant difference among the three models, because the
density dependence in the vector potential is nearly linear in the MS1 EoS\@.

\begin{figure}[tbhp]
  \includegraphics[width=8.0cm]{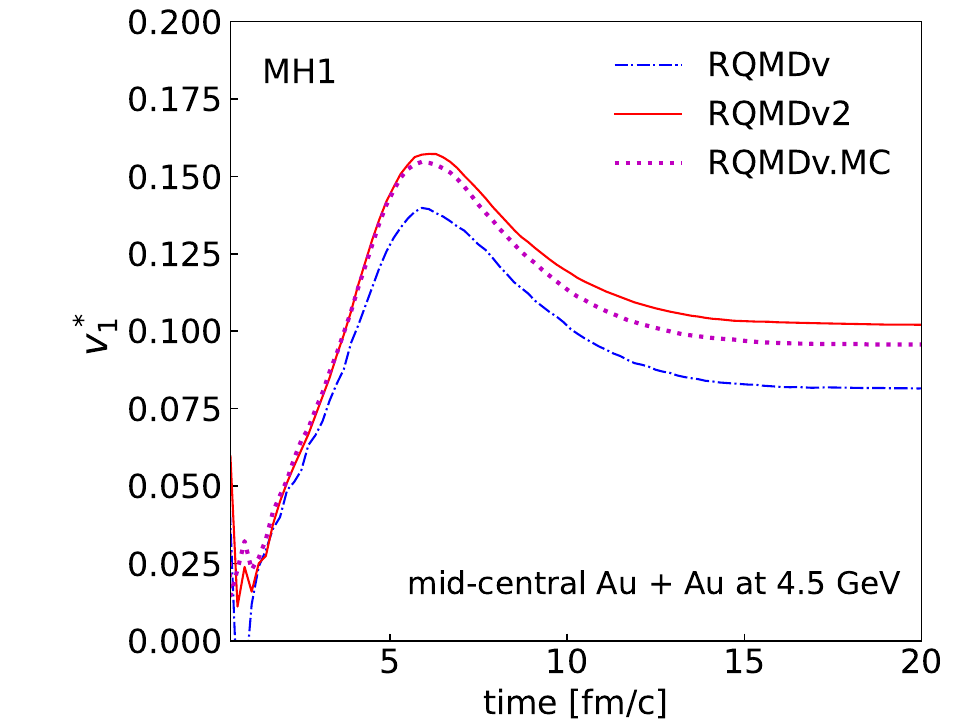}
  \includegraphics[width=8.0cm]{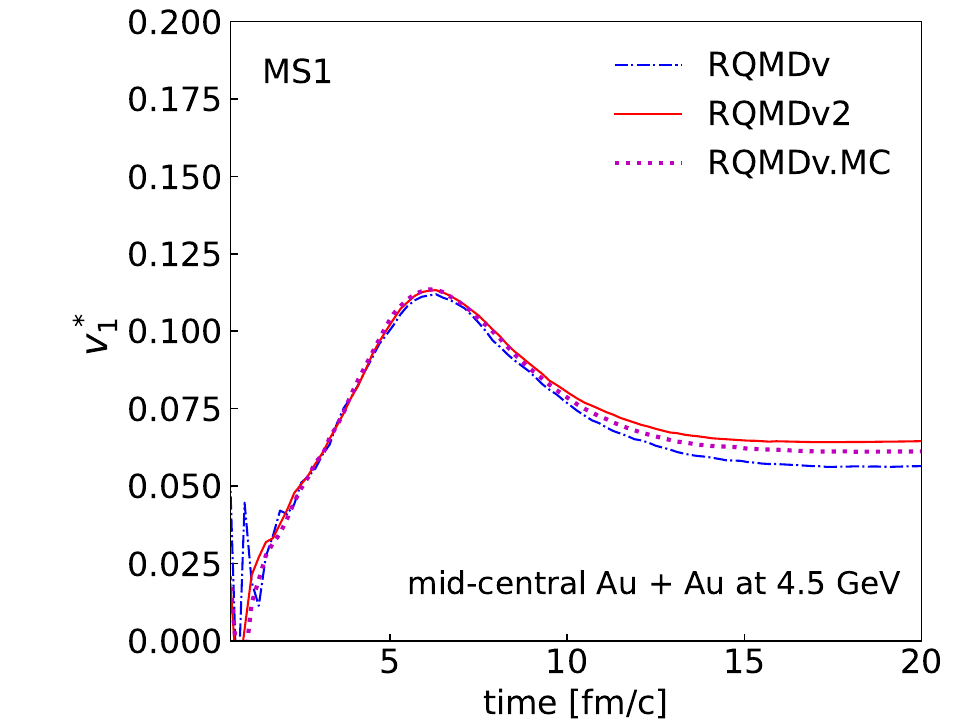}
  \caption{Time evolution of sign-weighted directed flows $v_1^*$ of baryons in
  mid-central Au + Au collisions at $\sqrt{s_{NN}}=4.5\,\text{GeV}$ is
  compared with different models.  Rapidity cut for the integration of the
  directed flow is $|y|<1.0$.  The meaning of the lines is the same as in
  Fig.~\ref{fig:dens45hard}.}
\label{fig:timeevolv1rqmd45}
\end{figure}

Figure~\ref{fig:timeevolv1rqmd45} shows the time evolutions of the
sign-weighted directed flow $v_1^*$~\eqref{eq:sign-weightedv1} of baryons
integrated in the rapidity range $|y|<1$.  We also confirm that RQMDv2 results
are in good agreement with the RQMDv.MC results for the directed flow, while
the traditional RQMDv predicts a smaller directed flow compared to the RQMDv.MC\@.

\begin{figure}[tbhp]
  \includegraphics[width=8.0cm]{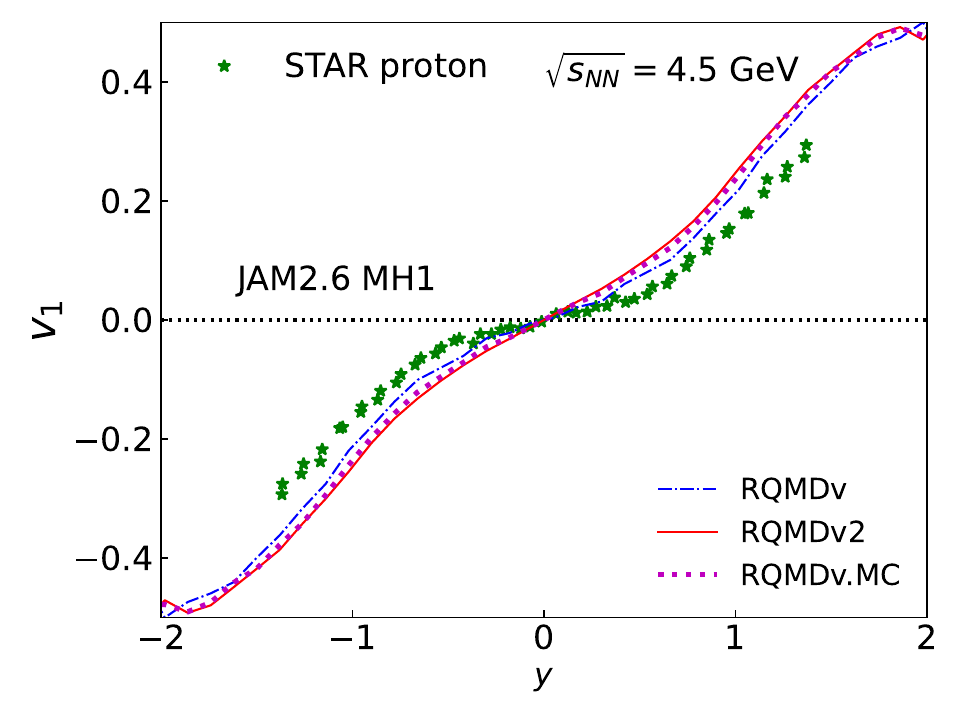}
  \includegraphics[width=8.0cm]{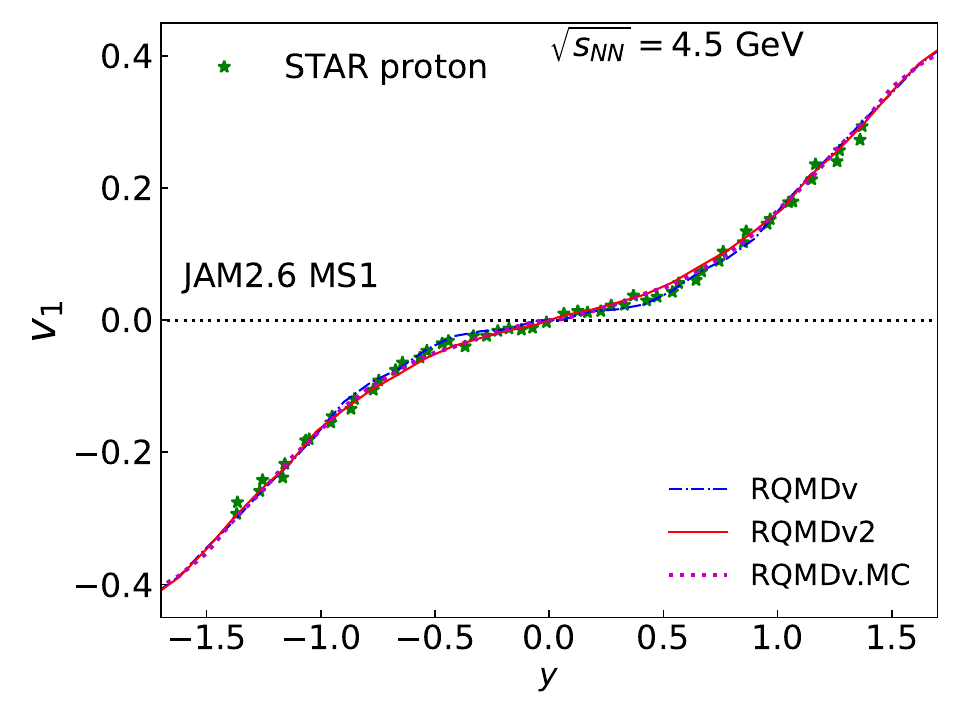}
  \caption{Directed flows of protons in mid-central Au + Au collisions at
  $\sqrt{s_{NN}}=4.5\,\text{GeV}$ are compared with different models.  The
  upper panel shows results obtained with the MS1 EoS, while the lower panel
  shows results from the MH1 EoS\@.  STAR data were taken
  from~\cite{STAR:2020dav}.  The meaning of the lines is the same as in
  Fig.~\ref{fig:dens45hard}.  }
  \label{fig:v1_rqmdv45}
\end{figure}

Finally, we show in Fig.~\ref{fig:v1_rqmdv45} the rapidity dependence of the
proton directed flow $v_1$ in mid-central Au + Au collisions at
$\sqrt{s_{NN}}=4.5$~GeV, compared with the STAR data~\cite{STAR:2020dav}.  For
the MS1 EoS, it is seen that all approaches are in good agreement, and the
directed flow is not sensitive to the treatment of the density-dependent
potential, as expected when the value of $\gamma$ is close to unity.
However, for the MH1 EoS, which uses a large $\gamma$ value, the differences are
observed between the RQMDv and RQMDv2 results.  For a large $\gamma$, RQMDv2
remains a good approximation to RQMDv.MC for describing the rapidity dependence
of the directed flow.

\subsection{Frame dependence, foliation dependence, and energy conservation}
\label{sec:validation}

To validate the Lorentz covariance of the present formulation,
we first demonstrate that the calculated results are independent of
the choice of inertial frame for a fixed foliation.
Furthermore, we also check the extent of the violation of 
the time-reparametrization invariance
and energy conservation by the new formulation and the approximations.

As discussed earlier, we employ a uniform foliation
$\hat{s}(x)=\hat{a}\cdot x$ with the Lorentz vector
$\hat{a}=(1,0,0,0)$ 
defined in the initial global center-of-mass (CM)
frame~\cite{Maruyama:1996rn}.
In this setup, the global CM frame is shown to remain to be the initial global
CM frame at later times after deriving the EoM; we will simply call this frame
the global CM frame hereafter.
When changing the reference frame, the components of $\hat{a}$
must be Lorentz-transformed accordingly to ensure the Lorentz covariance.

\begin{figure}[htbp]
\includegraphics[width=9.5cm]{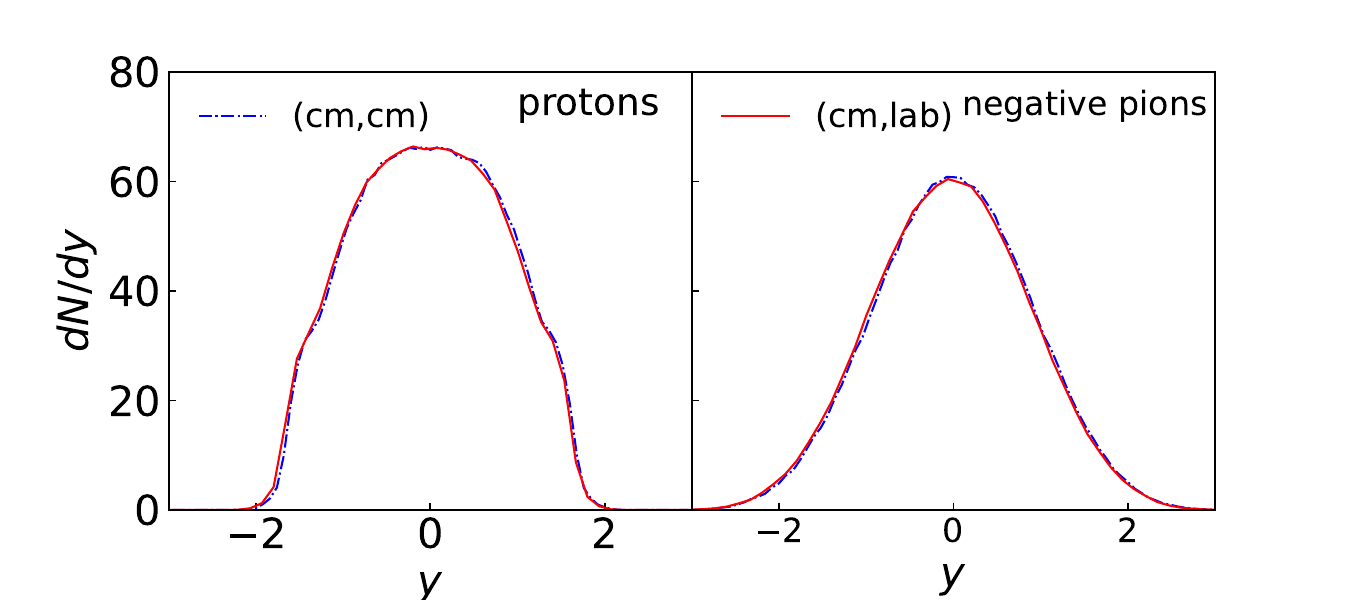}
\caption{Rapidity distributions of the protons (left panel) and negative pions
(right panel) in central Au + Au collisions at $\sqrt{s_{NN}}=4.5$ GeV
are compared for different reference frames:
the global CM frame (blue lines) and the laboratory frame (red lines).
The impact parameters are chosen as $0< b< 3.0$ fm.
}
\label{fig:frame}
\end{figure}

Figure~\ref{fig:frame} shows the rapidity distributions of protons and negative pions
in central Au+Au collisions at $\sqrt{s_{NN}}=4.5$ GeV
obtained using the RQMDv2 model to demonstrate reference-frame independence.
We have confirmed the same behavior with the RQMD.RMF2 model.
The label ``(cm,cm)'' denotes the case where both
$\hat{a}$ and the computational frame are the global CM frame,
shown by the blue curves.
In contrast, ``(cm,lab)'' corresponds to calculations performed
in the laboratory frame---where one gold nucleus is initially at rest---while keeping
$\hat{a}$ fixed to $(1,0,0,0)$ in the global CM frame and Lorentz transformed
to laboratory frame.  These results are shown by the red curves.
The two sets of calculations coincide exactly,
demonstrating frame independence for the same specified foliation.
This frame independence should be distinguished from
the residual dependence on the choice of foliation,
which is investigated in the following subsection.

Having established reference-frame independence for a fixed foliation,
we now examine the dependence on the choice of foliation.
The foliation $\hat{s}(x)$ is a concept independent of the computational reference frame.
Different choices of foliation are possible
as long as it satisfies certain requirements
within the covariant formulation,
but the dynamics itself may change because different foliations
define different evolution hypersurfaces.
A physically reasonable foliation should be determined by the physics
so that it effectively captures the dynamics of a real system.
For the foliation $\hat{s}(x) = \hat{a}\cdot x$,
a natural choice for the Lorentz vector $\hat{a}$
is the inertial frame of the global CM system
which ensures synchronization of particle clocks in the global CM frame%
~\cite{Maruyama:1996rn,Mancusi:2009zz,Marty:2012vs}.
It is shown by Refs.~\cite{Maruyama:1996rn,Nara:2023vrq} that
this foliation reproduces the results obtained with Sorge's
method given by the original RQMD paper~\cite{Sorge:1989dy},
in which the evolution parameters of particles are determined
by reflecting the velocities of surrounding particles.

\begin{figure}[htbp]
\includegraphics[width=9.5cm]{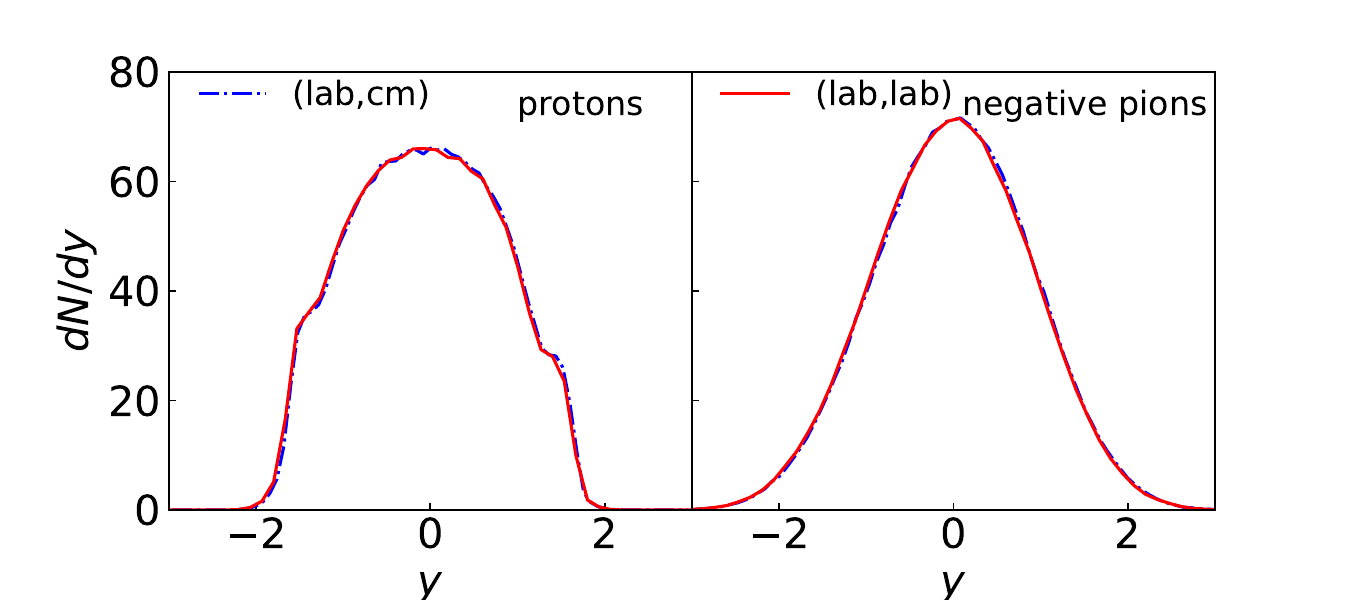}
\caption{Same as Fig.~\ref{fig:frame}, but the $\hat{a}$ is chosen
to be $(1,0,0,0)$ in the laboratory frame.
}
\label{fig:frame2}
\end{figure}

Although the foliation dependence is unavoidable as far as the system is
described by the $N$-body particle coordinates, it is important to examine
results obtained with different foliations to check that the extent of the
violation of the time-reparametrization invariance within the new formulation
is not large.
In Fig.~\ref{fig:frame2}, we present results obtained by choosing
a foliation $\hat s(x)=\hat a'\cdot x$ with $\hat{a}' = (1,0,0,0)$ in the laboratory frame.

The resulting rapidity distributions remain similar to those in
Fig.~\ref{fig:frame}, but are no longer perfectly symmetric,
reflecting the foliation dependence of the RQMD framework written by a finite
number of degrees of freedom.  It is noted that the violation is not a
reference-frame artifact, as confirmed by observing that results computed for a
given foliation remain invariant under the change of the reference frame
between the red solid and blue dot-dashed lines in Fig.~\ref{fig:frame2}.
This clearly distinguishes the residual foliation dependence
from the reference-frame independence demonstrated in Fig.~\ref{fig:frame}.
This also implies that the cluster separability can be approximately maintained
by determining the frame for each cluster~\cite{Nara:2023vrq} because the
results are not largely affected by whether $\hat a$ is determined based on the
global CM frame or the cluster frame.

\begin{figure}[hbtp]
\includegraphics[width=8.5cm]{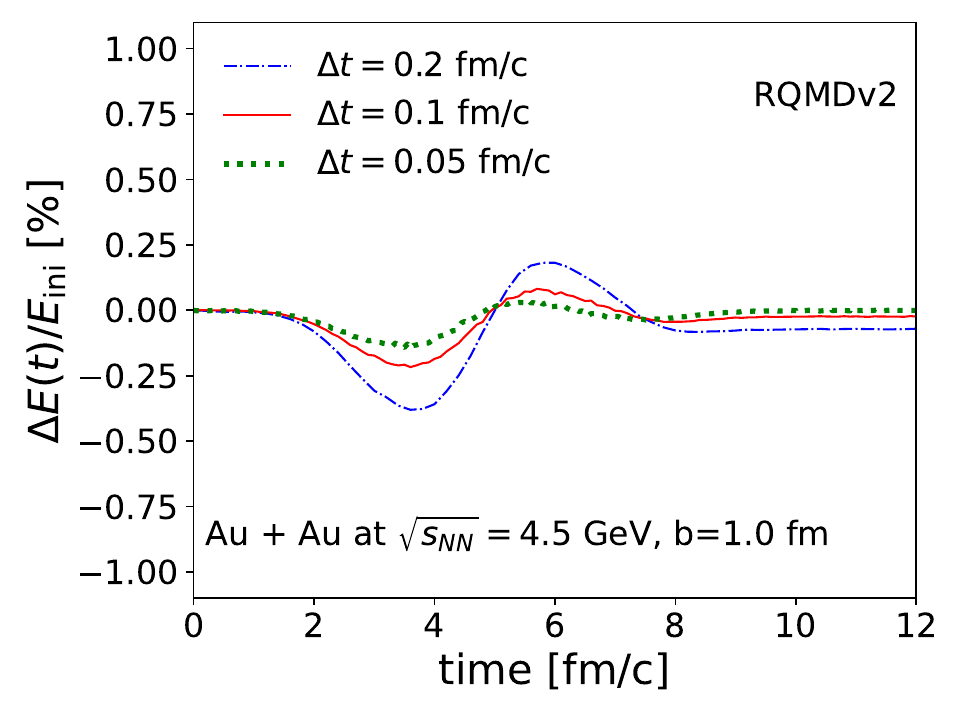}
\includegraphics[width=8.5cm]{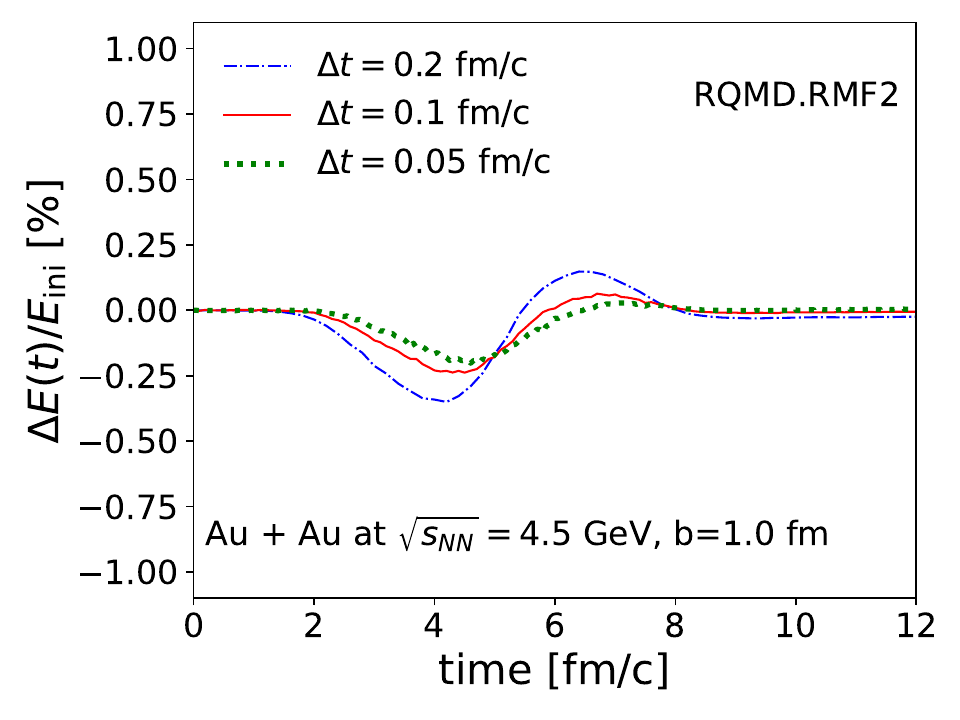}
\caption{Relative error to the initial total energy as a function of time
is plotted for RQMDv2 (upper panel) and RQMD.RMF2 (lower panel) in Au + Au collision at
$\sqrt{s_{NN}}=4.5$ GeV, and $b=1.0$ fm.
The three different time step-sizes of $dt=0.2, 0.1$ and 0.05 fm/$c$ are compared.}
\label{fig:econ}
\end{figure}

In our formulation, momentum is conserved independent of time step size,
because of the strict implementation of action-reaction law in our model.
We now turn to the validation of total energy conservation.
Although our EoMs with the QMD2 approximation exactly conserve momentum,
exact energy conservation is not guaranteed
due to the approximation of the spatial integral in the EoMs.
Thus, we need to check the degree of violation of the total energy
conservation.
Since our approximation to the force has been shown to reproduce well
the results of spatial integration, and since density evolution and directed flow
are consistent with the numerical integration,
we expect good energy conservation.
To see the degree of the total energy conservation,
the relative error of the total energy $E$ from its initial value $E_\mathrm{ini}$,
\begin{equation}
\frac{\Delta E}{E_\mathrm{ini}}
= \frac{\langle E - E_\mathrm{ini}\rangle}{\langle E_\mathrm{ini}\rangle},
\label{eq:rerror}
\end{equation}
is plotted in Fig.~\ref{fig:econ}
as a function of the evolution time in central Au + Au collisions
at $\sqrt{s_{NN}}=4.5$ GeV with impact parameter $b=1.0$~fm.
The results from RQMDv2 (RQMD.RMF2) are shown in the left (right) panel
for three different time-step sizes: $dt=0.2$, $0.1$ and 0.05~fm/$c$.
Even with a larger time-step size of $dt=0.2$~fm/$c$, the relative error remains within 1\%.
It is seen that
the total energy is conserved within 0.25\% when using $dt=0.1$~fm/$c$.
The accuracy of the energy conservation is even better for mid-central collisions.
We also confirm that physical observables such as $dN/dy$ and $v_1$ do not change
when varying the time-step size.
We finally mention that
the relative RMSD
\begin{equation}
  \frac{\sqrt{\langle (E - E_\mathrm{ini})^2\rangle}}{\langle E_\mathrm{ini}\rangle}
\end{equation}
agrees with the relative error~\eqref{eq:rerror} except the sign,
which means that event-by-event fluctuations
in the energy conservation are small.
In fact, we confirmed that the standard deviation of $E - E_\mathrm{ini}$ is
small compared to the typical size of $E - E_\mathrm{ini}$.

\subsection{Difference between RQMD and RQMD2}

Finally, we mention the differences between the previous RQMD
approach~\cite{Nara:2019qfd,Nara:2020ztb,Nara:2021fuu} and the new one.  In the
previous RQMD approach, the EoMs at the center-of-mass frame are obtained by
the Hamiltonian using one-particle potentials,
\begin{equation}
  H = \sum_{j=1}^N \sqrt{(m_j + V_{s,j})^2 + (\bm{p}_j- \bm{V}_j)^2} + V_j^0,
\end{equation}
where $m_j$ is the mass of the $j$th particle, and the scalar and vector
one-particle potentials, $V_{s,j} := V_s(x_j)$ and $V_j^\mu :=
V^\mu(x_j)$, are defined by
Eqs.~\eqref{eq:QMDscalar} and \eqref{eq:QMDvector} using the interaction
density with the relativistic Gaussian.  The momentum-dependent one-particle
potential is given by half of the single-particle potential in
Eqs.~\eqref{eq:Usm} and~\eqref{eq:Uvm}.  See Ref.~\cite{Nara:2021fuu} for the
details.  The structure of the EoMs for RQMD is similar to that of the new
approach (RQMD2). Their main difference is the same as the QMD case; the
density of the potential is evaluated by the real particle density in the new
approach instead of the interaction density in the RQMD approach.  There are
two differences in the relativistic approaches: the RQMD uses the relativistic
Gaussian for the interaction density, while the new approach explicitly
computes the overlap of the relativistic Gaussian.  The new RQMD2 uses the
mass-shell condition with the single-particle potential, while the previous one
uses the one-particle potential.  We will see that the difference in the mass-shell
condition does not affect much the dynamics of heavy-ion collisions.
However, the mass-shell condition with the single-particle potential may become
relevant when chiral models are used to generate effective mass.

\section{Summary}
\label{sec:summary}

We have presented a novel formulation for the mean-field propagation part of
RQMD based on the variational
principle for interacting Gaussian wave packets for the first time.  We derived
new covariant canonical EoMs for an $N$-body system
interacting via Lorentz scalar and vector potentials.
We have first derived BUU-like EoMs, which are expressed by the single-particle
potentials, in the external and dynamical fields cases.
Then,
assuming the local density approximation described in Sec.~\ref{sec:covariant.static},
we have rewritten the EoMs to obtain QMD-like
EoMs, which are expressed by the one-particle potentials.
Furthermore, we proposed a new approximation for the spatial integral,
and showed that it provides an excellent approximation to the full integral.

The main differences from the previous approaches are as follows: 1) Our
approach imposes mass-shell constraints with the single-particle potential,
whereas previous approaches employed one-particle potentials.  2) Our method
uses the overlap of relativistic Gaussians.  3) It provides an accurate
estimation of the underlying EoS.

We have implemented these equations into the event generator \texttt{JAM2} to
perform event-by-event simulations of heavy-ion collisions.  These covariant
EoMs enable numerical simulation of heavy-ion collisions with computational
costs comparable to those of nonrelativistic quantum molecular dynamics.  We
demonstrated that the new EoMs accurately simulate density-dependent potential
in heavy-ion collisions by comparing the results with Monte Carlo integration
for the three-dimensional spatial integral.

In this work, we assumed the local density approximation to evaluate the scalar
and vector potentials in the numerical implementation.  As a future extension,
dynamical meson fields must be included by solving explicitly the spacetime
evolution of meson fields.

\begin{acknowledgments}
This work was supported in part by the Grants-in-Aid for Scientific Research
from JSPS (Nos. JP21K03577, JP25K07284, 
JP23K13102, 
and JP25KJ1584). 
This work was also supported by JST SPRING (No. JPMJSP2110). 
\end{acknowledgments}

\appendix

\section{Proof of Eq.~(\ref{eq:totpot.cancel})}
\label{app:totpot.cancel}
In this Appendix, we show Eq.~\eqref{eq:totpot.cancel}.  We first consider the
contribution of the variation of the generalized potential $\delta W(x,p)$:
\begin{align}
  & \int d^4x d^4p\, \delta W(x,p) f(x,p) \notag \\
  &= \int d^4x d^4p[ \delta p^{*\mu}(x,p) p_\mu^*(x,p) \notag \\ &\qquad\qquad
    - \delta m^*(x,p) m^*(x,p)] f(x,p) \notag \\
  & = \int d^4x d^4p[ -\delta U(x,p) \phiv_\mu(x,p) - \delta U_s(x,p) \phis(x,p)].
  \label{eq:app.totpot.cancel.W}
\end{align}
In the second equality, we used Eqs.~\eqref{def:effective-mass}
and~\eqref{def:kinetic-momentum} to rewrite $\delta p^*(x,p)$ and $\delta
m^*(x,p)$ and Eqs.~\eqref{eq:self-consistent.s}
and~\eqref{eq:self-consistent.v} to obtain $\phis(x,p)$ and $\phiv_\mu(x,p)$.

The variation of the field energy can be rewritten as
\begin{align}
  & \delta V_\mathrm{field}
    = \delta F[\phis, \phiv_\mu] \notag \\
  &\quad - \int d^4x d^4p \biggl\{\delta \phis(x,p) \frac{\delta F[\phis, \phiv_\mu]}{\delta \phis(x,p)}
    + \delta \phiv_\mu(x,p) \frac{\delta F[\phis, \phiv_\mu]}{\delta \phiv_\mu(x,p)} \biggr\} \notag \\
  &\quad - \int d^4x d^4p \biggl\{\phis(x,p) \delta \frac{\delta F[\phis, \phiv_\mu]}{\delta \phis(x,p)}
    + \phiv_\mu(x,p) \delta \frac{\delta F[\phis, \phiv_\mu]}{\delta \phiv_\mu(x,p)}\biggr\} \notag \\
  &\quad = -\int d^4x d^4p [\phis(x,p) \delta U_s(x,p) + \phiv_\mu(x,p) \delta U^\mu(x,p)].
  \label{eq:app.totpot.cancel.field}
\end{align}
In the second equality, we used the fact that the first and second terms cancel
with each other, and the third term was rewritten in terms of $U_s(x,p)$ and
$U^\mu(x,p)$ using Eqs.~\eqref{eq:sec3.spot} and~\eqref{eq:sec3.vpot}.

It is now clear that Eqs.~\eqref{eq:app.totpot.cancel.W}
and~\eqref{eq:app.totpot.cancel.field} cancel with each other.


\section{Monte Carlo integration with a relativistic Gaussian weight}
\label{sec:MCrelgauss}

In this Appendix, we explain how we perform the Monte Carlo integration of the
integral:
\begin{equation}
  I_i = \int d^4 x f(x) G(x-x_i;u_i),
  \label{eq:Ii}
\end{equation}
where $f(x)$ is an arbitrary function of the coordinates $x$, and
$G(x-x_i;u_i)$ is the relativistic Gaussian~\eqref{eq:gx}.

Without loss of generality, we may choose new coordinates $x'^\mu$ in the
rest frame of $\hat a$ so that the particle position is the origin, $x_i' = 0$.
This can be achieved by performing the following Poincar\'e transformation:
\begin{gather}
  x^\mu\mapsto
  x'^\mu = \Lambda^\mu{}_\nu (x^\nu - x^\nu_i),
\end{gather}
where $\Lambda$ is the Lorentz boost with the four-velocity $\hat a^\mu = (a^0,
\bm{a})^\mathrm{T}$:
\begin{align}
  \Lambda =
  \left(
  \begin{array}{c|c}
    a^0     & -\bm{a}^\mathrm{T} \\ \hline
    -\bm{a} & \mathbbm{1}+\frac{\bm{a}\bm{a}^\mathrm{T}}{a^0+1}
  \end{array}
  \right),
\end{align}
where $\mathbbm{1}$ is the $3\times3$ identity matrix.  The integral is written
as
\begin{align}
  I_i &= \int d^4 x' f(\Lambda^{-1} x'+x_i) \delta(x'\cdot\hat{a}')
  \dfrac{u'\cdot \hat{a}'}{(2\pi L)^{3/2}}
     e^{x'_\mu \Delta'^{\mu\nu}_i x'_\nu/2L} \nonumber\\
  &= \int d^3 x' f(\Lambda^{-1} x'+x_i)|_{x'^0=0}
  \dfrac{\gamma_i}{(2\pi L)^{3/2}} e^{-x'^a A^{ab} x'^b/2L},
\end{align}
where $\hat{a}' = (1,0,0,0)$, $\gamma_i= u'_i\cdot \hat{a}'= u'^0_i$,
$\Delta'^{\mu\nu}_i = g^{\mu\nu} - u'^\mu_i u'^\nu_i$, $A^{ab} = -
\Delta'^{ab}_i$, and
\begin{align}
u'^\mu &=
\begin{pmatrix}
 a^\mu u_{i\mu} \\
 -u_i^0\bm{a} + \bigl(1+\frac{\bm{a}\bm{a}^\mathrm{T}}{a^0+1}\bigr)\bm{u}_i
\end{pmatrix}.
\end{align}

Next, we complete the square for the quadratic form in the exponent by
diagonalizing the $3\times 3$ coefficient matrix $A^{ab}$, ($a,b=1,2,3$).  The
matrix can be decomposed into the components of its eigenspaces as
\begin{gather}
  A = P_\perp + \gamma^2_i P_u,
\end{gather}
using the projectors,
\begin{gather}
  (P_\perp)^{ab} = \delta^{ab} - (P_u)^{ab}, \qquad
  (P_u)^{ab} = \hat{u}^a_i \hat{u}^b_i,
\end{gather}
with $\hat{u}^a_i := u'^a_i/\sqrt{\gamma^2_i - 1}$.  In the following argument,
duplicate indices $a$, $b$, or $c$ appearing in a term imply the sum over $1$,
$2$, $3$.  Using the properties of the projectors,
\begin{equation}
  P_\perp^2 = P_\perp, \qquad
  P_\perp P_u = 0, \qquad
  P_u^2 = P_u,
\end{equation}
one can confirm that the square root of the matrix $A$ and its inverse become
\begin{align}
  (A^{1/2})^{ab} &= (P_\perp + \gamma_i P_u)^{ab}, \\
  (A^{-1/2})^{ab} &= \left(P_\perp + \frac{P_u}{\gamma_i}\right)^{ab}
  = \delta^{ab} - \frac{u'^a_i u'^b_i}{\gamma_i(\gamma_i+1)}.
\end{align}
Noting that $\det(A^{-1/2})=1/\gamma_i$, we may simplify the expression of the
integral $I_i$ by the change of variables, $\bm{x}'^a\mapsto \bm{x}' =
\sqrt{L}A^{-1/2}\tilde{\bm{x}}$:
\begin{equation}
  \label{eq:MCint}
  I_i = \int d^3 \tilde{x} f(T(\tilde{x};x_i,u_i))
  \dfrac{1}{(2\pi)^{3/2}} e^{-\tilde{x}^2/2},
\end{equation}
where
\begin{equation}
  T(\tilde{x};x_i,u_i)= x_i + \sqrt{L}\Lambda^{-1}
  \begin{pmatrix}
  0 \\ A^{-1/2} \tilde{\bm{x}}
  \end{pmatrix}.
\end{equation}

Based on the above consideration, we implement the Monte Carlo integration in
the following procedure: First, we sample the coordinates
$\tilde{\bm{x}}=(\tilde{x}_1,\tilde{x}_2,\tilde{x}_3)$ according to the
Gaussian distribution $\exp(-\tilde{x}^2/2)$ and transform it as
\begin{equation}
 \bar{x}^a =
      \sqrt{L}(A^{-1/2})^{ab} \tilde{x}^b
  = \sqrt{L}\biggl[
     \delta^{ab} - \frac{u'^a_i u'^b_i}{\gamma_i(\gamma_i+1)}
     \biggr]\tilde{x}^b.
\end{equation}
To obtain the $n$th Monte Carlo sample point $x_{(n)}^\mu$ in the computational frame, we perform
the Poincar\'e transformation:
\begin{align}
 x_{(n)}^0 &= x^0_i + \hat{a}^a \bar{x}^a,\nonumber\\
 x_{(n)}^a &= x^a_i + \biggl(\delta^{ab} + \frac{\hat{a}^a \hat{a}^b}{\hat{a}^0+1}\biggr)\bar{x}^b.
\end{align}
We repeat this procedure $N_\mathrm{MC}$ times and take the average to obtain a
value for the integral:
\begin{equation}
  I_i \approx \frac{1}{N_\mathrm{MC}}
   \sum^{N_{\mathrm{MC}}}_{n=1} f(x_{(n)}).
\end{equation}

\section{Gaussian overlap integral}
\label{sec:GaussianOverLap}

We evaluate the integral of the overlap of two relativistic Gaussians,
\begin{equation}
  \tilde{g}_{ij}(x_i-x_j) = \int d^4x\,\delta((x-x_i)\cdot\hat{a}) \tilde{g}(x-x_j) \tilde{g}(x-x_i),
  \label{eq:gaussoverlap}
\end{equation}
where
\begin{equation}
  \tilde{g}(x-x_i)= \frac{1}{(2\pi L)^{3/2}} \exp\frac{R(x-x_i;u_i)^2}{2L}.
\end{equation}
Note that because of the translational invariance, $\tilde{g}_{ij}$ is a function of
the relative distance $x_i-x_j$.  We go to the frame $\hat{a}=(1,0,0,0)$ so
that the temporal components vanish: $x^0 - x^0_i = x^0 - x^0_j = 0$.  The
Gaussian overlap~\eqref{eq:gaussoverlap} becomes
\begin{align}
  \tilde{g}_{ij}(0,\bm{r})
  &= \frac{1}{(2\pi L)^3}
    \int d^3 x e^{ -\frac1{2L}Q},
\end{align}
where $\bm{r}=\bm{x}_i - \bm{x}_j$,
\begin{align}
  Q &= (\bm{x} - \bm{r})^\mathrm{T} A(\bm{x} - \bm{r}) + \bm{x}^\mathrm{T} B\bm{x},
\end{align}
and
\begin{align}
  A &= \mathbbm{1} + \bm{u}_i\bm{u}_i^\mathrm{T}, \qquad
  B = \mathbbm{1} + \bm{u}_j\bm{u}_j^\mathrm{T},
\end{align}
with $\mathbbm{1}$ being the $3\times3$ identity matrix.

The square in $Q$ with respect to $\bm{x}$ may be completed as
\begin{align}
  Q
  &= [\bm{x} - (A+B)^{-1}A \bm{r}]^\mathrm{T}(A+B)[\bm{x} - (A+B)^{-1}A \bm{r}] \notag \\
  &\quad + \bm{r}^\mathrm{T} A \bm{r} - \bm{r}^\mathrm{T} A(A+B)^{-1}A\bm{r} \notag \\
  &= [\bm{x} - (A+B)^{-1}A \bm{r}]^\mathrm{T}(A+B)[\bm{x} - (A+B)^{-1}A \bm{r}] \notag \\
  &\quad + \bm{r}^\mathrm{T} A(A+B)^{-1} B \bm{r}.
\end{align}
The relative kernel can be integrated as
\begin{align}
  \tilde{g}_{ij}(\bm{r}, 0)
  = \frac{e^{-\frac1{2L}\bm{r}^\mathrm{T} A(A+B)^{-1} B \bm{r}}}{(2\pi L)^{3/2}\det(A+B)^{1/2}}.
\end{align}

Let us consider the explicit form of $(A+B)^{-1}$ and $A(A+B)^{-1}B$.  The
$3\times3$ matrices $A$ and $B$ can be written as
\begin{align}
  A &:= \mathbbm{1} + a^2 \pi_a, \qquad B := \mathbbm{1} + b^2 \pi_b,
\end{align}
with $a = |\bm{u}_i|$ and $b = |\bm{u}_j|$, and the projectors onto the
direction of $\bm{u}_i$ and $\bm{u}_j$:
\begin{align}
  \pi_a &:= \frac{\bm{u}_i\bm{u}_i^\mathrm{T}}{a^2}, \qquad
  \pi_b := \frac{\bm{u}_j\bm{u}_j^\mathrm{T}}{b^2}.
\end{align}
Those projectors satisfy $\pi_a^2 = \pi_a$ and $\pi_b^2 = \pi_b$ but are not
orthogonal to each other, $\pi_a \pi_b \neq 0$. We define the anticommutator
of the projectors as
\begin{align}
  \rho &:= \pi_a\pi_b + \pi_b\pi_a = c\frac{\bm{u}_i \bm{u}_j^\mathrm{T} + \bm{u}_j \bm{u}_i^\mathrm{T}}{ab},
\end{align}
where $c := \cos\theta = \bm{u}_i\cdot\bm{u}_j/ab$ with $\theta$ being the
angle between $\bm{u}_i$ and $\bm{u}_j$.
It is useful to define
\begin{align}
  X &= \frac{A+B}2 = \mathbbm{1} + \frac{a^2\pi_a + b^2\pi_b}2.
\end{align}
To write down the components of $X$, we choose the basis as $\bm{e}_1 = \hat
u_i = \bm{u}_i/a$ and $\bm{e}_2 \propto (1-\pi_a) \hat u_j$. Here, $\bm{u}_i$
and $\bm{u}_j$ are written as
\begin{align}
  \bm{u}_i &= a\begin{pmatrix}
    1 \\ 0 \\ 0
  \end{pmatrix}, &
  \bm{u}_j &= b\begin{pmatrix}
    c \\ s \\ 0
  \end{pmatrix},
\end{align}
where
\begin{align}
  c &:= \cos\theta = \frac{\bm{u}_i^\mathrm{T}\bm{u}_j}{ab}, \qquad
  s := \sin\theta = \sqrt{1-c^2},
\end{align}
and $\theta$ is the angle between $\bm{u}_i$ and $\bm{u}_j$. In this basis, $X$
is written as
\begin{align}
  X &= \begin{pmatrix}
    1+\frac{a^2 + b^2c^2}2 & \frac{b^2cs}2 & 0 \\
    \frac{b^2cs}2 & 1+\frac{b^2s^2}2 & 0 \\
    0 & 0 & 1
  \end{pmatrix}.
\end{align}

Now, we consider the Cayley--Hamilton theorem for a $3\times3$ matrix $X$:
\begin{align}
  X^3 - t X^2 + u X - d\mathbbm{1} = 0,
  \label{eq:half.finite.CH3}
\end{align}
where $t$, $u$, and $d$ are the coefficients of the characteristic equation
$\det(\lambda\mathbbm{1} - X) = 0$:
  \begin{align}
    t = \tr X &= 3 + \frac{a^2 + b^2}2, \\
    d = \det X
      &= 1 + \frac{a^2 + b^2}2 + \frac{a^2b^2s^2}4 \notag \\
      &= \Bigl(1+ \frac{a^2}2\Bigr)\Bigl(1+ \frac{b^2}2\Bigr) - \frac{a^2b^2c^2}4,
  \end{align}
  \begin{align}
    u &= \begin{vmatrix}
        1+\frac{a^2 + b^2c^2}2 & \frac{b^2cs}2 \\
        \frac{b^2cs}2 & 1+\frac{b^2s^2}2
      \end{vmatrix}
      + \begin{vmatrix}
        1+\frac{b^2s^2}2 & 0 \\ 0 & 1
      \end{vmatrix} \notag \\
    &\quad
      + \begin{vmatrix}
        1 & 0 \\ 0 & 1+\frac{a^2 + b^2c^2}2
      \end{vmatrix} \notag \\
      &= d + X_{11} + X_{22} = d + t - 1.
  \end{align}
The inverse of $X$ can be obtained using Eq.~\eqref{eq:half.finite.CH3}:
\begin{align}
  X^{-1}
    &= \frac1d(X^2 - t X + u\mathbbm{1}) \notag \\
    &= \mathbbm{1} + \frac1d[X^2 - \mathbbm{1} + t (X-\mathbbm{1})] \notag \\
    &= \mathbbm{1} + \frac1d(X-\mathbbm{1})[X+(1-t)\mathbbm{1}] \notag \\
    &= \mathbbm{1} + \frac{a^2\pi_a + b^2\pi_b}{2d} \biggl[\frac{a^2\pi_a + b^2\pi_b}2 - \Bigl(1 + \frac{a^2+b^2}2\Bigr)\mathbbm{1}\biggr] \notag \\
    &= \mathbbm{1} - \frac1{2d}\biggl[
      a^2\pi_a + b^2\pi_b + \frac{a^2b^2}2(\pi_a + \pi_b - \rho)\biggr] \notag \\
    &= \mathbbm{1} - \frac1{2d}\tau
    + \frac{a^2 b^2}{4d}\rho.
  \label{eq:app.half.finite.rker.invX}
\end{align}

We next consider $A - A(A+B)^{-1}A = (1/2)AX^{-1}B$\@.  Since the result is
a symmetric matrix by definition, we can calculate it ignoring the difference of
the antisymmetric part and pick up the symmetric part of the final
result, i.e., $M = M^\mathrm{T} + (M-M^\mathrm{T}) \equiv M^\mathrm{T}$ for an
arbitrary matrix $M$, where the symbol `$\equiv$' represents equivalence
ignoring antisymmetric differences.  For example, we can use the relations $M_1 M_2
\equiv M_2^\mathrm{T} M_1^\mathrm{T}$ and $\pi_a\pi_b \equiv \rho/2$.
\begin{align}
  AX^{-1}B
  &\equiv (\underbrace{\mathbbm{1} + a^2\pi_a + b^2\pi_b}_{2X-1})X^{-1} + a^2b^2 \pi_a X^{-1} \pi_b \notag \\
  &= 2\mathbbm{1} - X^{-1} + a^2b^2 \pi_a X^{-1} \pi_b.
  \label{eq:app.half.finite.rker.AXB}
\end{align}
Applying $\pi_a(\pi_a, \pi_b, \rho)\pi_b = (1, 1, 1+c^2)\pi_a\pi_b$
to the second last line of Eq.~\eqref{eq:app.half.finite.rker.invX}, we obtain
\begin{align}
  \pi_a X^{-1}\pi_b
    &= \pi_a\pi_b \biggl\{1 - \frac1{2d}\biggl[
      a^2 + b^2 + \frac{a^2b^2}2(1-c^2)\biggr]\biggr\} \notag\\
    &=\pi_a\pi_b\frac{2d - a^2 - b^2 - a^2b^2s^2/2}{2d} \notag\\
    &= \pi_a\pi_b \frac{2}{2d} \equiv \frac{\rho}{2d}.
  \label{eq:app.half.finite.rker.aXb}
\end{align}
Substituting Eq.~\eqref{eq:app.half.finite.rker.aXb} in
Eq.~\eqref{eq:app.half.finite.rker.AXB}, we obtain
\begin{align}
  AX^{-1}B
    &\equiv 2\mathbbm{1} - X^{-1} + \frac{a^2b^2}{2d}\rho \notag \\
    &= \mathbbm{1} + \frac{1}{2d}\tau + \frac{a^2b^2}{4d} \rho.
\end{align}
The right-hand side is already in a symmetric form, so it matches $AX^{-1}B$.

Finally, since $(A+B)^{-1} = 2^{-1}[(A+B)/2]^{-1}$, we get
\begin{align}
  (A+B)^{-1} &= \frac12 X^{-1}, \\
  A(A+B)^{-1}B &= \frac12 AX^{-1}B.
\end{align}
Using those matrices, $(A+B)^{-1}$ and $A(A+B)^{-1}B$ can be written as
\begin{align}
  (A+B)^{-1}
    &= \frac12\biggl[\mathbbm{1} + \frac1{2d}\Bigr(-\tau + \frac{a^2 b^2}{2}\rho\Bigr)\biggr],
    \label{eq:half.finite.rker.cmat1}\\
  A(A+B)^{-1}B
    &= \frac12\biggl[\mathbbm{1} + \frac1{2d}\Bigr(\tau + \frac{a^2 b^2}{2}\rho\Bigr)\biggr],
    \label{eq:half.finite.rker.cmat2}
\end{align}
where
\begin{align}
  \tau &:= \Bigl(1+\frac{b^2}2\Bigr)a^2\pi_a + \Bigl(1+\frac{a^2}2\Bigr)b^2\pi_b.
\end{align}
The exponent is obtained as
\begin{multline}
  -\frac1{2L}\bm{r}^\mathrm{T} A(A+B)^{-1}B \bm{r}
  = -\frac1{4L}\biggl\{-r^2 + \frac1{2d} \\ \times \biggl[
    \Bigl(1+\frac{b^2}2\Bigr)t_a^2 +
    \Bigl(1+\frac{a^2}2\Bigr)t_b^2 +
    abc t_a t_b\biggr]\biggr\},
  \label{eq:half.finite.rker.exponent}
\end{multline}
where $r^2 = -\bm{r}^\mathrm{T}\bm{r}$, $t_a = -\bm{r}^\mathrm{T}\bm{u}_i$, and
$t_b = -\bm{r}^\mathrm{T} \bm{u}_j$.
Note that $d = \det[(A+B)/2]$, thus $\det(A+B) = 2^3 d$.

When we perform a simulation with the constraints $x_i\cdot \hat a = s$,
we can define the relative kernel in the $\hat a$ frame [i.e., the inertial
frame in which $\hat a = (1, 0, 0, 0)^\mathrm{T}$]. Since $r^0 = x_i^0 - x_j^0
= s - s = 0$ in the $\hat a$ frame, we may write
$\bm{r}^\mathrm{T}\bm{v}$ (with $v^\mu$ being an arbitrary Lorentz vector) in a
covariant form as $\bm{r}^\mathrm{T}\bm{v} = -(r^0v^0 -
\bm{r}^\mathrm{T}\bm{v}) = -r^\mu v_{\mu}$. Therefore, we can evaluate the
exponent~\eqref{eq:half.finite.rker.exponent} in a covariant way using
\begin{align}
  a^2 &= -(g_{\mu\nu} - \hat a_\mu \hat a_\nu) u_i^\mu u_i^\nu \notag \\
    &= -u_i^2 + (u_i\cdot \hat a)^2, \\
  b^2 &= -(g_{\mu\nu} - \hat a_\mu \hat a_\nu) u_j^\mu u_j^\nu \notag \\
    &= -u_j^2 + (u_j\cdot \hat a)^2, \\
  c &= -(g_{\mu\nu} - \hat a_\mu \hat a_\nu) u_i^\mu u_j^\nu/ab \notag \\
    &= \frac{-u_i\cdot u_j + (u_i\cdot\hat a)(u_j\cdot \hat a)}{ab}, \\
  r^2 &= r\cdot r, \\
  t_a &= r\cdot u_i, \\
  t_b &= r\cdot u_j.
\end{align}

\section{Nonlinear vector potential}
\label{sec:nlvector}

We consider the vector potential of the form
\begin{equation}
V(\omega)=-\frac{m_\omega^2}{2}\omega^2 - \frac{c_4}{4}\omega^4\,.
\end{equation}
The gap equation is
\begin{equation}
  m_\omega^2 \omega^\mu + c_4\omega^2\omega^\mu = g_\omega J^\mu.
\end{equation}
The differentiation with respect to $x_i$ yields
\begin{equation}
 [ (m_\omega^2+c_4\omega^2) g^{\mu\nu} + 2c_4\omega^\mu\omega^\nu]\frac{\partial
\omega_\nu}{\partial x_i} = g_\omega \frac{\partial J^\mu}{\partial x_i}.
\end{equation}
The inverse is given by
\begin{align}
&[(m^2 + c\omega^2)g^{\mu\nu} + 2c\omega^\mu \omega^\nu]^{-1}\nonumber\\
&= \frac{(m^2+3c\omega^2)g_{\mu\nu} - 2c\omega_\mu \omega_\nu}{(m^2+3c\omega^2)(m^2+c\omega^2)}\nonumber \\
&= \frac{1}{m^2+3c\omega^2} \biggl[
  g_{\mu\nu} + \frac{2c(\omega^2 g_{\mu\nu} - \omega_\mu \omega_\nu)}{m^2+c\omega^2}\biggr],
\end{align}
or by another expression,
\begin{align}
& [(m^2 + c\omega^2)g^{\mu\nu} + 2c\omega^\mu \omega^\nu]^{-1} \notag \\
&\quad = \frac{1}{m^2+c\omega^2} \biggl(
  g_{\mu\nu} - \frac{2c\omega_\mu\omega_\nu}{m^2+3c\omega^2}\biggr),
\end{align}
which is obtained by separating the transverse and longitudinal component.
When a matrix is decomposed into the transverse component $a$ and the
longitudinal component $b$,
\begin{equation}
A^{\mu\nu} = a\biggl(g^{\mu\nu} - \frac{k^\mu k^\nu}{k^2}\biggr) +b \frac{k^\mu k^\nu}{k^2},
\end{equation}
and its inverse is given by
\begin{equation}
A^{-1}_{\mu\nu}=\frac1a\biggl(g_{\mu\nu} - \frac{k_\mu k_\nu}{k^2}\biggr) +\frac1b \frac{k^\mu k^\nu}{k^2}.
\end{equation}
One can confirm $A^{\mu\lambda}A^{-1}_{\lambda \nu}
=g^{\mu\lambda}g_{\lambda\nu}$.  In our case, $a=m_\omega^2+c_4\omega^2$ and
$b=m_\omega^2+3c_4\omega^2$.  Thus, the derivative of the $\omega$ field is
expressed as the linear combination of the current:
\begin{equation}
\frac{\partial \omega_\mu}{\partial x_i}
= \frac{g_\omega}{m^2+3c\omega^2} \biggl[
  g_{\mu\nu} + \frac{2c(\omega^2 g_{\mu\nu} - \omega_\mu \omega_\nu)}{m^2+c\omega^2}\biggr]
   \frac{\partial J^\nu}{\partial x_i}.
\end{equation}

The derivative with respect to $J_\lambda$ yields
\begin{equation}
 [ (m_\omega^2+c_4\omega^2) g^{\mu\nu} + 2c_4\omega^\mu\omega^\nu]\frac{\partial
\omega_\nu}{\partial J_\lambda} = g_\omega \frac{\partial J^\mu}{\partial J_\lambda}
 = g_\omega g^{\mu\lambda},
\end{equation}
\begin{align}
\frac{\partial \omega_\nu}{\partial J_\lambda}
&= \frac{g_\omega}{m^2+c\omega^2} \biggl(
  g_{\mu\nu} - \frac{2c\omega_\mu \omega_\nu}{m^2+3c\omega^2}\biggr)
   g^{\mu\lambda}
 \nonumber\\
&= \frac{g_\omega}{m^2+c\omega^2} \biggl(
  \delta^\lambda_\nu - \frac{2c\omega^\lambda \omega_\nu}{m^2+3c\omega^2}\biggr).
\end{align}

\bibliography{crqmdref}

@article{Bertsch:1984gb,
    author = "Bertsch, G. F. and Kruse, H. and Gupta, S. D.",
    title = "{BOLTZMANN EQUATION FOR HEAVY ION COLLISIONS}",
    doi = "10.1103/PhysRevC.33.1107",
    journal = "Phys. Rev. C",
    volume = "29",
    pages = "673--675",
    year = "1984",
    note = "[Erratum: Phys.Rev.C 33, 1107--1108 (1986)]"
}

@article{Aichelin:1985zz,
    author = "Aichelin, J. and Bertsch, G.",
    title = "{Numerical simulation of medium energy heavy ion reactions}",
    doi = "10.1103/PhysRevC.31.1730",
    journal = "Phys. Rev. C",
    volume = "31",
    pages = "1730--1738",
    year = "1985"
}

@article{Kruse:1985hy,
    author = "Kruse, H. and Jacak, B. V. and Stoecker, Horst",
    title = "{Microscopic theory of pion production and sidewards flow in heavy ion collisions}",
    doi = "10.1103/PhysRevLett.54.289",
    journal = "Phys. Rev. Lett.",
    volume = "54",
    pages = "289--292",
    year = "1985"
}

@article{Kruse:1985pg,
    author = "Kruse, H. and Jacak, B. V. and Molitoris, J. J. and Westfall, G. D. and Stoecker, Horst",
    title = "{VLASOV-UEHLING-UHLENBECK THEORY OF MEDIUM-ENERGY HEAVY ION REACTIONS: ROLE OF MEAN FIELD DYNAMICS AND TWO-BODY COLLISIONS}",
    doi = "10.1103/PhysRevC.31.1770",
    journal = "Phys. Rev. C",
    volume = "31",
    pages = "1770--1774",
    year = "1985"
}

@article{Bertsch:1988ik,
    author = "Bertsch, G. F. and Das Gupta, S.",
    title = "{A Guide to microscopic models for intermediate-energy heavy ion collisions}",
    doi = "10.1016/0370-1573(88)90170-6",
    journal = "Phys. Rept.",
    volume = "160",
    pages = "189--233",
    year = "1988"
}

@article{Cassing:1990dr,
    author = "Cassing, W. and Metag, V. and Mosel, U. and Niita, K.",
    title = "{Production of energetic particles in heavy ion collisions}",
    doi = "10.1016/0370-1573(90)90164-W",
    journal = "Phys. Rept.",
    volume = "188",
    pages = "363--449",
    year = "1990"
}

@article{Danielewicz:1991dh,
    author = "Danielewicz, P. and Bertsch, G. F.",
    title = "{Production of deuterons and pions in a transport model of energetic heavy ion reactions}",
    reportNumber = "MSUCL-762",
    doi = "10.1016/0375-9474(91)90541-D",
    journal = "Nucl. Phys. A",
    volume = "533",
    pages = "712--748",
    year = "1991"
}

@article{SMASH:2016zqf,
    author = "Weil, J. and others",
    collaboration = "SMASH",
    title = "{Particle production and equilibrium properties within a new hadron transport approach for heavy-ion collisions}",
    eprint = "1606.06642",
    archivePrefix = "arXiv",
    primaryClass = "nucl-th",
    doi = "10.1103/PhysRevC.94.054905",
    journal = "Phys. Rev. C",
    volume = "94",
    number = "5",
    pages = "054905",
    year = "2016"
}

@article{Ko:1987gp,
    author = "Ko, C. M. and Li, Q. and Wang, Ren-Chuan",
    title = "{Relativistic Vlasov Equation for Heavy Ion Collisions}",
    doi = "10.1103/PhysRevLett.59.1084",
    journal = "Phys. Rev. Lett.",
    volume = "59",
    pages = "1084--1087",
    year = "1987"
}

@article{Blattel:1988zz,
    author = "Blattel, Bernhard and Koch, Volker and Cassing, Wolfgang and Mosel, Ulrich",
    title = "{Covariant Boltzmann-Uehling-Uhlenbeck approach for heavy-ion collisions}",
    doi = "10.1103/PhysRevC.38.1767",
    journal = "Phys. Rev. C",
    volume = "38",
    pages = "1767--1775",
    year = "1988"
}

@article{Fuchs:1995fa,
    author = "Fuchs, C. and Wolter, H. H.",
    title = "{The Relativistic Landau-Vlasov method in heavy ion collisions}",
    doi = "10.1016/0375-9474(95)00180-9",
    journal = "Nucl. Phys. A",
    volume = "589",
    pages = "732--756",
    year = "1995"
}

@article{Weber:1992qc,
    author = "Weber, K. and Blaettel, B. and Cassing, W. and Doenges, H. C. and Koch, V. and Lang, A. and Mosel, U.",
    title = "{A Relativistic effective interaction for heavy ion collisions}",
    doi = "10.1016/0375-9474(92)90134-6",
    journal = "Nucl. Phys. A",
    volume = "539",
    pages = "713--751",
    year = "1992"
}

@article{Blaettel:1993uz,
    author = "Blaettel, B. and Koch, V. and Mosel, U.",
    title = "{Transport theoretical analysis of relativistic heavy ion collisions}",
    doi = "10.1088/0034-4885/56/1/001",
    journal = "Rept. Prog. Phys.",
    volume = "56",
    pages = "1--62",
    year = "1993"
}

@article{Cassing:2009vt,
    author = "Cassing, W. and Bratkovskaya, E. L.",
    title = "{Parton-Hadron-String Dynamics: an off-shell transport approach for relativistic energies}",
    eprint = "0907.5331",
    archivePrefix = "arXiv",
    primaryClass = "nucl-th",
    doi = "10.1016/j.nuclphysa.2009.09.007",
    journal = "Nucl. Phys. A",
    volume = "831",
    pages = "215--242",
    year = "2009"
}

@article{Buss:2011mx,
    author = "Buss, O. and Gaitanos, T. and Gallmeister, K. and van Hees, H. and Kaskulov, M. and Lalakulich, O. and Larionov, A. B. and Leitner, T. and Weil, J. and Mosel, U.",
    title = "{Transport-theoretical Description of Nuclear Reactions}",
    eprint = "1106.1344",
    archivePrefix = "arXiv",
    primaryClass = "hep-ph",
    doi = "10.1016/j.physrep.2011.12.001",
    journal = "Phys. Rept.",
    volume = "512",
    pages = "1--124",
    year = "2012"
}

@article{Aichelin:1986wa,
    author = "Aichelin, J. and Stoecker, Horst",
    title = "{Quantum molecular dynamics. A Novel approach to N body correlations in heavy ion collisions}",
    doi = "10.1016/0370-2693(86)90916-0",
    journal = "Phys. Lett. B",
    volume = "176",
    pages = "14--19",
    year = "1986"
}

@article{Aichelin:1991xy,
    author = "Aichelin, J.",
    title = "{'Quantum' molecular dynamics: A Dynamical microscopic n body approach to investigate fragment formation and the nuclear equation of state in heavy ion collisions}",
    doi = "10.1016/0370-1573(91)90094-3",
    journal = "Phys. Rept.",
    volume = "202",
    pages = "233--360",
    year = "1991"
}

@article{Aichelin:2019tnk,
    author = "Aichelin, J. and Bratkovskaya, E. and Le F\`evre, A. and Kireyeu, V. and Kolesnikov, V. and Leifels, Y. and Voronyuk, V. and Coci, G.",
    title = "{Parton-hadron-quantum-molecular dynamics: A novel microscopic $n$ -body transport approach for heavy-ion collisions, dynamical cluster formation, and hypernuclei production}",
    eprint = "1907.03860",
    archivePrefix = "arXiv",
    primaryClass = "nucl-th",
    doi = "10.1103/PhysRevC.101.044905",
    journal = "Phys. Rev. C",
    volume = "101",
    number = "4",
    pages = "044905",
    year = "2020"
}

@article{Hartnack:1997ez,
    author = "Hartnack, C. and Puri, Rajeev K. and Aichelin, J. and Konopka, J. and Bass, S. A. and Stoecker, Horst and Greiner, W.",
    title = "{Modeling the many body dynamics of heavy ion collisions: Present status and future perspective}",
    eprint = "nucl-th/9811015",
    archivePrefix = "arXiv",
    reportNumber = "SUBATECH-96-13",
    doi = "10.1007/s100500050045",
    journal = "Eur. Phys. J. A",
    volume = "1",
    pages = "151--169",
    year = "1998"
}

@article{Bass:1998ca,
    author = "Bass, S. A. and others",
    title = "{Microscopic models for ultrarelativistic heavy ion collisions}",
    eprint = "nucl-th/9803035",
    archivePrefix = "arXiv",
    doi = "10.1016/S0146-6410(98)00058-1",
    journal = "Prog. Part. Nucl. Phys.",
    volume = "41",
    pages = "255--369",
    year = "1998"
}

@article{Maruyama:1990zz,
    author = "Maruyama, Toshiki and Ohnishi, Akira and Horiuchi, Hisashi",
    title = "{Quantum molecular dynamics study of fusion and its fade out in the O-16 + O-16 system}",
    doi = "10.1103/PhysRevC.42.386",
    journal = "Phys. Rev. C",
    volume = "42",
    pages = "386--394",
    year = "1990"
}

@article{Maruyama:1997rp,
    author = "Maruyama, Toshiki and Niita, Koji and Oyamatsu, Kazuhiro and Maruyama, Tomoyuki and Chiba, Satoshi and Iwamoto, Akira",
    title = "{Quantum molecular dynamics approach to the nuclear matter below the saturation density}",
    eprint = "nucl-th/9705039",
    archivePrefix = "arXiv",
    doi = "10.1103/PhysRevC.57.655",
    journal = "Phys. Rev. C",
    volume = "57",
    pages = "655--665",
    year = "1998"
}

@article{Sorge:1989dy,
    author = "Sorge, H. and Stoecker, Horst and Greiner, W.",
    title = "{Poincare Invariant Hamiltonian Dynamics: Modeling Multi - Hadronic Interactions in a Phase Space Approach}",
    doi = "10.1016/0003-4916(89)90136-X",
    journal = "Annals Phys.",
    volume = "192",
    pages = "266--306",
    year = "1989"
}

@article{Maruyama:1991bp,
    author = "Maruyama, T. and Huang, S. W. and Ohtsuka, N. and Li, Guo-Qiang and Faessler, A. and Aichelin, J.",
    title = "{Lorentz covariant description of intermediate-energy heavy ion reactions in relativistic quantum molecular dynamics}",
    doi = "10.1016/0375-9474(91)90468-L",
    journal = "Nucl. Phys. A",
    volume = "534",
    pages = "720--740",
    year = "1991"
}

@article{Maruyama:1996rn,
    author = "Maruyama, Tomoyuki and Niita, Koji and Maruyama, Toshiki and Chiba, Satoshi and Nakahara, Yasuaki and Iwamoto, Akira",
    title = "{Relativistic effects in the transverse flow in the molecular dynamics framework}",
    eprint = "nucl-th/9601010",
    archivePrefix = "arXiv",
    doi = "10.1143/PTP.96.263",
    journal = "Prog. Theor. Phys.",
    volume = "96",
    pages = "263--268",
    year = "1996"
}

@article{Mancusi:2009zz,
    author = "Mancusi, Davide and Niita, Koji and Maruyama, Tomoyuki and Sihver, Lembit",
    title = "{Stability of nuclei in peripheral collisions in the JAERI quantum molecular dynamics model}",
    doi = "10.1103/PhysRevC.79.014614",
    journal = "Phys. Rev. C",
    volume = "79",
    pages = "014614",
    year = "2009"
}

@article{Marty:2012vs,
    author = "Marty, Rudy and Aichelin, Jorg",
    title = "{Molecular dynamics description of an expanding $q$/$\bar{q}$ plasma with the Nambu\textendash{}Jona-Lasinio model and applications to heavy ion collisions at energies available at the BNL Relativistic Heavy Ion Collider and the CERN Large Hadron Collider}",
    eprint = "1210.3476",
    archivePrefix = "arXiv",
    primaryClass = "hep-ph",
    doi = "10.1103/PhysRevC.87.034912",
    journal = "Phys. Rev. C",
    volume = "87",
    number = "3",
    pages = "034912",
    year = "2013"
}

@article{Isse:2005nk,
    author = "Isse, M. and Ohnishi, A. and Otuka, N. and Sahu, P. K. and Nara, Y.",
    title = "{Mean-field effects on collective flows in high-energy heavy-ion collisions from AGS to SPS energies}",
    eprint = "nucl-th/0502058",
    archivePrefix = "arXiv",
    doi = "10.1103/PhysRevC.72.064908",
    journal = "Phys. Rev. C",
    volume = "72",
    pages = "064908",
    year = "2005"
}

@article{Fuchs:1996uv,
    author = "Fuchs, C. and Lehmann, E. and Sehn, L. and Scholz, F. and Kubo, T. and Zipprich, J. and Faessler, A.",
    title = "{Heavy ion collisions and the density dependence of the local mean field}",
    doi = "10.1016/0375-9474(96)80012-G",
    journal = "Nucl. Phys. A",
    volume = "603",
    pages = "471--485",
    year = "1996"
}

@article{Nara:2019qfd,
    author = "Nara, Yasushi and Stoecker, Horst",
    title = "{Sensitivity of the excitation functions of collective flow to relativistic scalar and vector meson interactions in the relativistic quantum molecular dynamics model RQMD.RMF}",
    eprint = "1906.03537",
    archivePrefix = "arXiv",
    primaryClass = "nucl-th",
    doi = "10.1103/PhysRevC.100.054902",
    journal = "Phys. Rev. C",
    volume = "100",
    number = "5",
    pages = "054902",
    year = "2019"
}

@article{Nara:2020ztb,
    author = "Nara, Yasushi and Maruyama, Tomoyuki and Stoecker, Horst",
    title = "{Momentum-dependent potential and collective flows within the relativistic quantum molecular dynamics approach based on relativistic mean-field theory}",
    eprint = "2004.05550",
    archivePrefix = "arXiv",
    primaryClass = "nucl-th",
    doi = "10.1103/PhysRevC.102.024913",
    journal = "Phys. Rev. C",
    volume = "102",
    number = "2",
    pages = "024913",
    year = "2020"
}

@article{Ono:1991uz,
    author = "Ono, A. and Horiuchi, H. and Maruyama, T. and Ohnishi, A.",
    title = "{Fragment formation studied with antisymmetrized version of molecular dynamics with two nucleon collisions}",
    reportNumber = "KUNS-1121",
    doi = "10.1103/PhysRevLett.68.2898",
    journal = "Phys. Rev. Lett.",
    volume = "68",
    pages = "2898--2900",
    year = "1992"
}

@article{Ono:1992uy,
    author = "Ono, Akira and Horiuchi, Hisashi and Maruyama, Toshiki and Ohnishi, Akira",
    title = "{Antisymmetrized version of molecular dynamics with two nucleon collisions and its application to heavy ion reactions}",
    reportNumber = "KUNS-1122",
    doi = "10.1143/PTP.87.1185",
    journal = "Prog. Theor. Phys.",
    volume = "87",
    pages = "1185--1206",
    year = "1992"
}

@article{TMEP:2016tup,
    author = "Xu, Jun and others",
    collaboration = "TMEP",
    title = "{Understanding transport simulations of heavy-ion collisions at 100A and 400A MeV: Comparison of heavy-ion transport codes 
under controlled conditions}",
    eprint = "1603.08149",
    archivePrefix = "arXiv",
    primaryClass = "nucl-th",
    doi = "10.1103/PhysRevC.93.044609",
    journal = "Phys. Rev. C",
    volume = "93",
    number = "4",
    pages = "044609",
    year = "2016"
}

@article{TMEP:2017mex,
    author = "Zhang, Ying-Xun and others",
    collaboration = "TMEP",
    title = "{Comparison of heavy-ion transport simulations: Collision integral in a box}",
    eprint = "1711.05950",
    archivePrefix = "arXiv",
    primaryClass = "nucl-th",
    doi = "10.1103/PhysRevC.97.034625",
    journal = "Phys. Rev. C",
    volume = "97",
    number = "3",
    pages = "034625",
    year = "2018"
}

@article{TMEP:2019yci,
    author = "Ono, Akira and others",
    collaboration = "TMEP",
    title = "{Comparison of heavy-ion transport simulations: Collision integral with pions and \ensuremath{\Delta} resonances in a box}",
    eprint = "1904.02888",
    archivePrefix = "arXiv",
    primaryClass = "nucl-th",
    doi = "10.1103/PhysRevC.100.044617",
    journal = "Phys. Rev. C",
    volume = "100",
    number = "4",
    pages = "044617",
    year = "2019"
}

@article{TMEP:2021ljz,
    author = "Colonna, Maria and others",
    collaboration = "TMEP",
    title = "{Comparison of heavy-ion transport simulations: Mean-field dynamics in a box}",
    eprint = "2106.12287",
    archivePrefix = "arXiv",
    primaryClass = "nucl-th",
    doi = "10.1103/PhysRevC.104.024603",
    journal = "Phys. Rev. C",
    volume = "104",
    number = "2",
    pages = "024603",
    year = "2021"
}

@article{TMEP:2022xjg,
    author = "Wolter, Hermann and others",
    collaboration = "TMEP",
    title = "{Transport model comparison studies of intermediate-energy heavy-ion collisions}",
    eprint = "2202.06672",
    archivePrefix = "arXiv",
    primaryClass = "nucl-th",
    doi = "10.1016/j.ppnp.2022.103962",
    journal = "Prog. Part. Nucl. Phys.",
    volume = "125",
    pages = "103962",
    year = "2022"
}

@article{TMEP:2023ifw,
    author = "Xu, Jun and others",
    collaboration = "TMEP",
    title = "{Comparing pion production in transport simulations of heavy-ion collisions at 270AMeV under controlled conditions}",
    eprint = "2308.05347",
    archivePrefix = "arXiv",
    primaryClass = "nucl-th",
    doi = "10.1103/PhysRevC.109.044609",
    journal = "Phys. Rev. C",
    volume = "109",
    number = "4",
    pages = "044609",
    year = "2024"
}

@article{Sorensen:2020ygf,
    author = "Sorensen, Agnieszka and Koch, Volker",
    title = "{Phase transitions and critical behavior in hadronic transport with a relativistic density functional equation of state}",
    eprint = "2011.06635",
    archivePrefix = "arXiv",
    primaryClass = "nucl-th",
    doi = "10.1103/PhysRevC.104.034904",
    journal = "Phys. Rev. C",
    volume = "104",
    number = "3",
    pages = "034904",
    year = "2021"
}

@article{Oliinychenko:2022uvy,
    author = "Oliinychenko, Dmytro and Sorensen, Agnieszka and Koch, Volker and McLerran, Larry",
    title = "{Sensitivity of Au+Au collisions to the symmetric nuclear matter equation~ of state at 2\textendash{}5 nuclear saturation densities}",
    eprint = "2208.11996",
    archivePrefix = "arXiv",
    primaryClass = "nucl-th",
    doi = "10.1103/PhysRevC.108.034908",
    journal = "Phys. Rev. C",
    volume = "108",
    number = "3",
    pages = "034908",
    year = "2023"
}

@article{Sudarshan:1981pp,
    author = "Sudarshan, E. C. G. and Mukunda, N. and Goldberg, J. N.",
    title = "{Constraint Dynamics of Particle World Lines}",
    reportNumber = "DOE-ER-03992-400",
    doi = "10.1103/PhysRevD.23.2218",
    journal = "Phys. Rev. D",
    volume = "23",
    pages = "2218",
    year = "1981"
}

@article{Samuel:1982jk,
    author = "Samuel, Joseph",
    title = "{Constraints in Relativistic Hamiltonian Mechanics}",
    reportNumber = "Print-82-0509 (BANGALORE)",
    doi = "10.1103/PhysRevD.26.3475",
    journal = "Phys. Rev. D",
    volume = "26",
    pages = "3475",
    year = "1982"
}

@article{Samuel:1982jn,
    author = "Samuel, Joseph",
    title = "{Relativistic Particle Models With Separable Interactions}",
    reportNumber = "Print-82-0508 (BANGALORE)",
    doi = "10.1103/PhysRevD.26.3482",
    journal = "Phys. Rev. D",
    volume = "26",
    pages = "3482",
    year = "1982"
}

@article{Ono:2019jxm,
    author = "Ono, Akira",
    title = "{Dynamics of clusters and fragments in heavy-ion collisions}",
    eprint = "1903.00608",
    archivePrefix = "arXiv",
    primaryClass = "nucl-th",
    doi = "10.1016/j.ppnp.2018.11.001",
    journal = "Prog. Part. Nucl. Phys.",
    volume = "105",
    pages = "139--179",
    year = "2019"
}

@article{Lenk:1989zz,
    author = "Lenk, R. J. and Pandharipande, V. R.",
    title = "{Nuclear mean field dynamics in the lattice Hamiltonian Vlasov method}",
    doi = "10.1103/PhysRevC.39.2242",
    journal = "Phys. Rev. C",
    volume = "39",
    pages = "2242--2249",
    year = "1989"
}

@preprint{Persram:1999ec,
    author = "Persram, Declan and Gale, Charles",
    title = "{A Study of splintering central nuclear collisions with a momentum dependent lattice Hamiltonian theory}",
    eprint = "nucl-th/9901019",
    archivePrefix = "arXiv",
    reportNumber = "MCGILL-98-44",
    month = "1",
    year = "1999"
}

@article{Persram:2001dg,
    author = "Persram, Declan and Gale, Charles",
    title = "{Elliptic flow in intermediate-energy heavy ion collisions and in-medium effects}",
    eprint = "nucl-th/0111035",
    archivePrefix = "arXiv",
    doi = "10.1103/PhysRevC.65.064611",
    journal = "Phys. Rev. C",
    volume = "65",
    pages = "064611",
    year = "2002"
}

@article{Wang:2019ghr,
    author = "Wang, Rui and Chen, Lie-Wen and Zhang, Zhen",
    title = "{Nuclear collective dynamics in the lattice Hamiltonian Vlasov method}",
    eprint = "1902.01256",
    archivePrefix = "arXiv",
    primaryClass = "nucl-th",
    doi = "10.1103/PhysRevC.99.044609",
    journal = "Phys. Rev. C",
    volume = "99",
    number = "4",
    pages = "044609",
    year = "2019"
}

@article{Yang:2021gwa,
    author = "Yang, Junping and Zhang, Yingxun and Wang, Ning and Li, Zhuxia",
    title = "{Influence of the treatment of initialization and mean-field potential on the neutron to proton yield ratios}",
    eprint = "2103.13132",
    archivePrefix = "arXiv",
    primaryClass = "nucl-th",
    doi = "10.1103/PhysRevC.104.024605",
    journal = "Phys. Rev. C",
    volume = "104",
    number = "2",
    pages = "024605",
    year = "2021"
}

@article{Ono:1993ac,
    author = "Ono, Akira and Horiuchi, Hisashi and Maruyama, Toshiki",
    title = "{Nucleon flow and fragment flow in heavy ion reactions}",
    eprint = "nucl-th/9308004",
    archivePrefix = "arXiv",
    reportNumber = "KUNS-1213",
    doi = "10.1103/PhysRevC.48.2946",
    journal = "Phys. Rev. C",
    volume = "48",
    pages = "2946--2955",
    year = "1994"
}

@article{Zhang:2014sva,
    author = "Zhang, Yingxun and Tsang, M. B. and Li, Zhuxia and Liu, Hang",
    title = "{Constraints on nucleon effective mass splitting with heavy ion collisions}",
    eprint = "1402.3790",
    archivePrefix = "arXiv",
    primaryClass = "nucl-th",
    doi = "10.1016/j.physletb.2014.03.030",
    journal = "Phys. Lett. B",
    volume = "732",
    pages = "186--190",
    year = "2014"
}

@article{Zhang:2020dvn,
    author = "Zhang, Yingxun and Wang, Ning and Li, Qing-Feng and Ou, Li and Tian, Jun-Long and Liu, Min and Zhao, Kai and Wu, Xi-Zhen and Li, Zhu-Xia",
    title = "{Progress of quantum molecular dynamics model and its applications in heavy ion collisions}",
    eprint = "2005.12877",
    archivePrefix = "arXiv",
    primaryClass = "nucl-th",
    doi = "10.1007/s11467-020-0961-9",
    journal = "Front. Phys. (Beijing)",
    volume = "15",
    pages = "54301",
    year = "2020"
}

@article{OmanaKuttan:2022the,
    author = "Omana Kuttan, Manjunath and Motornenko, Anton and Steinheimer, Jan and Stoecker, Horst and Nara, Yasushi and Bleicher, Marcus",
    title = "{A chiral mean-field equation-of-state in UrQMD: effects on the heavy ion compression stage}",
    eprint = "2201.01622",
    archivePrefix = "arXiv",
    primaryClass = "nucl-th",
    doi = "10.1140/epjc/s10052-022-10400-2",
    journal = "Eur. Phys. J. C",
    volume = "82",
    number = "5",
    pages = "427",
    year = "2022"
}

@article{Steinheimer:2022gqb,
    author = "Steinheimer, Jan and Motornenko, Anton and Sorensen, Agnieszka and Nara, Yasushi and Koch, Volker and Bleicher, Marcus",
    title = "{The high-density equation of state in heavy-ion collisions: constraints from proton flow}",
    eprint = "2208.12091",
    archivePrefix = "arXiv",
    primaryClass = "nucl-th",
    doi = "10.1140/epjc/s10052-022-10894-w",
    journal = "Eur. Phys. J. C",
    volume = "82",
    number = "10",
    pages = "911",
    year = "2022"
}

@article{Nara:2021fuu,
    author = "Nara, Yasushi and Ohnishi, Akira",
    title = "{Mean-field update in the JAM microscopic transport model: Mean-field effects on collective flow in high-energy heavy-ion collisions at sNN=2\textendash{}20 GeV energies}",
    eprint = "2109.07594",
    archivePrefix = "arXiv",
    primaryClass = "nucl-th",
    reportNumber = "YITP-21-97",
    doi = "10.1103/PhysRevC.105.014911",
    journal = "Phys. Rev. C",
    volume = "105",
    number = "1",
    pages = "014911",
    year = "2022"
}

@article{Nara:2023vrq,
    author = "Nara, Yasushi and Jinno, Asanosuke and Maruyama, Tomoyuki and Murase, Koichi and Ohnishi, Akira",
    title = "{Poincar\'e covariant cascade method for high-energy nuclear collisions}",
    eprint = "2306.12131",
    archivePrefix = "arXiv",
    primaryClass = "nucl-th",
    reportNumber = "YITP-23-77, KUNS-2969",
    doi = "10.1103/PhysRevC.108.024910",
    journal = "Phys. Rev. C",
    volume = "108",
    number = "2",
    pages = "024910",
    year = "2023"
}

@article{Nara:1999dz,
    author = "Nara, Y. and Otuka, N. and Ohnishi, A. and Niita, K. and Chiba, S.",
    title = "{Study of relativistic nuclear collisions at AGS energies from p + Be to Au + Au with hadronic cascade model}",
    eprint = "nucl-th/9904059",
    archivePrefix = "arXiv",
    doi = "10.1103/PhysRevC.61.024901",
    journal = "Phys. Rev. C",
    volume = "61",
    pages = "024901",
    year = "2000"
}

@software{JAM2,
  author = "Nara, Yasushi",
  title = {{JAM2: https://gitlab.com/transportmodel/jam2}},
  url ={https://gitlab.com/transportmodel/jam2}
}

@article{STAR:2020dav,
    author = "Adam, J. and others",
    collaboration = "STAR",
    title = "{Flow and interferometry results from Au+Au collisions at $\sqrt{s_{NN}} = 4.5$ GeV}",
    eprint = "2007.14005",
    archivePrefix = "arXiv",
    primaryClass = "nucl-ex",
    doi = "10.1103/PhysRevC.103.034908",
    journal = "Phys. Rev. C",
    volume = "103",
    number = "3",
    pages = "034908",
    year = "2021"
}

@article{Sjostrand:2014zea,
    author = {Sj\"ostrand, Torbj\"orn and Ask, Stefan and Christiansen, Jesper R. and Corke, Richard and Desai, Nishita and Ilten, Philip and Mrenna, Stephen and Prestel, Stefan and Rasmussen, Christine O. and Skands, Peter Z.},
    title = "{An introduction to PYTHIA 8.2}",
    eprint = "1410.3012",
    archivePrefix = "arXiv",
    primaryClass = "hep-ph",
    reportNumber = "LU-TP-14-36, MCNET-14-22, CERN-PH-TH-2014-190, FERMILAB-PUB-14-316-CD, DESY-14-178, SLAC-PUB-16122",
    doi = "10.1016/j.cpc.2015.01.024",
    journal = "Comput. Phys. Commun.",
    volume = "191",
    pages = "159--177",
    year = "2015"
}

@software{PYTHIA8,
  author = {Sj\"ostrand},
  title = "{{PYTHIA8}}",
  url ={https://www.pythia.org/}
}
\end{document}